# Roadmap on Photovoltaic Absorber Materials for Sustainable Energy Conversion

*James C. Blakesley,[1,†] Ruy S. Bonilla,[2,†] Marina Freitag,[3,†] Alex M. Ganose,[4, †] Nicola Gasparini,[4, †] Pascal Kaienburg,[5,†] George Koutsourakis,[1,†] Jonathan D. Major,[6, †] Jenny Nelson,[7,8 †] Nakita K. Noel,[5, †] Bart Roose,[9, †] Jae Sung Yun,[11, †] Simon Aliwell,[12] Pietro P. Altermatt[2,13] Tayebeh Ameri,[14,15] Virgil Andrei,[16,23] Ardalan Armin,[17] Diego Bagnis,[18] Jenny Baker,[19] Hamish Beath,[7,8] Mathieu Bellanger,[12] Philippe Berrouard,[20] Jochen Blumberger,[21] Stuart A. Boden,[22] Hugo Bronstein,[16,23] Matthew J. Carnie,[24,25] Chris Case,[26] Fernando A. Castro,[1] Yi-Ming Chang,[27] Elmer Chao,[28] Tracey M. Clarke,[29] Graeme Cooke,[30] Pablo Docampo,[30] Ken Durose,[6] James R. Durrant,[4,25] Marina R. Filip,[5] Richard H. Friend,[23] Jarvist M. Frost,[4] Elizabeth A. Gibson,[2] Alexander J. Gillett,[23] Pooja Goddard,[31] Severin N. Habisreutinger,[26] Martin Heeney,[4] Arthur D. Hendsbee,[20] Louise C. Hirst,[23,32] M. Saiful Islam,[2] K. D. G. Imalka Jayawardena,[11] Michael B. Johnston,[5] Matthias Kauer,[12] Jeff Kettle,[33] Ji-Seon Kim,[34] Dan Lamb,[35] David Lidzey,[36] Jihoo Lim,[11,37] Roderick MacKenzie,[38] Nigel Mason,[39] Iain McCulloch,[40] Keith P. McKenna,[41] Sebastian B. Meier,[42] Paul Meredith,[17] Graham Morse,[43] John D. Murphy,[44] Chris Nicklin,[45] Paloma Ortega-Arriaga,[7,8] Thomas Osterberg,[46] Jay B. Patel,[47] Anthony Peaker,[48] Moritz Riede,[5] Martyn Rush,[49] James W. Ryan,[50] David O. Scanlon,[29,51] Peter J. Skabara,[30] Franky So,[52,53] Henry J. Snaith,[5] Ludmilla Steier,[10] Jarla Thiesbrummel,[5] Alessandro Troisi,[54] Craig Underwood,[55] Karsten Walzer,[56] Trystan Watson,[25] J. Michael Walls,[57] Aron Walsh,[58] Lucy D. Whalley,[59] Benedict Winchester,[7,8] Samuel D. Stranks[9,*] and Robert L. Z. Hoye[10,58,*]*

[1.] National Physical Laboratory, Hampton Rd, Teddington, UK

[2.] Department of Materials, University of Oxford, Oxford OX1 3PH, UK

[3.] School of Natural and Environmental Sciences, Bedson Building, Newcastle University, NE1 7RU, Newcastle upon Tyne, UK

[4.] Department of Chemistry, Imperial College London, 80 Wood Lane, London W12 0BZ, UK

[5.] Department of Physics, University of Oxford, Clarendon Laboratory, Parks Road, Oxford OX1 3PU, UK

[6.] Stephenson Institute for Renewable Energy, Department of Physics, University of Liverpool, Chadwick Building, Peach Street, Liverpool L69 7ZF, United Kingdom

[7.] Department of Physics and Centre for Processable Electronics, Imperial College London, London SW7 2AZ, UK

[8.] Grantham Institute – Climate Change and the Environment, Imperial College London, London SW7 2AZ, UK

[9.] Department of Chemical Engineering and Biotechnology, University of Cambridge, Phillipa Fawcett Road, Cambridge CB3 0AS, United Kingdom

[10.] Inorganic Chemistry Laboratory, Department of Chemistry, University of Oxford, South Parks Road, Oxford OX1 3QR, United Kingdom

[11.] School of Computer Science and Electronic Engineering, Advanced Technology Institute (ATI), University of Surrey, Guildford GU2 7XH, United Kingdom

[12.] Lightricity Ltd., Edmund Halley Road, Oxford OX4 4GB, United Kingdom

[13.] Trina Solar Co Ltd, Changzhou, China.

[14.] Institute for Materials and Processes, School of Engineering, University of Edinburgh, Sanderson Building, Robert Stevenson Road, Edinburgh EH9 3FB, UK

[15] Chair for Composite Materials, Department of Materials Science, Faculty of Engineering, Christian-Albrechts-Universität zu Kiel, Kaiserstrasse 2, 24143 Kiel, Germany

[16] Yusuf Hamied Department of Chemistry, University of Cambridge, Cambridge CB2 1EW, UK

[17] Centre for Integrative Semiconductor Materials (CISM), Department of Physics, Swansea University Bay Campus, Swansea SA1 8EN, UK

[18] Oninn – Centro de Inovação, Av. José Candido da Silveira 2000 - Horto Florestal, Belo Horizonte - MG, 31035-536, Brazil

[19] Faculty of Science and Engineering, Swansea University, UK

[20] Brilliant Matters, 237 Rue de Liverpool, Saint-Augustin-de-Desmaures, QC G3A 2C8, Canada

[21] Department of Physics and Astronomy and Thomas Young Centre, University College London, Gower Street, London WC1E 6BT, UK

[22] School of Electronics and Computer Science, University of Southampton, United Kingdom

[23] Cavendish Laboratory, University of Cambridge, JJ Thomson Ave, Cambridge CB3 0HE, UK

[24] Department of Materials Science and Engineering, Swansea University, Bay Campus, Fabian Way, Crymlyn Burrows, Swansea SA1 8EN, UK

[25] SPECIFIC, Swansea University, Bay Campus, Swansea SA1 8EN, UK

[26] Oxford Photovoltaics Ltd., Unit 7-8, Oxford Pioneer Park, Mead Road, Oxford OX5 1QU, United Kingdom

[27] Raynergy Tek Incorporation, 60, Park Ave. 2, Hsinchu Science Park, Hsinchu 30844, Taiwan

[28] Luminescence Technology Corp., Hsinchu 30076, Taiwan

[29] Department of Chemistry, University College London, Christopher Ingold Building, London WC1H 0AJ, UK

[30] School of Chemistry, University of Glasgow, Joseph Black Building, University Place, Glasgow G12 8QQ, United Kingdom

[31] School of Science, Loughborough University, LE11 3TU, UK

[32] Department of Materials Science and Metallurgy, University of Cambridge, 27 Charles Babbage Road, Cambridge CB3 0FS, UK

[33] James Watt School of Engineering, University of Glasgow, Glasgow G12 8QQ, UK

[35] Oxide and Chalcogenide Facility, Centre for Integrative Semiconductor Materials, Faculty of Science and Engineering, Swansea, University, SA1 8EN United Kingdom

[36] Department of Physics and Astronomy, The University of Sheffield, Hicks Building, Hounsfield Road, Sheffield S3 7RH, UK

[37] Australian Centre for Advanced Photovoltaics (ACAP), School of Photovoltaic and Renewable Energy Engineering, University of New South Wales, Sydney, NSW 2052, Australia

[38] Department of Engineering, Durham University, Lower Mount Joy, South Road, Durham DH1 3LE, UK

[39] Formerly BP Solar and PV Consulting, United Kingdom.

[40] Department of Chemistry, University of Oxford, Mansfield Road, Oxford OX1 3QR, UK

[41] School of Physics, Engineering and Technology, University of York, Heslington, York YO10 5DD, UK

[42] ASCA GmbH & Co. KG Steigweg 24, 97318 Kitzingen, Germany

[43] Merck Chemicals Ltd, Chilworth Technical Centre, University Parkway, Southampton SO16 7QD, UK

[44] School of Engineering, University of Warwick, Coventry CV4 7AL, United Kingdom

[45] Diamond Light Source, Harwell Science and Innovation Campus, Didcot, Oxfordshire, OX11 0DE, UK

[46] Epishine AB, 582 13 Linköping, Sweden

[47] Department of Physics, King's College London, Strand, London WC2R 2LS, United Kingdom

[48] School of Electrical Engineering and Electronics, University of Manchester, United Kingdom.

[49] Polysolar Ltd, Aurora BAS, High Cross, Meadingley Road, Cambridge, CB3 0ET, UK

[50] Department of Chemistry, Swansea University, Swansea SA2 8PP, UK

[51] Thomas Young Centre, University College London, Gower St, London, UK

[52] Materials Science and Engineering, North Carolina State University, Raleigh, NC, 27695 USA

[53] Organic and Carbon Electronics Laboratories (ORaCEL), North Carolina State University, Raleigh, NC, 27695 USA

[54] Department of Chemistry and Materials Innovation Factory, University of Liverpool, Liverpool L69 7ZD, UK

[55] Surrey Space Centre, School of Computer Science and Electronic Engineering, University of Surrey, Guildford, Surrey, GU2 7XH, United Kingdom

[56] Heliatek GmbH, Treidlerstrasse 3, 01139 Dresden, Germany

[57] Centre for Renewable Energy Systems Technology (CREST), Loughborough University, Loughborough LE11 3TU, United Kingdom

[58] Department of Materials, Imperial College London, Exhibition Road, London SW7 2AZ, United Kingdom

[59] Department of Maths, Physics and Electrical Engineering, Northumbria University, Ellison Place, Newcastle Upon Tyne, NE1 8ST, UK

[†] These authors contributed equally to this roadmap article. Arranged alphabetically by surname. The remaining co-authors are also arranged alphabetically by surname

**Email:** sds65@cam.ac.uk (S.D.S), robert.hoye@chem.ox.ac.uk (R.L.Z.H.)

**Abstract**

Photovoltaics (PVs) are a critical technology for curbing growing levels of anthropogenic greenhouse gas emissions, and meeting increases in future demand for low-carbon electricity. In order to fulfil ambitions for net-zero carbon dioxide equivalent ($CO_2$eq) emissions worldwide, the global cumulative capacity of solar PVs must increase by an order of magnitude from 0.9 $TW_p$ in 2021 to 8.5 $TW_p$ by 2050 according to the International Renewable Energy Agency, which is considered to be a highly conservative estimate. In 2020, the Henry Royce Institute brought together the UK PV community to discuss the critical technological and infrastructure challenges that need to be overcome to address the vast challenges in accelerating PV deployment. Herein, we examine the key developments in the global community, especially the progress made in the field since this earlier roadmap, bringing together experts primarily from the UK across the breadth of the photovoltaics community. The focus is both on the challenges in improving the efficiency, stability and levelized cost of electricity of current technologies for utility-scale PVs, as well as the fundamental questions in novel technologies that can have a significant impact on emerging markets, such as indoor PVs, space PVs, and agrivoltaics. We discuss challenges in advanced metrology and computational tools, as well as the growing synergies between PVs and solar fuels, and offer a perspective on the environmental sustainability of the PV industry. Through this roadmap, we emphasize promising pathways forward in both the short- and long-term, and for communities working on technologies across a range of maturity levels to learn from each other.

**Contents**

## 1. Introduction

*Robert L. Z. Hoye,[1,2] Henry J. Snaith,[3] and Samuel D. Stranks[4]*

[1] Inorganic Chemistry Laboratory, Department of Chemistry, University of Oxford, South Parks Road, Oxford OX1 3QR, United Kingdom

[2] Department of Materials, Imperial College London, Exhibition Road, London SW7 2AZ, United Kingdom

[3] Department of Physics, University of Oxford, Clarendon Laboratory, Parks Road, Oxford OX1 3PU, UK

[4] Department of Chemical Engineering and Biotechnology, University of Cambridge, Phillipa Fawcett Road, Cambridge CB3 0AS, United Kingdom

Curbing the rise in anthropogenic greenhouse gas (GHG) emissions has become one of the most pressing challenges facing society, and 88 nations across the world have now pledged to achieve net-zero carbon dioxide equivalent ($CO_2$eq) emissions within a set timeframe [1]. Fulfilling these net-zero emissions ambitions requires a substantial increase in the deployment of electricity generation sources with no $CO_2$eq emissions during operation, which will be central to decarbonizing the power generation sector and meet the increased demand for low-carbon electricity in other sectors, especially surface transport and the heating and cooling of buildings [2]. Solar photovoltaics (PVs) are one of the most promising renewable energy technologies, and are the lowest cost source of electricity in many parts of the world already [3]. Long term price predictions have always underestimated the realized trends. It is clear however, that since the levelized cost of electricity (LCOE) for installed PVs today ranges from 0.06 to 0.01 € $kWh^{-1}$, depending upon the location in the world, the future cost of PV electricity will asymptotically approach the land cost. Although PV energy production now exceeds $10^3$ TWh per year worldwide [4], the world's electricity consumption increased by 690 TWh per year from 2021 to 2022, and is on the order of $3 \times 10^4$ TWh in total [4], [5]. But each year, the Earth receives approximately $1.5 \times 10^9$ TWh of solar irradiation at the top of the atmosphere [6], which is reliably and ubiquitously available. The critical challenge will be to accelerate the scale-up of the deployment of PVs worldwide [7]–[9], whilst bringing online electricity storage and redistribution/conversion technologies at the same rate. In this respect, one may consider the storage of sustainable electricity as high energy density fuels, which are compatible with the existing energy infrastructure. To this end, PV materials and technologies must be integrated with selective catalysts towards solar fuel synthesis, for instance green hydrogen production from water electrolysis, or $CO_2$ reduction into (multi)carbon fuels [10]–[16].

The scale of the challenge of accelerating the deployment of PVs and associated technologies is vast. For example, the International Renewable Energy Agency (IRENA) predicted that in order to keep global temperature rises within 2 ˚C of pre-industrial levels, solar PVs need to contribute to 25% of total electricity generation [17], requiring cumulative PV capacity to increase from 0.9 $TW_p$ in 2021 [18] to 2.8 $TW_p$ in 2030 and 8.5 $TW_p$ in 2050 [17]. This order of magnitude increase in cumulative capacity would require nearly €6 trillion of investment between now and 2050 [17], as well as technological developments to enable multiple routes to meet these scaling challenges. Other models have involved a stronger contribution from solar PVs, up to 80 $TW_p$ cumulative capacity by 2050 [19]–[21], and the UK government has a target of increasing PV deployment from 14 $GW_p$ in 2021 to 70 $GW_p$ by 2035 [22], [23]. But in turn, fully implementing the deployment of PVs, as well as effecting other measures to enabling a low $CO_2$eq (and eventually net-zero) economy, is estimated to bring in €150 trillion in cumulative benefits globally over a 30 year period, in terms of health, subsidy and climate-related savings [17].

Accelerating the deployment of PVs requires further reductions in the levelized cost of electricity (LCOE), increased investment in production capacity and technology development, reductions in the capital-intensity and environmental impact of manufacturing, as well as increased security of supply chains, among many other factors [7]–[9], [24]. These requirements can be fulfilled by further improvements in established technologies in commercial production, as well as new technologies that can lead to step-wise increases in power conversion efficiency (PCE) or reductions in the capital-intensity of manufacturing. The PVs market is currently dominated by crystalline silicon (c-Si) PVs, which accounted for 95% of all solar cell production worldwide in 2021 [18]. Although c-Si PVs were historically considered to be an expensive technology (*e.g.*, 0.40 € $kWh^{-1}$ in 2005 in Germany) [25], electricity generation costs have consistently reduced as production capacity increased [25], reaching 0.04 € $kWh^{-1}$ in Germany in 2018 [18], which is lower than onshore wind (0.06–0.09 € $kWh^{-1}$) or coal-fired power plants (0.07–0.11 € $kWh^{-1}$) [25], as shown in figure 1a. These trends are due to a large extent to reductions in the cost of polysilicon, reductions in wafer thickness and waste during slicing from ingots, as well as increases in PCE from below 18% in 2018 to 21% today for mainstream large modules due to reductions in impurity levels, improved cell architectures, and improved production technology [19]. At the same time, further developments are needed to overcome current technological challenges in increasing the efficiencies of commercial modules, reducing operational degradation processes to improve device lifetimes to 40 years and beyond, developing low-cost and sustainable metallization, increasing the "bifaciality" (ability to harvest light from the front and back) from 70% towards 100%, as well as improved strategies for module recycling [19], [26]. It is also important to consider the end-of-life of these modules, especially as their deployment rapidly increases, with predictions that global PVs panel waste would reach 54–160 million tonnes by 2050 (depending on many factors, such as panel lifetime) [20], [27]. Whilst substantial, it is important to bear in mind that this is significantly smaller than the waste from the continued use of fossil fuels, or from municipal waste [20].

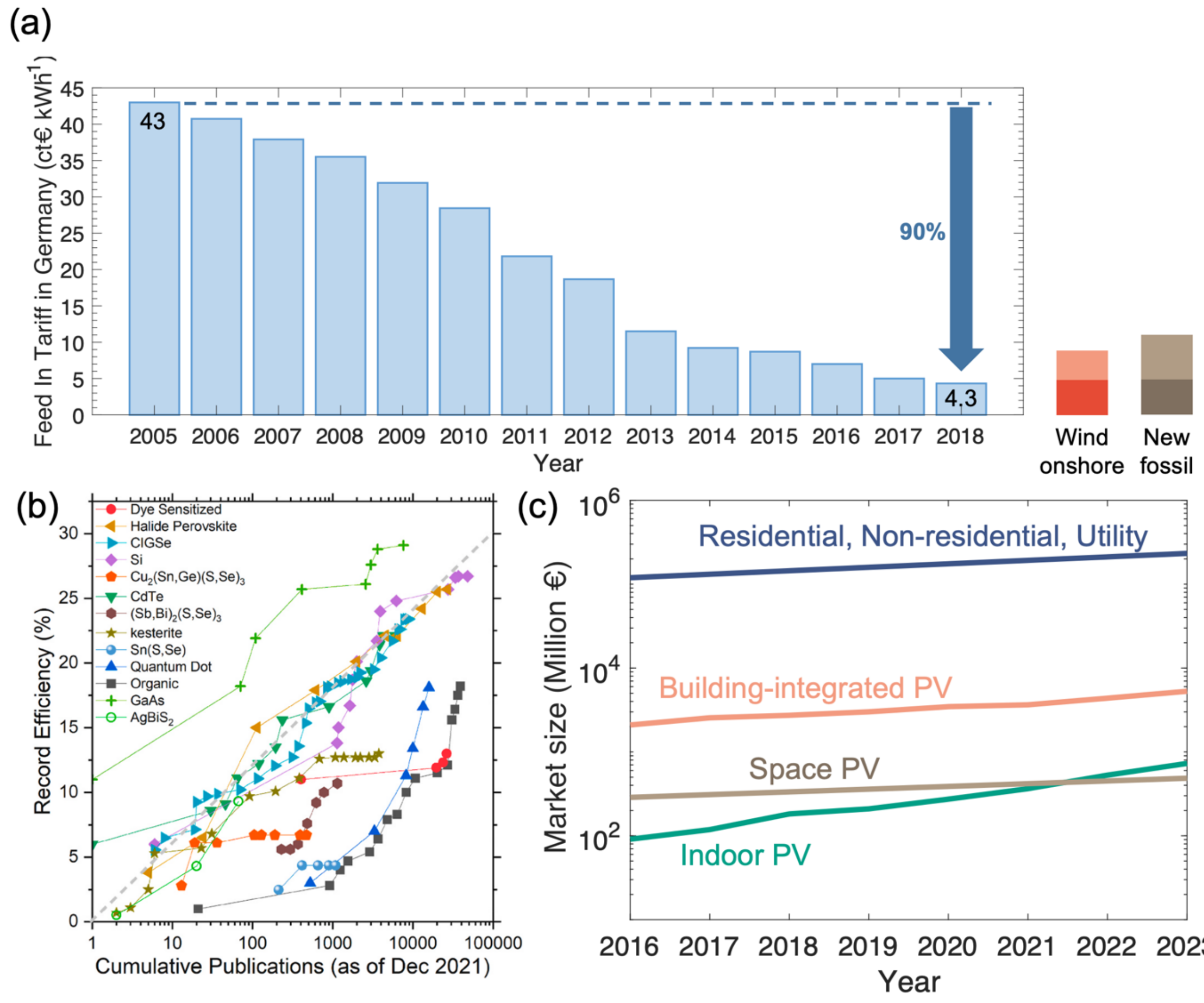

**Figure 1.** (a) Reduction in the cost of electricity from photovoltaics over time, illustrated by showing the progression in the lowest feed-in tariff for solar photovoltaics each year in Germany compared to onshore wind and new fossil fuel sources. Adapted from Ref. [25], with further data from Ref. [18] and [28]. (b) Record power conversion efficiency of single-junction photovoltaic materials against the cumulative number of publications. Reproduced under the terms of the CC-BY license from Ref. [29]. Copyright The Authors, 2022. Published by Elsevier. (c) Growth in market for photovoltaics used at the utility, residential and non-residential scale, building-integrated PVs, indoor PVs and space PVs. Data obtained from Ref. [30], [31] and [32].

Beyond c-Si PVs, CdTe PVs offer a credible commercial alternative, with the important advantage of lower materials usage and costs, owing to its higher optical absorption coefficient [19]. CdTe PVs currently comprise 5% of the world's PVs market, and modules have an efficiency (19.5±1.4%) close to that of c-Si PV modules [18], [32]. Further improvements in PCE to an expected value of 25% are achievable through passivation and the manipulation of defect states [26], [33], with further increases in efficiency achievable with bifacial modules [26]. However, a critical challenge is the scarcity of tellurium, which is insufficient to meet future demand as the industry aims to reach terawatt-scale production capacity [19], and advanced recycling strategies are essential. Other thin-film alternatives include GaAs, which achieves higher PCEs than c-Si or CdTe, but are limited by their high production costs and limited availability of Ga [19], [34].

The only other material currently considered to have potential to directly break into the utility-scale PVs market in the near-term are lead-halide perovskites [19]. These materials, which can have a range of compositions, have demonstrated surprisingly rapid increases in PCE, reaching a certified record value of 26.1% (as of mid-2023) in lab-size single-junction PV cells after only a decade of research [35]. This is because of the facile processability of lead-halide perovskites, their tolerance towards point and extrinsic defects, as well as their amenability to a wide range of defect-passivation techniques [36], [37]. As a result, a large number of groups worldwide could access the highest-quality films, leading to a rapid cycle of learning. Indeed, a plot of the record efficiency against cumulative number of publications places the learning rate of lead-halide perovskites on the same trend as c-Si and CdTe (figure 1b), yet the overall development time of lead-halide perovskite PVs is significantly shorter than either of these technologies because of the much higher cycle of learning and explosion of research activity [29]. Importantly, through composition changes, the bandgap of halide perovskites can be tuned to absorb in complementary parts of the solar spectrum (visible) as c-Si (near-infrared), and can therefore be combined to form tandem PVs. Since the first certified efficiency for a monolithic perovskite-silicon tandem in 2017 [35], [38], record values have already reached 33.7% as of mid-2023 [35]. Not only does this exceed the practical limit (27%) [39] and the current record efficiency for c-Si PVs (26.8%) [32], it approaches the practical limit considered for perovskite-silicon modules (35%) [26]. Given that the cost of a cell is less than half the cost of a module, and the cost of deployed solar PV systems are dominated by factors other than the module itself (*e.g.*, balance of systems), increasing cell efficiencies without substantially increasing manufacturing cost will be critical. Perovskite-silicon tandems are therefore promising, and it is expected that the modification of industrial c-Si production lines to fabricate tandem PVs is feasible [19]. However, as of yet, this is a new and unproven technology, and hence the reliability and "bankability" need to be achieved. The critical challenges with perovskite-based PVs are their reliability, scalability and toxicity. The module lifetime needs to at least match that of c-Si, and it is important to elucidate the degradation pathways to develop reliable accelerated degradation tests. It is also important to transition from slow, small-scale lab-based techniques towards methods for manufacturing large-area perovskite modules rapidly at scale [37]. Such efforts are indeed being undertaken by the growing perovskite industry, with tens of companies, both start-ups and established industry, ramping up efforts to manufacture at scale and develop stable materials to realize a reliable technology. A further concern which dominated early research on lead-halide perovskites was the reliance upon lead as an essential component within the material composition. However, full lifecycle assessments and thorough ecotoxicological studies indicate that the use of lead in metal-halide perovskites in PV manufacturing and deployment is entirely manageable, with low embodied energy, low toxicological impact, and extremely sustainable. Nevertheless, the sustainability, environmental and toxicological impact and the circularity of the whole PV manufacturing, use and end-of-life process is of critical importance to study and adapt every aspect of the processes to maximize sustainability and minimize any negative impacts.

Furthermore, beyond the use of PVs in the utility, residential and non-residential markets, their applications in building-integrated PVs (BIPVs), indoor PVs, agrivoltaics and others (*e.g.*, space PVs and on automobiles) are also important (figure 1c). BIPVs are expected to help enable more sustainable buildings with lower $CO_2$eq footprint and generate electricity closer to where it is needed [26], whilst indoor PVs are necessary to create a more sustainable Internet of Things ecosystem, which is expected to be worth trillions of dollars by 2025 [40]. In these emerging markets, c-Si PVs are not usually suitable, due to its lack of translucency, difficulty in tuning the bandgap over a large range, and low efficiencies under indoor lighting, thus opening up space for novel materials to have an impact [40]. Potential materials include organic PVs, which have recently found a resurgence of interest and promise with the advent of non-fullerene acceptors, leading to a rapid efficiency increase reaching nearly 20% on a cell level [41]. Dye-sensitized solar cells and lead-free alternatives to halide perovskites, which may be favorable for consumer electronics applications, also hold significant promise for BIPVs and indoor PVs [40], [42], and III-V compounds also have commercial potential for indoor PVs. There is still much to be discovered in the process-property-structure relationships of these materials, setting the scene for rich opportunities in fundamental research, and tailoring these materials towards applications with significant potential to reduce the $CO_2$eq footprint of electricity generation. Advanced metrology and computational techniques will therefore be critical. Finally, the photoelectrochemical field is benefiting from adopting the learnings from the PVs field, such as device architectures and materials [10]. Closer synergies between the two fields could lead to important future breakthroughs in the sustainable production of solar fuels.

In 2020, the Henry Royce Institute organized the UK PVs community to produce a roadmap on the current status and future challenges in advancing the photovoltaics field to reach net-zero $CO_2$eq emissions, emphasizing new capabilities needed locally in the UK [26]. Herein, we examine how the global field has changed (particularly over the time since the Royce PV roadmap came out in 2020), and look to the future with opportunities in newer, faster-paced sectors. The focus is on scientific and technological challenges faced by the whole community for both the utility-scale application of PVs, as well as for emerging markets, and for solar fuel production. By bringing together the insights from researchers (primarily in the UK) working across the vast breadth of photovoltaics, from academia and industry, and across technologies spanning a wide range of maturity levels, we hope to offer broad perspectives on the critical developments needed in the short-to medium-term to accelerate the deployment and sustainability of PVs, as well as emerging horizons in PVs research that will offer longer-term benefit.

**Acknowledgements**

For the roadmap as a whole, the authors would like to thank the support of Prof. Neil Alford (Imperial College London) and Dr. Suman-Lata Sahonta (University of Cambridge).

R.L.Z.H. would like to thank support from the Royal Academy of Engineering through the Research Fellowships scheme (no. RF\201718\17101), a UK Research and Innovation (UKRI) Frontier Grant (no. EP/X022900/1) awarded via the European Research Council Starting Grant 2021 scheme, and the Engineering and Physical Sciences Research Council (EPSRC; no. EP/V014498/2). SDS acknowledges support from the Royal Society and Tata Group (UF150033), EPSRC (no. EP/T02030X/1, EP/V027131/1), and the European Research Council (ERC) (no. 756962 [HYPERION]).

## 2. Crystalline Silicon Photovoltaics

*Ruy S Bonilla[1], Pietro P Altermatt[1,2], Stuart Boden[3], Nigel Mason[4], John D Murphy[5], Anthony Peaker[6]*

[1] Department of Materials, University of Oxford, Oxford OX1 3PH, United Kingdom.

[2] Trina Solar Co Ltd, Changzhou, China.

[3] School of Electronics and Computer Science, University of Southampton, United Kingdom

[4] Formerly BP Solar and PV Consulting, United Kingdom.

[5] School of Engineering, University of Warwick, Coventry CV4 7AL, United Kingdom.

[6] School of Electrical Engineering and Electronics, University of Manchester, United Kingdom.

### *2.1 Status*

Crystalline silicon (c-Si) solar cells continue to evolve as the fastest-growing and most widespread photovoltaic technology (figure 2a) due to: (i) the high conversion efficiencies attainable from its 1.12 eV bandgap that is close to the optimum for harvesting terrestrial solar irradiation. With a fundamental limit of 29.4% [43], c-Si PVs have demonstrated a certified record PCE of 26.8% at the research scale [32], and an increase of ~0.5%/year in industry (figure 2c), from 18% to over 23.5% in the last decade [44]. (ii) Silicon PVs are an established technology and has seen a marked cost reduction (figure 2b), down to US$ 48 $MWhr^{-1}$ [45] of the global weighted average "levelized cost of electricity" (LCOE). This is well below grid parity in most places [45], with energy payback times now below one year [18], [46]. (iii) Silicon is abundant and widely available, making it practical and affordable. (iv) Silicon PV

cells are stable, allow warranties of 25-30 years, and are therefore bankable [47]. (v) Manufacturing has been ramped up with an annual growth rate of 25-40%, accounting for around 95% of the overall PVs market today (figure 2a) [48], [49], and installing as much as 255 GW in 2022 alone equivalent to adding 30 average-sized nuclear power plants in one year .

The current workhorse of PVs deployment is the Passivated Emitter and Rear Cell (PERC). Its name comes from the dielectric layers on the front and rear sides that prevent electrons from losing their energy gained from sunlight. PERC is simple to fabricate: the front *pn*-junction is made by phosphorus doping the front surface of a *p*-type wafer, and the rear metal contacts are made by laser ablation of the dielectric layer, with an aluminium paste contacting silicon in a short firing step. Front contacts are made simultaneously during the short firing step via a screen printed fire-through silver metal paste. The average cell efficiency in mass production is 23.5% [44]. Currently, PERC is being overtaken by a new cell called 'Tunnelling Oxide Passivated Contact' (TOPCon). Thanks to a 1.5 nm thin $SiO_2$ layer, electrons flowing across the contact do not lose their energy. TOPCon uses the inherently superior *n*-type silicon wafers (PERC cells use *p*-type wafers), with a boron-doped front surface, which have reduced losses compared to phosphorus doping in PERC. The efficiency in mass production is 1% higher (in terms of absolute PCE) than PERC reaching about 24.5% [44]. It is unclear which cell concepts and manufacturing technologies will supersede TOPCon cells. One candidate is heterojunction technology (HJT), where amorphous silicon layers replace the dielectric thin films in PERC and TOPCon, and delivers the current certified lab-record PCE for c-Si PVs of 26.8% [35]. Another possibility is to move both metal contacts to the back in an Interdigitated Back Contact (IBC) design, giving efficiencies >26.0 % [50], [51]. Finally, a hybrid approach combining these designs could be the future workhorse before silicon cells are adapted to a tandem configuration.

The silicon PV field remains a crucial area of development in most parts of the world thanks to its relevance to the delivery of net-zero [9]. At present, industrial and academic institutions continue to work together to push the limits of silicon solar technology exploiting its advantages and looking towards multi terawatt production in the coming decade.

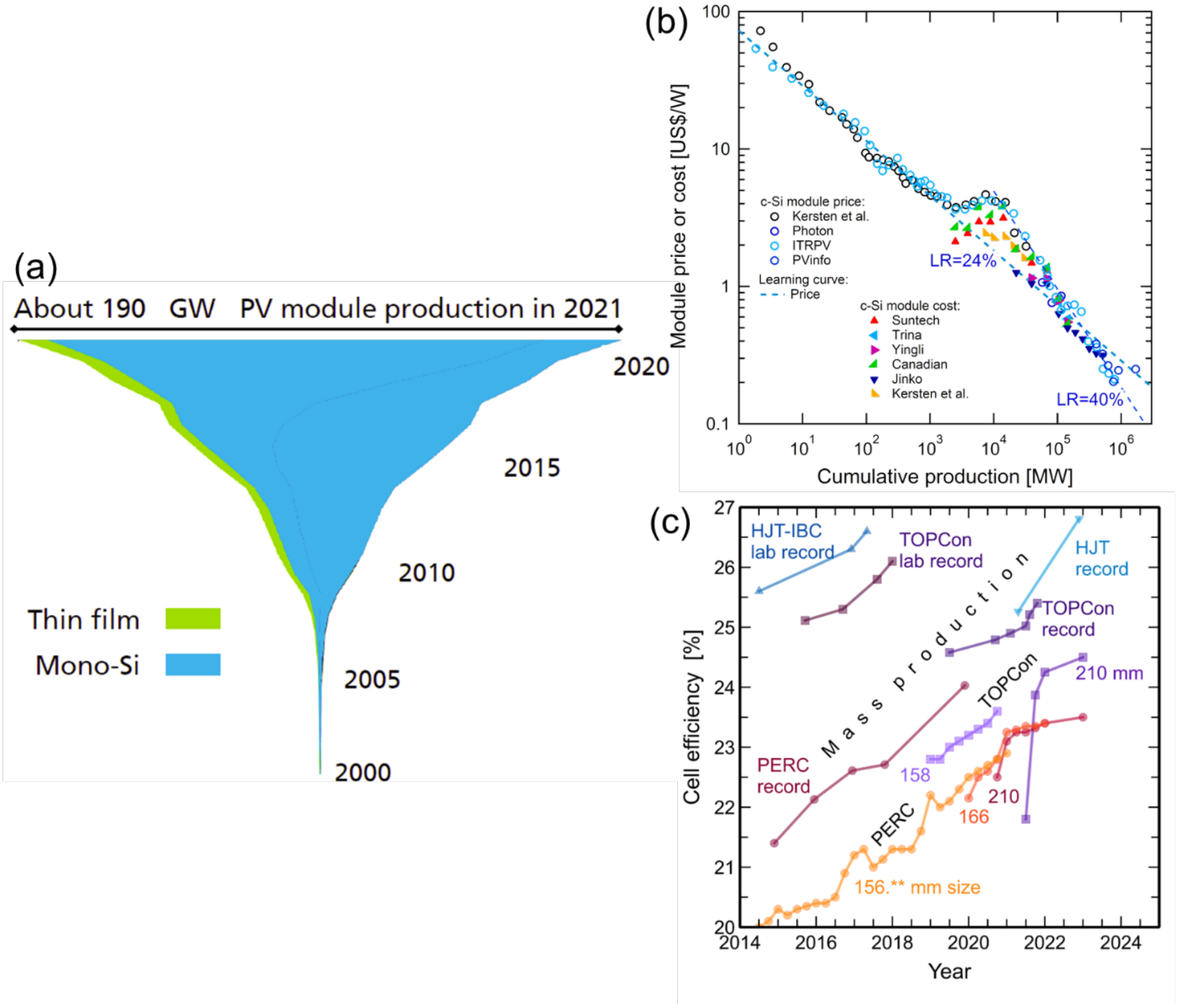

**Figure 2.** (a) Market share of different technologies in the solar energy market. Adapted with permission from Fraunhofer Institute for Solar Energy. Copyright 2023, Simon Philipps. (b) technology learning curve for Si PVs, (c) evolution of cell and module efficiency at laboratory and industrial scale.

## *2.2 Current and Future Challenges*

Although silicon PV is a mature technology, it faces disadvantages and hurdles for future progress. We divide these into intrinsic and extrinsic. Intrinsic drawbacks include silicon's fixed bandgap making it unsuitable for spectrum tuning and its relatively small absorption coefficient requires 160 μm thick absorbing wafers. The sensitivity to impurities and defects requires clean production and an energy intensive melt-based process at >1410 °C. Lastly, silicon is brittle and does not easily allow for mechanically flexible or curved shape solar cells. These intrinsic disadvantages provide guidance for niches other PV materials can fill. Extrinsic yet surmountable hurdles include:

1. The overarching challenge of c-Si PVs is to continue to drive efficiency gains at 0.2-0.5% per year, deployable to the hundreds of GWs production scales. Efficiency is the primary driver since it impacts the entire value chain, significantly reducing LCOE [52]–[54]. Innovations are essential in design, materials, and processing, which has come in the past and must continue to come from the research community, despite the relative maturity of c-Si PVs [55].
2. In parallel, c-Si PV technology must find suitable paths to extend its fundamental efficiency limit. At present, the practical limit of Si sits at ~27% [39], but enhancing partner tandem technologies can increase it to 32-35% (*e.g.*, with lead-halide perovskites – see Section 5), thus ensuring growth in the long-term, exploiting the already successful c-Si PVs industry and value chain.
3. Although silicon cells are reliable, we must also continue to improve their stability. Any degradation mechanisms (carrier-induced, light, elevated temperature, electrical potential, contact corrosion) must be fully understood and mitigated in a way that provides more confidence in accelerated life testing and a demonstrated scientific basis for extending the warranty period.

4. Net-zero scenarios imply scaling up the supply chain to a few $TW_p$ $yr^{-1}$ of production capacity. It can only occur if silver is replaced by copper and aluminium, scarce materials like indium and bismuth are avoided in HJT and tandem technologies, and material consumption is engineered to achieve minimal gram/kW use.
5. A demonstrated and efficient end-of-life strategy will be required to prove that c-Si PVs can not only lead the transition to a decarbonised electricity grid, but also satisfy the increasing demands on sustainability, recycling, and reuse [56].

### *2.3 Advances in Science and Technology to Meet Challenges*

Although silicon PVs is a mature technology, extensive research and development is needed to scale it up to 40-70 $TW_p$ (>20-fold) within the next two decades as required by the IPCC integrated assessment models [57]. Here we summarise the progress needed, and illustrate it in figure 3a:

1) To exploit the maximum efficiency limit of silicon cells, losses need to be further reduced:

    a) Increasing excess carrier lifetime ($\tau_{\mathrm{eff}}$) during silicon and cell production requires further fundamental understanding of defects and gettering processes.

    b) Highly passivating contacts are required for both electron and hole extraction, with saturation-current $J_{0s} < 5$ fA $cm^{-2}$ and contact resistivity $\rho_c < 100$ m$\Omega$ $cm^2$, without compromising optical properties needed for short-circuit currents ($J_{sc}$) ~42 mA $cm^{-2}$. Efficiency improvement forecast indicated in figure 3b.

    c) Improved texturing is necessary to allow for thinner wafers and still obtain high photocurrents. Si material accounts for about half the CAPEX and energy in the production chain.

d) Narrower metal fingers <25 μm with a higher aspect ratio are required to reduce shadowing and line resistance.

e) In the module, narrow strings or other contacting schemes are required, with low temperature soldering or other methods. Also cheap ways to remove iron from quartz helps to maintain high glass transparency, so does improved surface hardness to withstand sandblasting in dry climates.

2) Tandem cell architectures with >30% efficiency require suitable interconnect layers between the silicon bottom silicon cell and the top cell that fulfil the requirements of (1), contain only materials suitable for TW scale use, are passivating, selective, highly conductive, and stable.

3) Currently, module efficiency degrades by ~0.5%/yr and is affected by Light- and Temperature-Induced Degradation (LeTID) and is not yet fully understood [58]. A fundamental understanding and mitigation of all degradation mechanisms is required, as well as contact corrosion and laminate degradation. This boosts confidence in accelerated lifetime testing and thus financing, which accounts for 20-50% of the electricity costs of industrial-scale PVs projects [59].

4) Four critical advances are needed to ensure scalability of c-Si PVs to TW levels [60]:

a) The metallization of solar cells must be almost completely silver-free [61]. Promising alternatives are Al and Cu applied by screen printing, plating or other technologies.

b) Indium must be avoided in transparent conductive electrodes (TCEs) because of the limited availability and high price [62]. Promising alternatives are doped metal oxide films, yet no TCE has yet been found that performs at the same level as indium-based ones.

c) Bismuth in low temperature soldering must be avoided by further developing alternative technologies to soldering, given the limited quantities in which bismuth is produced commercially [63].

d) Copper in cables should be reduced. While copper and aluminium are being developed as replacements for silver metallization in silicon photovoltaics (see point a, above), this amount is tiny compared to copper in cables. But the mass of copper needed is estimated to equate to approximately 6 years worth of current production. There is the added complication that photovoltaics would be competing with other electronics and electricity generation technologies, as well as grid expansion and electric vehicles, for copper supply. Reducing the use of copper would alleviate supply chain challenges as the deployment of silicon photovoltaics is accelerated [64].

5) Reuse of module or system components, and recycling of materials to save energy, waste and mining activities. Materials circularity and extended module life will both contribute to lower LCOE and expanded c-Si PV adoption.

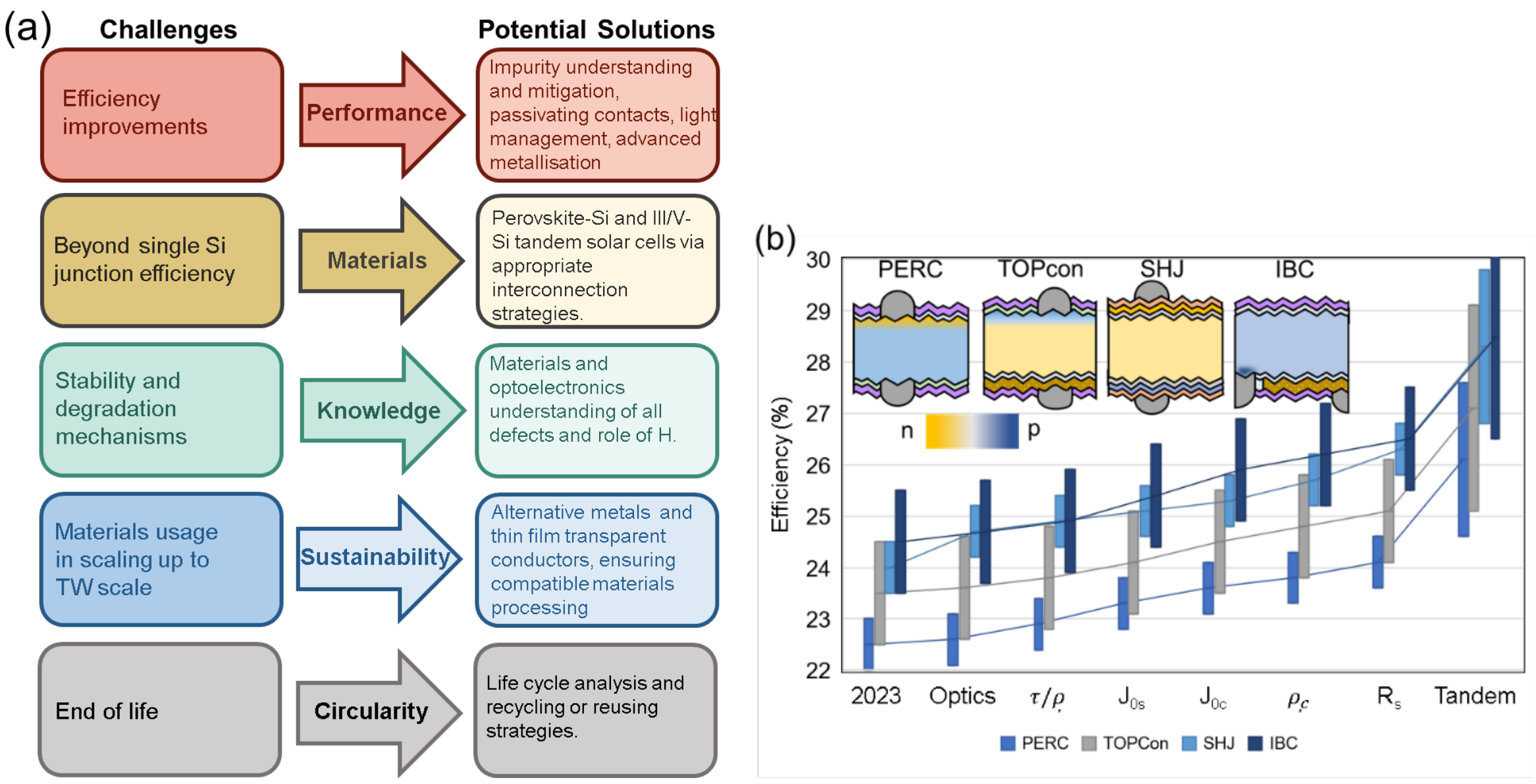

**Figure 3.** (a) Summary of most pressing challenges and potential solutions. (b) Future improvements in cell efficiency when challenges (summarized in horizontal axis) can be surmounted.

### *2.4 Concluding Remarks*

Silicon PV has enormous potential to grow to the scale required to replace the burning of fossil fuels for the generation of electricity. Moreover, Si PV can deliver such growth in the short time span required to limit the worst effects of climate change. There are clear and challenging scientific aims that need to be delivered through materials, engineering, and process innovation in order realize such potential. The nature of the Si PV industry requires concerted and resolute efforts to address these challenges. Industry, academia, and governments around the world must therefore continue to strengthen their research and development activity around these technological challenges. A booming solar PV industry will lead to jobs and economic prosperity all around the world since it allows for a decentralised, equitable, and flexible electricity grid, as well as huge mitigation of greenhouse gasses, and sustainable development of disadvantaged nations.

***Acknowledgements***

R.S.B was supported by the Royal Academy of Engineering under the Research Fellowship scheme (no. RF\201819\18\38). This work was supported by EPSRC (no. EP/V038605/1). J.D.M. was supported by the Leverhulme Trust (no. RPG-2020-377) and EPSRC (no. EP/V037749/1).

## 3. Cadmium Telluride and Related Chalcogenides

Jonathan D. Major,[1] and Ken Durose[1]

[1] Stephenson Institute for Renewable Energy, Department of Physics, University of Liverpool, Chadwick Building, Peach Street, Liverpool L69 7ZF, United Kingdom

### *3.1 Status*

CdTe photovoltaics are now firmly established as a technology second only to wafer silicon, making CdTe the most successful of the thin film materials to date. In 2021 thin film technologies represented 5% of the PV market (figure 2a) with CdTe being the vast majority of that with approximately 10 $GW_p$ of production [18]. Lab-scale power conversion efficiencies (PCEs) have now reached >21% [65] with modules of up to 19% PCE now being available [32], having increased from ~9% PCE only a decade earlier (figure 4). The fundamental advantage of thin film PV modules over wafer silicon is that they are formed from direct bandgap semiconductors having high optical absorption coefficients. Much lower thicknesses of absorber material are therefore required than for wafer silicon, and the modules can be produced directly on low-cost glass substrates using in-line manufacturing methods. Hence CdTe modules have the lowest LCOE of any PV technology [66] and even undercut fossil fuel-based energy production. The low-cost manufacturing uses the "superstrate" structure, where functional device layers are fabricated on directly onto the protective glass front surface (figure 5). A significant cost saving also comes from the large scale continuous float glass production process for the integrated manufacture of a $SnO_2$:F (FTO) front contact (transparent electrode). This makes the overall production process both rapid and inherently scalable, with no wafer-induced size limitation of active areas. Market-leader First Solar Inc. increased their panel size to ~2.5 $m^2$ for their Series 6 modules [67] in order to reduce module production and balance of system costs. This same scaling benefit is not possible for wafer-based technologies. Recently alternatives to CdTe that employ similar device architectures have emerged, including $Sb_2Se_3$, but they are at a much earlier stage of development and some way off production.

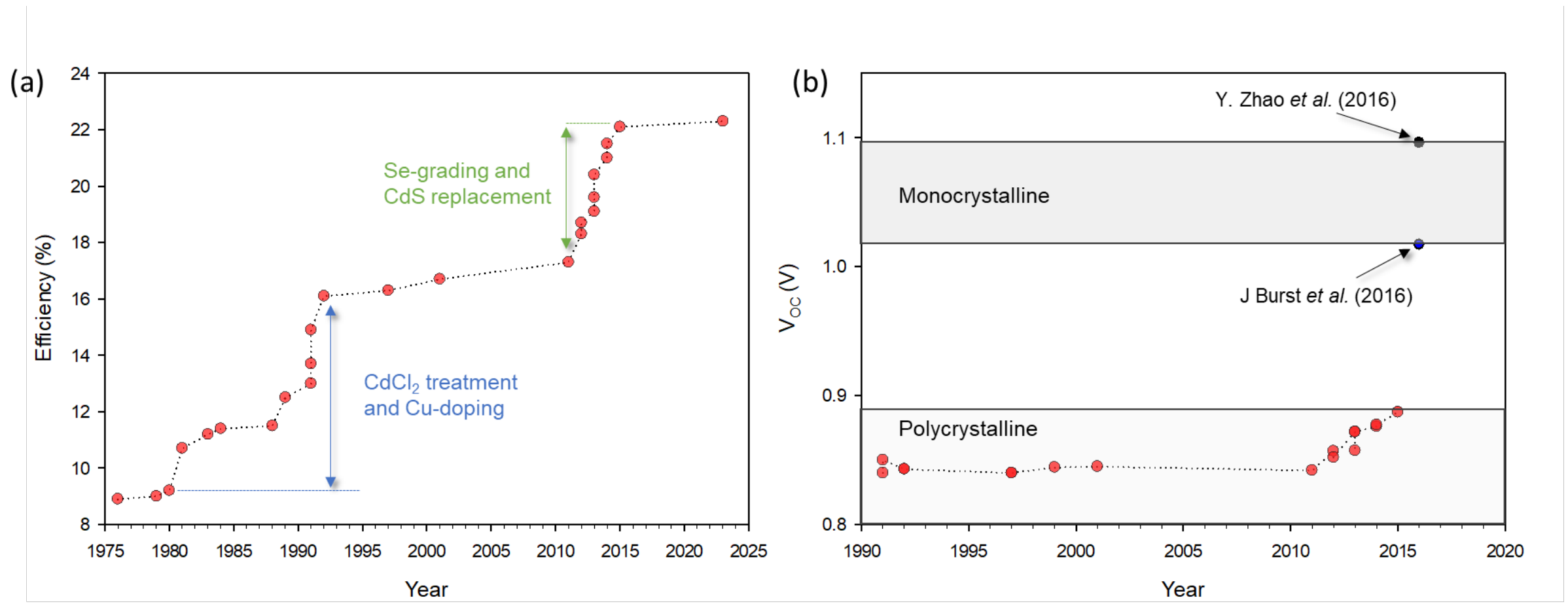

**Figure 4.** a) Historical record efficiency of CdTe cells and modules using data from NREL efficiency charts, and (b) $V_{OC}$ values for CdTe record cells from 1990 onwards compared with $V_{OC}$ records for monocrystalline based cells.

### *3.2 Current and Future Challenges*

The following description of the evolution of thin film polycrystalline CdTe PV technologies serves to highlight the ongoing challenge of increasing the open circuit voltage, $V_{OC}$, of the devices, which remains the most severe performance limitation after many years of development. CdTe PV technology has seen a high degree of evolution over the preceding decade, and while this yielded significant increases in short circuit currents, $J_{SC}$, and hence efficiency, $V_{OC}$ remained substantially unchanged.

Since its invention in the 1970s until about 2010, the state-of-the-art device structure remained a CdS/CdTe heterojunction, with Cu-doping [68] and a Te-rich back contact layer [69]. This suffered from parasitic optical absorption losses in the CdS partner layer [70] and inherent instability caused by Cu doping. Moving on from this produced dramatic results: Alternative window layers were introduced, replacing CdS with wider bandgap oxides such as $Mg_{1-x}Zn_xO$ (MZO) [71] - and these were coupled with the use of a lower bandgap absorber region at the near interface comprising $CdSe_{1-x}Te_x$ (CST) [72]. Both of these innovations increased light utilization, and hence $J_{SC}$, leading to a succession of record efficiency improvements. The incorporation of Se has further been suggested to play an additional role in passivating defects within the material [73], and the reduced bandgap has allowed increased current generation without compromising $V_{OC}$. In parallel to these developments, Cu-doping has also been supplanted by the introduction of group V doping, *i.e.*, As-doping and P-doping [74], [75] which, while more challenging to work with, offer greater shallow doping densities and increased operational stability.

As shown in figure 4, these innovations, beginning *ca.* 2010, served to increase the previously stagnated efficiency from 16.7% to 22.1% by 2016. However, these gains arose primarily from increases in $J_{SC}$. On the contrary, $V_{OC}$ represents the present limit, and has been a barrier to reaching the ultimate efficiency in CdTe PVs for decades. The open-circuit voltage of record devices currently stands at 887 mV representing only ~61% of the theoretical maximum approximated by the bandgap. Comparing to the previous voltage record of 848 mV reported in 2001 [76], or even 850 mV in 1992 [77], the technology has acquired less than 40 mV in over 30 years. This is in stark contrast to the over 1 V $V_{OC}$ reported for single crystal CdTe devices [78] and points to the absorber polycrystallinity as the key challenge to overcome. This attainment gap in $V_{OC}$ is the primary barrier remaining for CdTe. Minimizing this $V_{OC}$ deficit to the single crystal level and attaining 1 V $V_{OC}$ for thin film devices would equate to a PCE of ~25% (all other parameters assuming to remain constant).

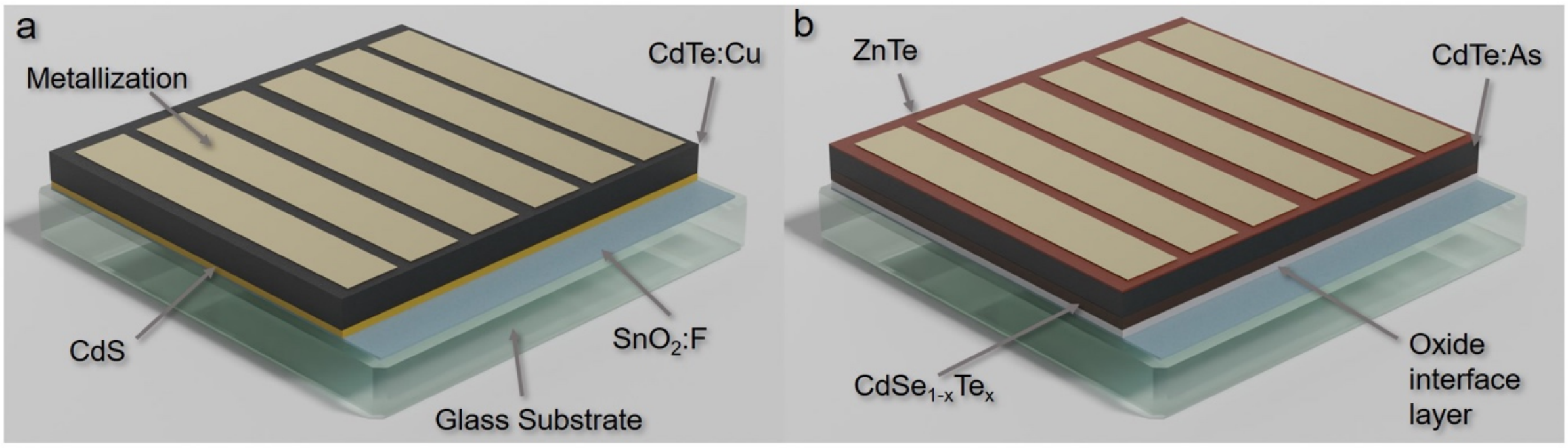

**Figure 5.** a) Traditional FTO/CdS/CdTe:Cu/metal cell structure and b) current high efficiency FTO/Oxide/$CdSe_{1-x}Te_x$/CdTe:As/ZnTe/metal cell structure.

In addition to these challenges, it is important to address challenges associated with the scarcity of Te. Presently, global production of Te occurs as a recovered by-product from mining of other metals and is currently of the order of 100s of tons per year, a significant from of which is for CdTe module production. However, at present, the efficiency of Te recovery is low at ~26%. Improvements to the refining process, along with purposeful mining for tellurium could significantly increase global production and sustain increased deployment [79]. The long-term sustainability is also aided by module recycling. First Solar incorporate a dedicated module recycling program with a 90% recovery rate of semiconductor materials from end-of-life modules, and analysis suggests 100 GW $yr^{-1}$ of production should be supported by mid-century [80]. The toxicity of CdTe is often cited as an issue due to the cadmium content, however this fails to recognize the high stability of CdTe, which has low solubility. Multiple studies have shown there to be essentially zero cadmium contamination risk during typical module operation and insignificant risks even in the event of extreme weather events or catastrophic fires [80].

Coupled with protections put in place for safe manufacture and analysis identifying CdTe to have advantages across all environmental impact categories, the materials related risks

### *3.3 Advances in Science and Technology to Meet Challenges*

As mentioned above, overcoming the voltage limit of CdTe technologies is the key to their continued performance development. There is a need for continued materials science work to understand why the voltage gap between crystals and films arises. Advanced characterization needs to be paired with lab scale device analysis to better understand the roles that grain boundaries, stoichiometry, doping and interface band alignments play. Because of the nature of the deposition processes used, it is possible to readily transfer lab scale innovations to large-scale production. The research community therefore has a real opportunity to surmount this remaining barrier and to see the results translated into established production processes, further driving down the cost of energy production.

A divergent area of interest related to CdTe is the emergence of the antimony chalcogenide absorbers $Sb_2Se_3$, $Sb_2S_3$ and $Sb_2(S,Se)_3$ [81], [82]. Whilst a relatively recent technology, they have already surpassed the 10% efficiency barrier and continued performance development could see them become industrialized. The current state of the art utilizes much the same device structure and deposition approaches as CdTe, but with a much thinner and cheaper absorber material. Theoretical predictions suggest that based on their core materials properties, the antimony chalcogenides could eventually surpass the performance of CdTe [83]. Because of the similarity of device design and manufacturing, there is a real possibility that established industrial processes used for CdTe could easily be adapted to this new material.

Achieving this aim will require the identification of the performance limiting factors through advanced characterization and continued development of $Sb_2Se_3$ doping and heterostructure designs to overcome them. Progress over the next decade will decide the future viability of $Sb_2Se_3$ in a landscape of new, emerging and competing PV technologies.

### *3.4 Concluding Remarks*

The inherent low cost and scalability of thin film PV designs has seen CdTe achieve the lowest LCOE of all PV technologies. Nevertheless, there remains a significant opportunity to further increase its competitiveness by overcoming the one remaining barrier to high efficiency, namely the $V_{OC}$ limit. Moreover, there are opportunities to diversify CdTe module deployment. While it is already successful in utility-scale power generation, it could find applications for which silicon is less well suited. These include bifacial PVs and building or automotive integrated PVs which remain relatively untapped but will likely see significant growth in the push towards net-zero. Alternatives to thin film CdTe, such as $Sb_2Se_3$, have the potential to become important but are at a much earlier stage of technical development.

Despite the lower LCOE of CdTe, silicon unquestionably remains the dominant market force. To realize the true market benefits, CdTe and other thin film PV technologies will need to follow silicon into the realm of very large-scale production. Here CdTe has an inherent advantage over other emerging technologies: it has been proven in the industrial environment and it is certain that in-line thin film processing is scalable. At the present time First Solar Inc is the only company to have begun to exploit CdTe at large scale. The challenge for European industry is therefore to develop their own technologies to the same level and to contribute to the expansion of manufacturing that will ultimately reduce the costs and challenge wafer-based technologies.

**Acknowledgements**

J.M. and K.D. acknowledge funding from EPSRC (no. EP/N014057/1, EP/T006188/1 and EP/W03445X/1).

## 4. Lead-Halide Perovskite Single-Junction Photovoltaics

Nakita K. Noel,[1] Severin N. Habisreutinger,[2] Jay B. Patel,[3] Michael B. Johnston,[1] Pablo Docampo,[4] Matthew J. Carnie,[5] Trystan Watson[5]

[1] Department of Physics, University of Oxford, Clarendon Laboratory, Parks Road, Oxford OX1 3PU, UK

[2] Oxford Photovoltaics Ltd., Unit 7-8, Oxford Pioneer Park, Mead Road, Oxford OX5 1QU, United Kingdom

[3] Department of Physics, King's college London, Strand, London WC2R 2LS, United Kingdom

[4] School of Chemistry, University of Glasgow, Joseph Black Building, University Place, Glasgow G12 8QQ, United Kingdom

[5] Department of Materials Science and Engineering, Swansea University, Bay Campus, Fabian Way, Crymlyn Burrows, Swansea SA1 8EN, United Kingdom

### *4.1 Status*

Lead-halide perovskites (LHPs) are an emerging class of semiconductor materials which have tremendous potential to revolutionize the PV industry. With many desirable characteristics, such as high absorption coefficients, long charge-carrier diffusion lengths, and largely benign defect chemistry, these materials are excellent candidates for PV absorbers. One additional benefit of these materials is their compositional tunability which allows for the fabrication of LHPs with a wide range of different bandgaps: a property which is ideal for the development of multijunction solar cells, a strategy for improving the power conversion efficiencies (PCEs) beyond that of single junction solar cells.

Since the first publication on a LHP solar cell in 2009 [84], their PCEs have steadily increased - starting with the shift from a liquid electrolyte dye-sensitised solar cell structure to a solid-state hole-conductor [85], [86]. As of mid-2023, perovskite solar cells have achieved certified efficiencies of up to 26.1% [35], [87]. This unprecedentedly rapid performance gain is largely due to a confluence of factors, including but not limited to improvements in film formation, evolution in material composition [35], [88], interface passivation [89], [90], and the introduction of more suitable charge transport layers [91]. LHP thin-films can also be deposited through a variety of simple deposition techniques such as solution-processing or vapour deposition [92], [93], making perovskite solar cells particularly attractive for low-cost, large-scale deployment of PVs.

While the efficiencies of these devices are still climbing, there is yet work to be done if perovskite PV is to be commercialized and deliver the energy revolution that many hope to see. Two of the major issues that must be addressed are scalability and stability, with increasing attention being placed on tackling these issues in a sustainable way. On a more fundamental level the challenge is rooted in gaining a deeper understanding of how the crystallization kinetics of these materials impacts compositional heterogeneity on the nanoscale, the formation of grain boundary and interface defects- and importantly, how we can reduce defect density and passivate the remaining defects, such that the long-term stability of the perovskite material itself is improved. For perovskite solar cells to move ‘from lab to fab’ and be a part of delivering terawatt-scale PV deployment, understanding these fundamental scientific issues is pivotal (figure 6). Not only will better fundamental understanding underlie further improvements in the efficiency and reproducibility of perovskite solar cells, but also be critical for the development of better materials and more scalable processes for stable perovskite-based optoelectronic devices.

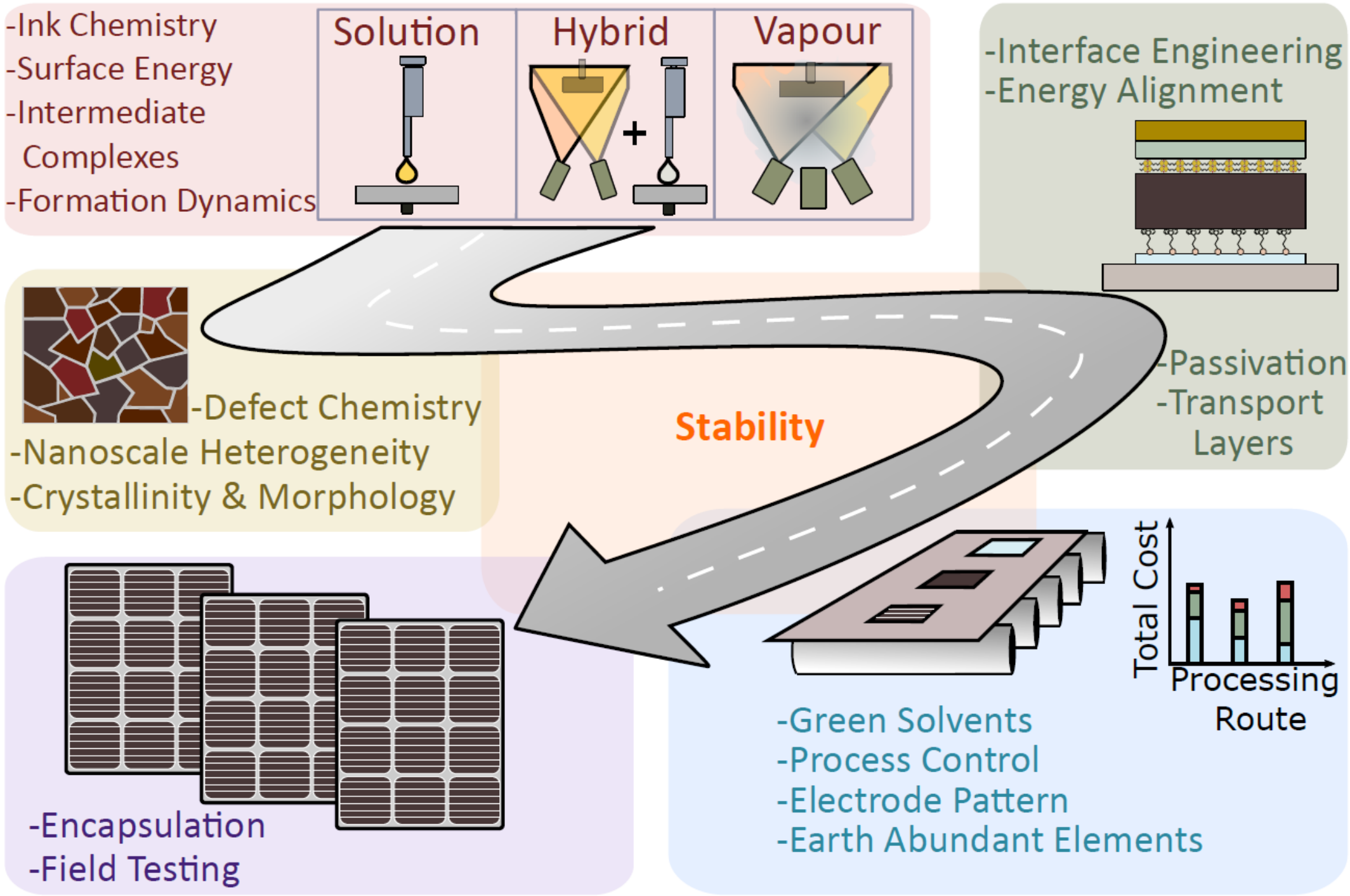

**Figure 6.** The path to the commercialization of perovskite photovoltaics

### *4.2 Current and Future Challenges*

Perovskite solar cells have now reached a level of maturity at which their single-junction laboratory efficiencies exceed all of their competing thin-film technologies, such as CdTe, $CuInGaSe_2$, etc. [35]. Improving thin-film quality through modulating crystallization kinetics, as well as increasingly effective surface passivation strategies have resulted in continuous efficiency improvements. Looking to the future, while there are many issues to be addressed, there are two main fundamental challenges.

The first is the macro-challenge which targets the rapid transition of this technology from the research lab to industrial-scale fabrication and real-world deployment. This can in turn be broken up into two essential aspects: i) scalability- how can we most effectively cover industry sized substrates at the rate required for at-scale manufacturing; and ii) long-term operational stability- can this technology operate without a significant reduction in performance over relevant timescales (figure 6)?

The other challenge is the micro-challenge which moves away from the current highly empirical approaches to improving device efficiency, and instead, re-examines the fundamental processes governing the crystallization and thin-film formation dynamics, and importantly, their relationship to the resulting optoelectronic properties of the perovskite thin-film (figure 6). An enormous driver of innovation and progress in the field has been the fact that LHPs are easily deposited and readily formed at low processing temperatures (often ≤ 150 °C), and that despite their low-tech fabrication, they still exhibit optoelectronic properties rivaling conventional semiconductors. This has been largely attributed to their inherent defect tolerance, where the presence of defects in the crystal do not immediately impact the degree of non-radiative recombination or mobility of free charge-carriers [94]. As a result of this, rapid iteration cycles, particularly in solution-processed LHPs, have allowed for remarkable device advances which often exclude a deeper understanding of material crystallization or the nature or formation of defects. Consequently, many open questions remain, such as the role of crystallite size and grain boundaries, or which surface passivation strategies yield not just short-term performance improvement, but are stable under long-term operation?

Importantly, providing a more comprehensive framework for understanding how the material crystallizes on a molecular or even atomic level will be important, if not essential, to this as it is conceivable that for detailed understanding of the possible degradation pathways and failure modes, fundamental understanding of the formation of the material is indispensable.

Another potentially contentious issue is the toxicity of halide perovskite materials. Currently, the best-performing metal-halide perovskite solar cells all contain Pb, whether entirely Pb-based, or in the case of mixed-metal perovskites, Pb-Sn based. To date, while some approaches at entirely removing Pb from LHPs have shown some promise [95], no devices based on Pb-free perovskites has been able to rival the current performance of Pb-based perovskites. More fundamental materials research must be done in this area to determine, on one hand, the viability of replacing Pb in halide perovskites, and on the other hand more thoroughly evaluating the how harmful the Pb content in LHPs is, especially since the total mass of Pb per unit area in LHP devices is an order of magnitude smaller than that found in Si PVs (which use Pb/Sn solder) [96]. Nonetheless, preventing or mitigating the potential effects of Pb-leakage, such as through sequestration [97], will be important in the large-scale deployment of LHP PVs.

### *4.3 Advances in Science and Technology to Meet Challenges*

*Scalability-* Currently, LHP thin-films are most commonly deposited through spin-coating a precursor ink onto substrates whose area is typically on the order of a few square centimeters. To become a commercially viable technology, the community must continue developing alternative deposition techniques which can rapidly deposit layers on hundreds to thousands of square meters. This should neither come at the expense of performance nor stability and cannot affect the cost-competitiveness of LHPs. In the case of solution-processing, the solvents used for perovskite inks are largely incompatible with industrial scale production due to health and safety and regulatory restrictions. To be a viable avenue, greener solvent alternatives need to be demonstrated [98], [99], and while some progress has been made in this regard, there is much more to be done. Importantly, this consideration applies to all other layers in the device stack as well [100]. This must also be done without incurring prohibitively large performance losses in these devices.

An alternative route is to use physical-vapor deposition (PVD) techniques [92], [93]. While substantial advances have recently been made using this technique, PVD has received less attention due to the need for more sophisticated hardware. Nevertheless, the performance of PVD-based solar cells are catching up with their solvent-based counterparts at around 24% [101]. An inherent advantage of such techniques is the pre-existing industrial knowledge base with regards to scaling and throughput.

*Long-term stability*- With the ISOS protocols, a more standardized set of stability tests has been introduced and is increasingly gaining a foothold in the research community [102]. However, a major open question is how well the described tests map onto operational stability in the field. For instance, thousands of hours of continuous operation under illumination are now being reported. However, this may not correspond to the same effective period of daytime, but instead might constitute an accelerated stability test. What the exact acceleration factor would be compared to a cell tested under diurnal cycling conditions is not yet established. This means that more studies that approximate real-world operation are needed, and that at the same time, develop meaningful acceleration factors with the goal of being able to compress multiple years of operation into a timeframe of months or weeks.

*Environmental Impact*- As it stands, lead will likely remain an essential component of perovskite PV, therefore, prior to terawatt scale deployment, the possible impact on environmental and human health needs to be established in conjunction with the development of strategies to stop lead leakage. This research is currently gaining traction, for example by looking at the scenarios in which the maximum amount of lead is leached into the environment and its bioavailability tracked, in order to assess consequences in the worst-case scenario [103]. The initial results are quite encouraging showing that the bio-availability of Pb is extraordinarily short [104]. However, more studies are needed to fully assess possible down-stream consequences. Simultaneously, there are studies looking at additives in the absorber itself, or adjacent layers which can chelate or otherwise bind the lead before it can be leached into the environment [97]. These studies are also very encouraging showing a very high degree of effectiveness, however, they have not yet been fully tested in leaching tests emulating extreme outdoor conditions. Ideally, controlled, outdoor testing would examine the total impact and the mitigating effects of the most promising additives to establish the actual risk profile of this technology. It will also be important to carefully devise and assess end-of-life strategies, such as recycling approaches that allow Pb-containing compounds to be recovered.

*Formation and Surface Defects* - The transition from a precursor to semiconductor layer requires more in-depth exploration. New and more sophisticated characterization approaches are needed to tackle these questions. Such questions exist for both PVD and solution-based techniques. Understanding and modulating the processes which govern the formation dynamics of LHPs, the exact nature of the defects formed and how to passivate them with stable agents using industrially relevant processes will be critical for further developments of both PVD and solution-processed LHP PVs.

The defect tolerance in the bulk of perovskite absorbers makes the effect of the surface defects even more prominent. In LHPs, defects have multiple effects, impacting mobile charge-carriers through trapping and non-radiative recombination; affecting the likelihood of ion migration, and determining the availability of reaction sites which participate in degradation due to interactions with light, charges, water, and/or oxygen. Therefore, a good understanding of perovskite formation and its impact on the makeup of the defect landscape, in particular on the surface, will be critical in order to develop comprehensive and rational strategies to passivate and eliminate those defects, or prevent their emergence later in the operation cycle.

### *4.4 Concluding Remarks*

In summary, metal-halide perovskites are incredibly promising semiconductor materials with the potential to make a significant contribution to producing terawatt-scale PV deployment. This can take many forms: single-junction perovskite PV, all-perovskite multi-junction tandems, and perovskite-silicon tandems. However, for the potential of this material to be realized, more focus needs to be placed on not simply further improvement in the efficiencies of perovskite PV, but a deeper understanding of the formation dynamics of these materials and how this affects defect densities and stability. Importantly, this cannot be done in isolation, but also needs to be done in the context of scalability and sustainability. Solving these material focused problems, will be beneficial not just for single-junction perovskite PV, but will greatly improve perovskite-containing multijunction devices. There are, of course, other tandem-specific issues which arise when fabricating multijunction devices, and these are discussed in detail in Section 5. With continued effort, and by creating an experimental feedback loop where these considerations of scalability and environmental impact inform the approaches used to improve performance, perovskite PV is poised to be a disruptive technology which fundamentally changes the energy landscape.

**Acknowledgements**

N.K.N, J.B.P, M.B.J, P.D., M. J. C, and T.W acknowledge financial support from EPSRC (no. EP/V011197/1, EP/T025077/1 and EP/W007975/1).

## 5. Lead-Halide Perovskite Multi-Junction Devices

Bart Roose,[1] Chris Case,[2] K. D. G. Imalka Jayawardena,[3] John D. Murphy,[4] Henry J. Snaith,[5] Samuel D. Stranks,[1] Jarla Thiesbrummel,[5] J. Michael Walls[6]

[1] Department of Chemical Engineering & Biotechnology, University of Cambridge, Philippa Fawcett Drive, Cambridge CB3 0AS, United Kingdom

[2] Oxford Photovoltaics Ltd., Unit 7-8, Oxford Pioneer Park, Mead Road, Oxford OX5 1QU, United Kingdom

[3] Advanced Technology Institute, Department of Electrical and Electronic Engineering, University of Surrey, Guilford GU2 7XH, United Kingdom

[4] School of Engineering, University of Warwick, Coventry CV4 7AL, United Kingdom

[5] Department of Physics, University of Oxford, Clarendon Laboratory, Parks Road, Oxford OX1 3PU, UK

[6] Centre for Renewable Energy Systems Technology (CREST), Loughborough University, Loughborough LE11 3TU, United Kingdom

### *5.1 Status*

In a multijunction solar cell, two or more sub-cells are stacked together. This way, by using light absorbing materials with different bandgaps in each sub-cell, a broader range of the solar spectrum can be harvested more efficiently than in single junction (SJ) solar cells by minimalizing thermalization losses. The theoretical efficiency limit for a single-junction device is 33% under 1-sun illumination [105], whereas for a double-junction (tandem) this increases to 45% [106]. Considering that the extra cost for processing an additional sub-cell can be low compared to the full systems costs for the installed PV modules, the increased efficiency can significantly reduce the system's LCOE.

Varieties of multijunction solar cells include monolithic 2-terminal (2T), three-terminal (3T), and 4-terminal (4T) configurations. In the 4T configuration, each sub-cell can be operated independently (figure 7a). For 2T, the sub-cells are connected in series (figure 7b), so the current from both sub-cells has to be matched to maximize efficiency, which is non-trivial. Current matching is not required in 4T systems, but modules can be more complex and expensive due to the need for additional components. 3T uses a back contact sub-cell, avoiding the drawbacks of 2T and 4T, but creates challenges with interconnections and scaling [107].

The first perovskite/Si tandem, reported in 2015, used a 4T configuration (17.0% efficiency) [108], but in recent years the focus has shifted more towards 2T tandems. Large advancements have been made by using better quality silicon bottom cells, optical management, optimizing fabrication processes for tandem applications and improving interfaces and contact layers [109]. This has resulted in certified efficiencies reaching as high as 33.7% by mid-2023 [35], [110], [111], which far exceeds the efficiency of both silicon (which has a practical efficiency limit of ~27%) and perovskite single-junction PV [39]. Recently, a 3T tandem was reported, which was shown to be much less susceptible to spectral variations than 2T tandems. The efficiency was 24.9% but could be improved to >30% with moderately improved sub-cells [112]. Another interesting development coming from the silicon field is a bifacial tandem, which can yield as much as 20% extra current entering the module from the back. Bifacial perovskite-silicon tandems have therefore been considered [113]. An added benefit of bifacial systems is that the optimal bandgap of the top cell is reduced, potentially avoiding higher ratios of bromide:iodide that are prone to instabilities such as halide segregation, thereby increasing stability [113].

Perovskite/perovskite tandems are not far behind perovskite/Si tandem PV when it comes to efficiency, which has improved from 17% in 2016 [114], to 29% in 2023 [32]. Yet, significant progress still has to be made in stability and scale up, particularly for tin-containing narrow bandgap perovskites, which are used as bottom cell materials in perovskite/perovskite tandems.

Notable companies commercializing perovskite tandems (perovskite/silicon, or perovskite/perovskite) are CubicPV, Tandem PV, UK-based Oxford PV, along with Hanwha Qcells building on established Si manufacturing processes, and Swift Solar and Renshine Solar developing all-perovskite tandems. Oxford PV built a pilot production line in 2016 for qualification, and are planning to commence production at their site in Brandenburg (Germany).

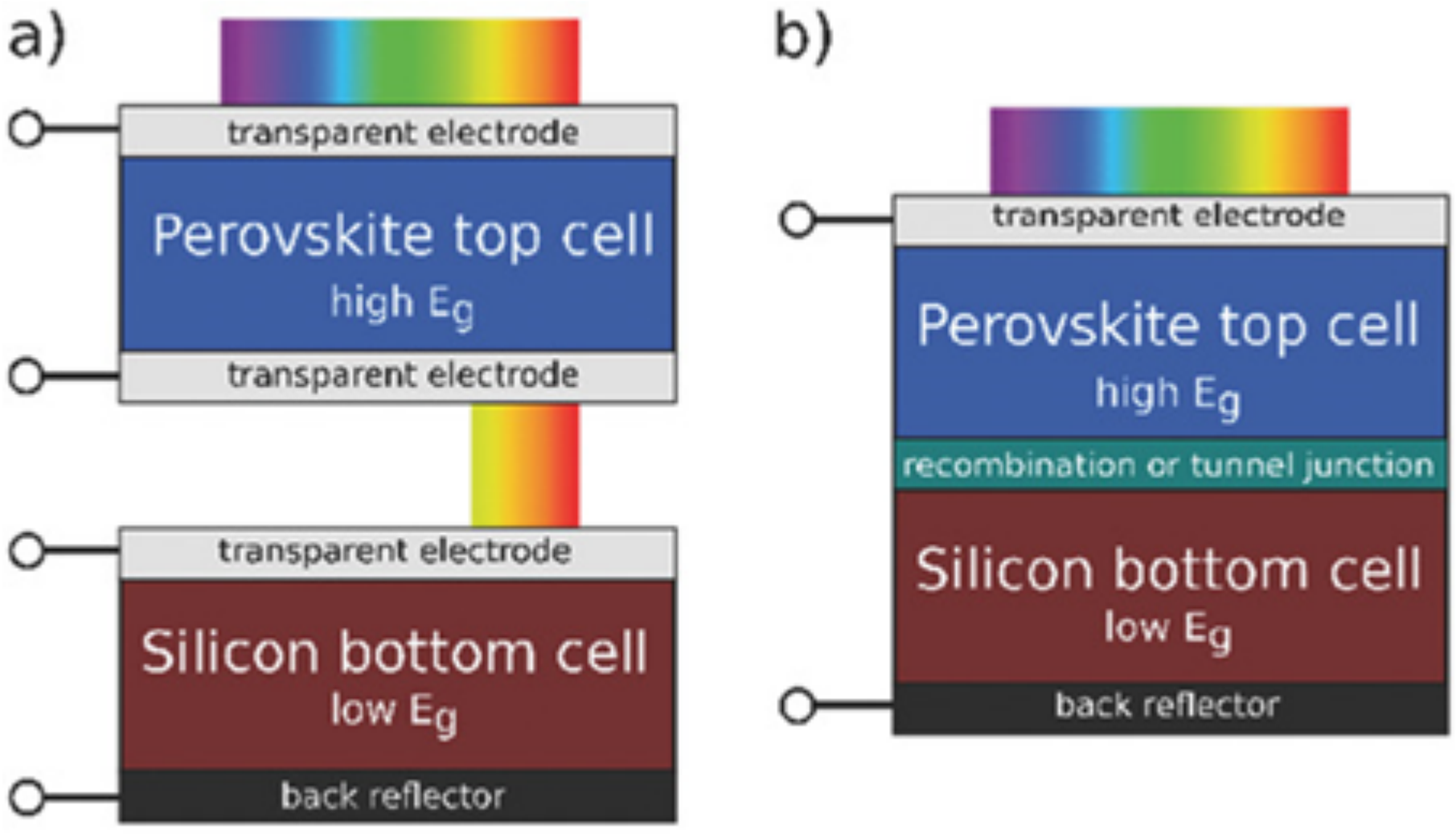

**Figure 7.** Schematics of the two main perovskite/silicon tandem architectures: a) four-terminal mechanically stacked; b) two-terminal monolithically integrated. The perovskite top cell has a higher bandgap ($E_g$) than the crystalline silicon bottom cell with $E_g$ = 1.1 eV. Adapted with permission from Ref. [115] under the terms of the CC-BY license. Copyright 2018, The Authors.

### *5.2 Current and Future Challenges*

Multijunction technologies face many of the same challenges as their single-junction counterparts (see Section 4); Progress on single-junction technologies will therefore also benefit multijunction technologies, as discussed in Section 4. Here, the focus is on multijunction specific challenges.

Challenges for perovskite tandems can be broadly divided into four categories: increasing efficiency, improving stability, designing effective recycling strategies, and commercialization. Finding effective solutions to these challenges will reduce the LCOE, which is the most important parameter for large-scale implementation of solar technologies.

Recent progress on tandem efficiencies is impressive, but still fall below theoretical (45%) [106] and practical (38%) [115] maximum lab-based efficiencies, and a number of tandem-specific challenges need to be addressed. Effective deposition techniques for textured surfaces are required to enable light trapping schemes and increase light absorption. For bromide-rich perovskite top cells, it is key to increase the open-circuit voltage towards their potential. The alloyed lead-tin perovskite used as narrow bandgap absorbers in perovskite bottom cells are more sensitive to precursor purity and atmospheric conditions during fabrication than neat lead systems, since tin easily oxidizes from desirable Sn(II) to problematic Sn(IV), lowering device performance. Currently, $SnO_x$/Au/PEDOT:PSS is the most popular recombination layer for perovskite/perovskite tandems in the academic literature [116]. However, both Au and PEDOT:PSS cause significant parasitic absorption, limiting current generated in the bottom cell. Alternatively, TCOs have been used, but the high lateral conductivity severely affects the fill factor in the presence of pinholes [116]. Furthermore, thicker TCO layers can introduce internal reflectance losses. Current matching requirements in 2T tandems may result in lower efficiencies under real-world spectral variations, as compared to lab-certified efficiencies using standardized illumination spectrum and intensity. However, modeling has shown that the loss in performance due to spectral variations is relatively small [117], [118].

Perovskite/Si tandems have passed many industry standard stability tests, but the suitability of stability tests to perovskites and the true lifetime of perovskite/Si will not be known until these devices have been extensively deployed in the field. Both types of perovskites used in perovskite/perovskite tandems face challenges. Mixed-halide lead-based perovskites used in the top cells suffer from halide segregation, and more voltage-losses originating from trap assisted and surface recombination [119]. while the oxidation of tin remains a problem for lead/tin perovskites even after fabrication of the devices. Due to current matching requirements in 2T tandems, degradation of just one of the sub-cells will severely affect tandem efficiency, and device lifetime is limited by the stability of the least stable sub-cell.

Recycling is not a tandem-specific challenge, but increased complexity of tandems may make recycling more challenging than for single-junction PVs. The only components from silicon modules that have been recycled cost effectively are aluminium frames, copper wiring and glass. However, scarce materials such as indium and silver, or toxic materials such as lead and halides should also be recovered to prevent them from ending up in landfill and to make the industry truly sustainable.

A common barrier to commercializing PV technology is scaling up from lab-scale to industrial processing. This involves using scalable depositions methods and replacing costly and scarce materials (TCOs, Ag, solvents). Furthermore, setting up PV manufacturing capacity is capital intensive. The UK government has so far not been supportive towards new initiatives, causing companies such as Oxford PV to move pilot plants to countries that offer more financial support, and more predictable support in light of the uncertainties over Brexit. Another issue for Si/ perovskite tandems is that they tend to use more niche heterojunction silicon base cells rather than industry-standard silicon PERC devices.

### *5.3 Advances in Science and Technology to Meet Challenges*

Lead-halide perovskite deposition on highly textured silicon surfaces has been achieved by using vapour deposition, or hybrid vapour/solution methods. However, upon slightly reduced texture, solution processing of thicker perovskite films has also been demonstrated to fully coat the underlying Si, and the present world record efficiency from KAUST uses this approach. Scalable methods will have to be further developed to make the process compatible with large-scale production, which required high yielding deposition methodologies which are compatible with high throughput coating techniques. The voltage of the top cell is limited by interfacial recombination [120] and phase segregation during processing, and further improvements can come from addressing these problems [121]. For real-world applications, the ideal bandgap of the light absorbing materials will depend on the local solar spectrum. Through compositional engineering, the bandgap of perovskites could be fine-tuned to match spectral variations. A more significant difference in required top cell band gap comes from the consideration of mono- versus bi-facial. If Bi-facial deployment is intended, then dropping the top cell band gap to as low as 1.5eV may be required to maximize energy yield.

Halide segregation has been studied intensively; although the mechanism is still debated, the process is understood to relate to defects [122]. Passivating these defects may therefore be key to eliminating halide segregation and improving stability of top cell perovskites [123]. Advances in post-processing of bottom cells will be key to prevent oxidation of tin, even more so than for neat lead-perovskites, the lead/tin perovskite film has to be shielded from oxygen and moisture completely during deposition of subsequent layers, and finished devices have to be encapsulated robustly, to prevent oxygen and moisture ingress over the entire lifetime of the tandem module. More research is needed to understand how degradation of individual sub-cells affects tandem performance, particularly in 2T and 3T configurations.

Most materials in tandem modules are present in small amounts and integrated with other materials, making material recovery challenging. Nevertheless, it is crucial that modules are recycled effectively to establish a circular industry, preserve scarce materials and capture toxic components, in line with industry-wide concerns about circularity. The technology exists to recycle tandem modules, but is currently not sufficiently incentivized. Further research is needed to design cost-effective recycling methods, but interest from industry is limited, as landfill is often the cheapest option. Legislation and financial incentives could help to force the industry to design better recycling solutions.

Advances in metrology are also desired to better predict efficiency and stability of perovskite tandems under real-world conditions. Tandem efficiencies (especially 2T) are susceptible to spectral variations, so tests with standardized illumination may not accurately reflect real-world efficiency. Bifacial tandems can yield significantly higher power output, but there are currently no standardized protocols to define the spectrum and intensity of the light entering from the back of the device. It is likely that standard stability tests are not representative for perovskite devices, considering their ionic character, and additional test may be required. Extended field tests will have to be conducted to refine stability tests for these devices to better predict their lifetime. Tandem specific measurement protocols will help to mature the field.

Companies like Oxford PV have shown that the technical upscaling challenges faced by perovskite tandems can be overcome, both in sourcing raw materials and in processing methods. A key hurdle now is financial support to establish GW-scale production facilities. If the UK government wants to achieve energy security and sustainable energy generation, it needs to act now, or risk missing an important opportunity to capitalize on a UK-driven technology.

### *5.4 Concluding Remarks*

Perovskite tandems are a very promising technology, and the first products are close to hitting the market. Perovskite/Si tandems are an excellent gateway technology, building on existing silicon technology to gain a foothold in the solar sector, and could initially be applied in areas where high performance is required (rooftops, space applications, electric vehicles). On the longer term, perovskite/perovskite multijunctions will likely build further on this foundation and become the dominant technology, as material usage, cost and carbon footprint are all potentially lower for perovskite/perovskite tandems than for perovskite/Si tandems. All-perovskite tandems also side-step the silicon materials supply and value chain, offering a unique opportunity for energy security. To get there, a continued research effort is required into understanding the materials properties of perovskites, so that efficiency and stability can be further understood and improved, and real world operating lifetimes can be reliably predicted. More research is also needed to replace scarce and expensive components, such as indium used in the TCO, and to find effective recycling methods.

To establish a favorable climate for perovskite tandem technologies, government action is required. Legislation will have to be put in place to enforce recycling of the whole tandem module, to establish a truly circular industry. In addition, barriers will need to be removed and financial help has to be provided to establish large-scale domestic production capabilities. This will be a significant step towards improving energy security and achieving a sustainable future.

**Acknowledgements**

All authors arranged alphabetically by surname, apart from the first author. BR and SDS acknowledge the EPSRC (no. EP/T02030X/1) for funding. S.D.S. acknowledges the Royal Society and Tata Group (no. UF150033). JDM acknowledges the support of a Leverhulme Trust Research Project Grant (no. RPG-2020-377). C. C. and H.J.S acknowledge the EPSRC Prosperity Partnership Project (EP/S004947/1) and J.T. is funded by the Deutsche Forschungsgemeinshaft (DFG) under the HIPSTER project.

## 6. Organic Photovoltaics

Pascal Kaienburg[1], Tayebeh Ameri[2,3], Ardalan Armin[4], Diego Bagnis[5], Philippe Berrouard[6], Jochen Blumberger[7], Hugo Bronstein[8,9], Yi-Ming Chang[10], Elmer Chao[11], Tracey M. Clarke[12], Graeme Cooke[13], James R. Durrant[14,15], Richard H. Friend[9], Alexander J. Gillett[9], Martin Heeney[14], Arthur D Hendsbee[6], Ji-Seon Kim[16], David Lidzey[17], Roderick MacKenzie[18], Iain McCulloch[19], Sebastian B. Meier[20], Paul Meredith[3], Graham Morse[21], Jenny Nelson[16], Chris Nicklin[22], Thomas Osterberg[23], Moritz Riede[1], Martyn Rush[24], Peter J. Skabara[13], Franky So[25,26], Alessandro Troisi[27], Karsten Walzer[28], Trystan Watson[15]

[1] Department of Physics, University of Oxford, Clarendon Laboratory, Parks Road, Oxford OX1 3PU, UK

[2] Institute for Materials and Processes, School of Engineering, University of Edinburgh, Sanderson Building, Robert Stevenson Road, Edinburgh EH9 3FB, UK

[3] Chair for Composite Materials, Department of Materials Science, Faculty of Engineering, Christian-Albrechts-Universität zu Kiel, Kaiserstrasse 2, 24143 Kiel, Germany

[4] Centre for Integrative Semiconductor Materials (CISM), Department of Physics, Swansea University Bay Campus, Swansea SA1 8EN, UK

[5] Oninn – Centro de Inovação, Av. José Candido da Silveira 2000 - Horto Florestal, Belo Horizonte - MG, 31035-536, Brazil

[6] Brilliant Matters, 237 Rue de Liverpool, Saint-Augustin-de-Desmaures, QC G3A 2C8, Canada

[7] Department of Physics and Astronomy and Thomas Young Centre, University College London, Gower Street, London WC1E 6BT, UK

[8] Yusuf Hamied Department of Chemistry, University of Cambridge, Cambridge CB2 1EW, UK

[9] Cavendish Laboratory, University of Cambridge, JJ Thomson Avenue, Cambridge, CB3 0HE, U.K.

[10] Raynergy Tek Incorporation, 60, Park Ave. 2, Hsinchu Science Park, Hsinchu 30844, Taiwan

[11] Luminescence Technology Corp., Hsinchu 30076, Taiwan

[12] Department of Chemistry, University College London, Christopher Ingold Building, London WC1H 0AJ, UK

[13] School of Chemistry, University of Glasgow, Glasgow, G12 8QQ

[14] Department of Chemistry, Imperial College London, Molecular Sciences Research Hub, 80 Wood Lane, London, W12 0BZ, UK

[15] SPECIFIC, Swansea University, Bay Campus, Swansea SA1 8EN, UK

[16] Department of Physics and Centre for Processable Electronics, Imperial College London, London SW7 2AZ, UK

[17] Department of Physics and Astronomy, The University of Sheffield, Hicks Building, Hounsfield Road, Sheffield S3 7RH, UK

[18] Department of Engineering, Durham University, Lower Mount Joy, South Road, Durham DH1 3LE, UK

[19] Department of Chemistry, University of Oxford, Mansfield Road, Oxford OX1 3QR, UK

[20] ASCA GmbH & Co. KG, Steigweg 24, 97318 Kitzingen, Germany

[21] Merck Chemicals Ltd, Chilworth Technical Centre, University Parkway, Southampton SO16 7QD, UK

[22] Diamond Light Source, Harwell Science and Innovation Campus, Didcot, Oxfordshire, OX11 0DE, UK

[23] Epishine AB, 582 13 Linköping, Sweden

[24] Polysolar Ltd, Aurora BAS, High Cross, Meadingley Road, Cambridge, CB3 0ET, UK

[25] Materials Science and Engineering, North Carolina State University, Raleigh, NC, 27695 USA

[26] Organic and Carbon Electronics Laboratories (ORaCEL), North Carolina State University, Raleigh, NC, 27695 USA

[27] Department of Chemistry and Materials Innovation Factory, University of Liverpool, Liverpool L69 7ZD, UK

[28] Heliatek GmbH, Treidlerstrasse 3, 01139 Dresden, Germany

### *6.1 Status*

As of today, organic photovoltaics (OPVs) have reached certified power conversion efficiency (PCE) exceeding 19% [124], with extrapolated intrinsic devices lifetimes of well above 20 years [125] for laboratory devices. Module efficiencies on industrial scale are as high as 8%, featuring several years of proven outdoor stability [126], which gives confidence for at least more than 15 years of outdoor PV lifetime. Companies like Heliatek, ASCA, and Sunew run large scale roll-to-roll production facilities with reported one million square meters annual capacity each, and have realized various outdoor OPV installations around the world (figure 8). Indoor OPV products are commercially available from companies like Epishine and ASCA (see Section 7 for an in-depth discussion of indoor PV).

OPV has several unique selling points that are a direct consequence of the soft and molecular nature of the absorber materials, enabling lightweight and flexible, easy and cost-effective to install, (semi)transparent and color-selective (including UV and NIR) modules. The high degree of freedom in organic molecule design allows tailoring to specific application requirements, such as tuning absorption bands for aesthetic appearance. Adaptive manufacturing allows for customized products that are highly designable and integrable into almost any kind of surface. Life Cycle Analyses of OPV show the technology to be highly appealing from a sustainability perspective [127], with low embodied energy and $CO_2$ emissions, earth-abundant raw materials and non-toxic end products, and easy end-of-life treatment. Existing and near-future applications of OPV exploit these unique properties to unlock vast surface areas for electricity generation and occupy markets that c-Si PVs cannot serve well. Prominent examples include an already available market for building-integrated and building-applied PVs (BIPVs, BAPVs) covering primarily façades and low load-bearing roofs, but also windows, the emerging agricultural PV market (see Section 11 on Agrivoltaics), as well as indoor PVs and local power sources. OPV module for BIPV/BAPV has thus reached the highest Technology Readiness Level 9, which newer applications, including automotive, transport, and aerospace, are yet to demonstrate.

The most common fabrication routes for OPV are solution-processing (SP) [128], [129] and vacuum thermal evaporation (VTE) [130], [131]. The development of OPV shown in figure 9 reflects the versatility and importance of molecule design and synthesis, which has been assisted by the optimization of device architecture and processing, continuous improvements in characterization methods, computational modelling, and an understanding of the physical processes in molecules, films and devices. Arguably the most critical recent advance for OPV has been the discovery of high-performance non-fullerene acceptors (NFAs) [132], such as the ‘Y-series’ [133], [134]. Their long exciton-diffusion lengths and exciton self-dissociation also enable bilayer devices with promising PCEs. OPV materials and technologies have also found applications in related fields such as solar fuels (see Section 12), photon up- and downconversion [135], photodetectors and other sensors.

While the development of the absorber and interlayers is critical and the focus of this perspective, the cost and quality of auxiliary components and technologies are also essential to the success of commercial OPV. These include module encapsulation, electrodes, flexible and sustainable plastic substrates, and manufacturing equipment. A major focus of industrial OPV manufacturers is automating and simplifying processes to achieve higher yields, efficient material use, and high throughput production.

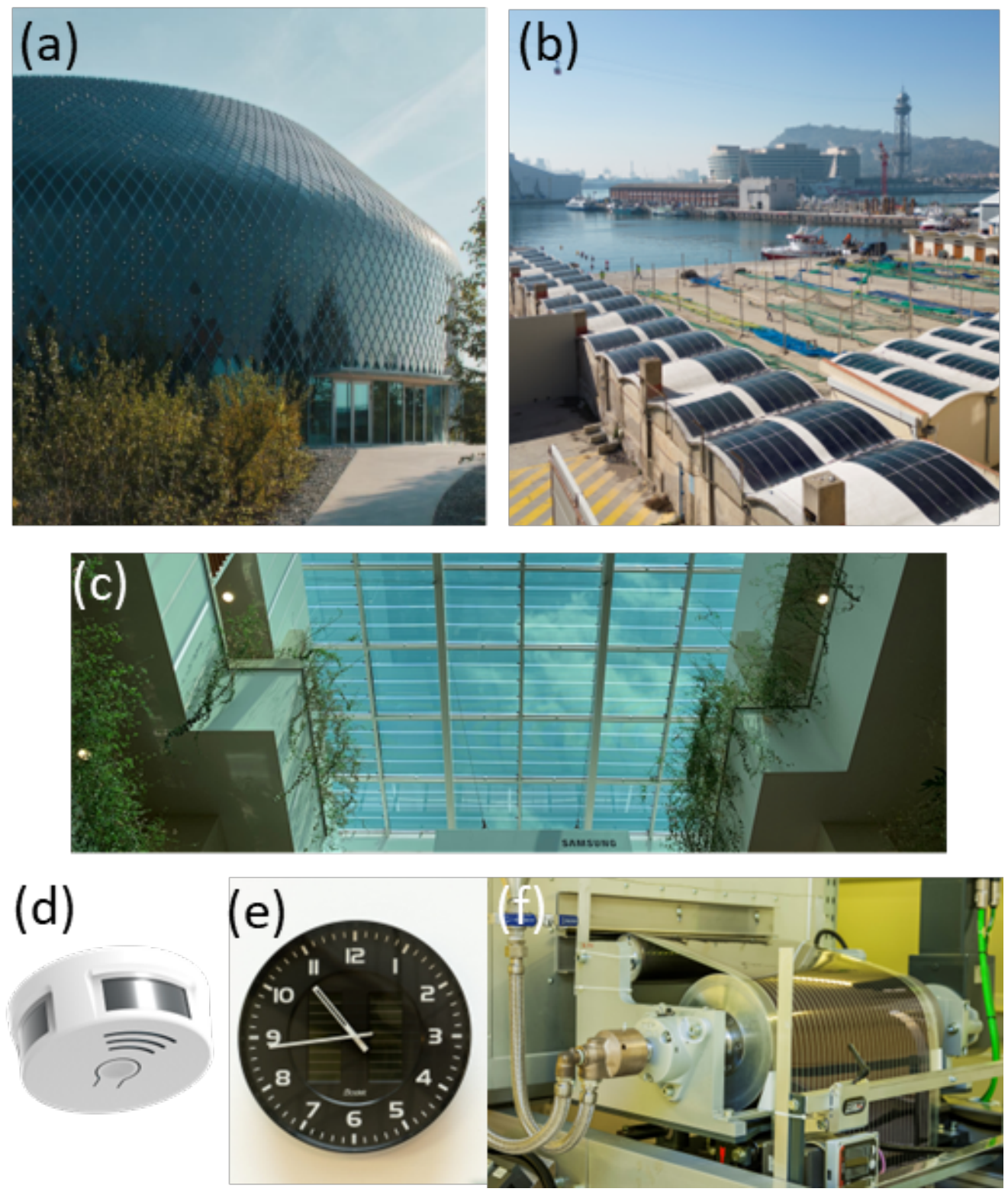

**Figure 8.** Examples of OPV projects, applications, and large-area fabrication. (a) Self-powered media façade at the Novartis Pavillon (Basel, Switzerland) with OPV modules by ASCA (SP), ©Laurits Jensen. (b) Heliatek (VTE) undulating rooftop installation at Port of Barcelona, ©comsa. (c) Semi-transparent roof by Sunew (SP) at a mall in São Paulo, Brazil, ©Diego Bagnis. (d) Smoke detector design with OPV by Epishine (SP lamination), removing the need to change the battery, ©Epishine. (e) 100% autonomous eco-designed clock by Bodet using ASCA OPV modules, ©ASCA. (f) Roll-to-roll coating process at Oninn/Sunew, ©Diego Bagnis.

### *6.2 Current and Future Challenges*

Early efforts to commercialize OPVs targeted the same markets as c-Si PVs and were eventually outcompeted by c-Si PV (see figure 2a). Since then, the appeal of and demand for renewable technologies has increased, which is now opening new markets for PV technologies. Also, materials producing OPV devices with much higher PCEs have become available. When applications are multi-purpose, *e.g.*, generating electricity while transmitting light to grow crops in protected environments (Agrivoltaics, see Section 11) or adding aesthetic appearance to a building, the business opportunities for OPVs are different from conventional PVs. That is, different regulations and sector-specific standards apply, sector-specific policy incentives would be beneficial, performance metrics beyond $ $W_p^{-1}$ are required [136], and charging a premium for installed PV capacity is possible. However, currently low production volumes and early times of technological learning imply higher cost. A clear-cut return on investment scenario, potentially supported by a diversification of refinancing scenarios, would motivate a broader implementation of OPVs in our everyday life. This would trigger the upscaling and diversification of the entire value chain, bringing down OPV module and balance-of-systems cost. The sooner economy of scale effects (Swanson's law) are unlocked, the larger the benefit of OPV technology to society, from ubiquitous application to lowest environmental impact of all current PV technologies.

Academic PV research has emphasized increasing PCE on the cell level [32], [137], even though device lifetime contributes equally to the total energy yield of a solar panel. The latter needs to be evaluated with respect to cost for a given application, which defines the specific technical requirements, such as the target absorption spectrum (see, *e.g.*, Section 7 on Indoor PVs). We note that the main challenges for SP and VTE OPVs differ. After reaching >19% PCE, improving cell efficiency is arguably no longer the primary target for SP OPVs. Instead, high PCE needs to be achieved with low-cost molecules, while also meeting requirements of stability and scalable fabrication. Achievements of cell PCE need to be transferred to industrially relevant materials, device stacks, and processing conditions. Milestones for industrial development like large area deposition and (mini)modules are a topic for academic research if they present new concepts or reveal design rules. VTE OPVs on the other hand has demonstrated sufficiently long device lifetime >20 years with comparably simple molecules and scalable fabrication [125], shown by the OLED industry. However, VTE OPV still faces the challenge of obtaining higher PCE for at least 15% for single-junction devices under 1-sun illumination.

### *6.3 Advances in Science and Technology to Meet Challenges*

One OPV threshold challenge is improving module lifetime under realistic operating conditions [126]. Here, we need to advance the understanding of degradation mechanisms [128], [138] in the active layer and at interfaces, covering morphological and (photo-)chemical degradation pathways and their influence on optoelectronic properties. This includes studying different stresses and early failure, burn-in, and longer-term degradation, rendering *in-operando* studies important. Intrinsic stability under heat, illumination, and varying loads (open-circuit conditions being the harshest) can be studied on the cell level. Extrinsic stability towards humidity and oxygen is critical for ambient coating of SP OPVs and for encapsulated modules in general. Here, mechanical stress is an additional factor primarily affecting delamination. The resulting design principles need to be applied to stack design (*n*-*i*-*p* is favourable), ink formulation (additives and ternary blends), and design of new molecules (particularly NFAs). Accelerated lifetime testing protocols with predictive capabilities need to be identified and reported consistently, with such reporting potentially enforced by journals [139], [140].

Industrial, scalable SP of OPVs (printing, see figure 8f) requires non-toxic solvents, fast processing in air, and large processing windows to achieve high yields. Solvents have to be non-halogenated and while xylene seems an acceptable choice, ‘greener’ solvents must be explored further [141]. Ideally, the mechanics of the coating process and the kinetics of the drying process in the laboratory match those of the manufacturing line, arguably favouring blade coating over spin-coating. Given slight variations in fabrication conditions, large area SP targets thicker layer stacks, including active layers >200nm, than typical work in academic laboratories. The synthesis of today's best performing SP materials is complex (although exact metrics are hard to define), typically involving many steps and the use of toxic reagents and chromatographic purification. Such factors are not commensurate with cost-efficient upscaling. The development of less complex absorbers is one solution, along with reducing the number of low yielding steps. In short, we need molecules that carry the benefits of today's high performing molecules but have low chemical complexity, are intrinsically stable and processable in air, have the right solubility in 'green' solvents to yield favourable microstructure, and maintain high performance in thick layers.

PCE is particularly critical to VTE OPV, where discovering efficient low-molecular weight, evaporable NFAs is imperative. Increasing the optimum absorber thickness brings further PCE gains and likely requires novel molecule design and better morphology control. While not the main bottleneck for SP OPV, higher cell PCEs of 23-25% seem feasible in the longer term. Encouraged by the endless possibilities of molecule design and having defied predicted PCE limits in the past, OPV becomes interesting as tandem partner for other PV technologies. Approaches to higher PCE of OPV include higher photoluminescence quantum efficiency of molecules, higher absorption coefficient, reduction of non-radiative losses, better coupling between molecules assisting charge transport and excited state delocalization, and further improving interlayers, morphology control, and charge collection. These are coupled to a better understanding of working mechanisms in OPV [142], including charge generation processes in low-offset NFA systems and the role of charge transfer, singlet, and triplet states in non-radiative decay [143], [144], and energetic and microstructural disorder. OPV molecule design is currently semi-heuristic and takes years of minor iterations to optimize a structure of interest [134]. Rational design requires further improvements of molecule structure – function - performance relations. This not only includes performance and optoelectronic properties, but also predicting microstructure; here sidechains and the influence on processing conditions present a major challenge. Given the synthetic variety of OPV, the ease of sample processing, and building on advances in multi-scale computational models 'from atoms to devices' as well as high-throughput computational screening and machine learning approaches for novel materials discovery, accelerated materials design and testing is highly promising for OPV. In the quest of simultaneously meeting all OPV performance targets, exploring diverse classes of molecules and different absorber concepts such as ternary and multi-component systems, all-polymer, block co-polymers, and all-small-molecule, seems worthwhile [145]. To make OPV cost-competitive, the involvement of companies focussing on upscaled and high purity synthesis of the most successful molecules is critical.

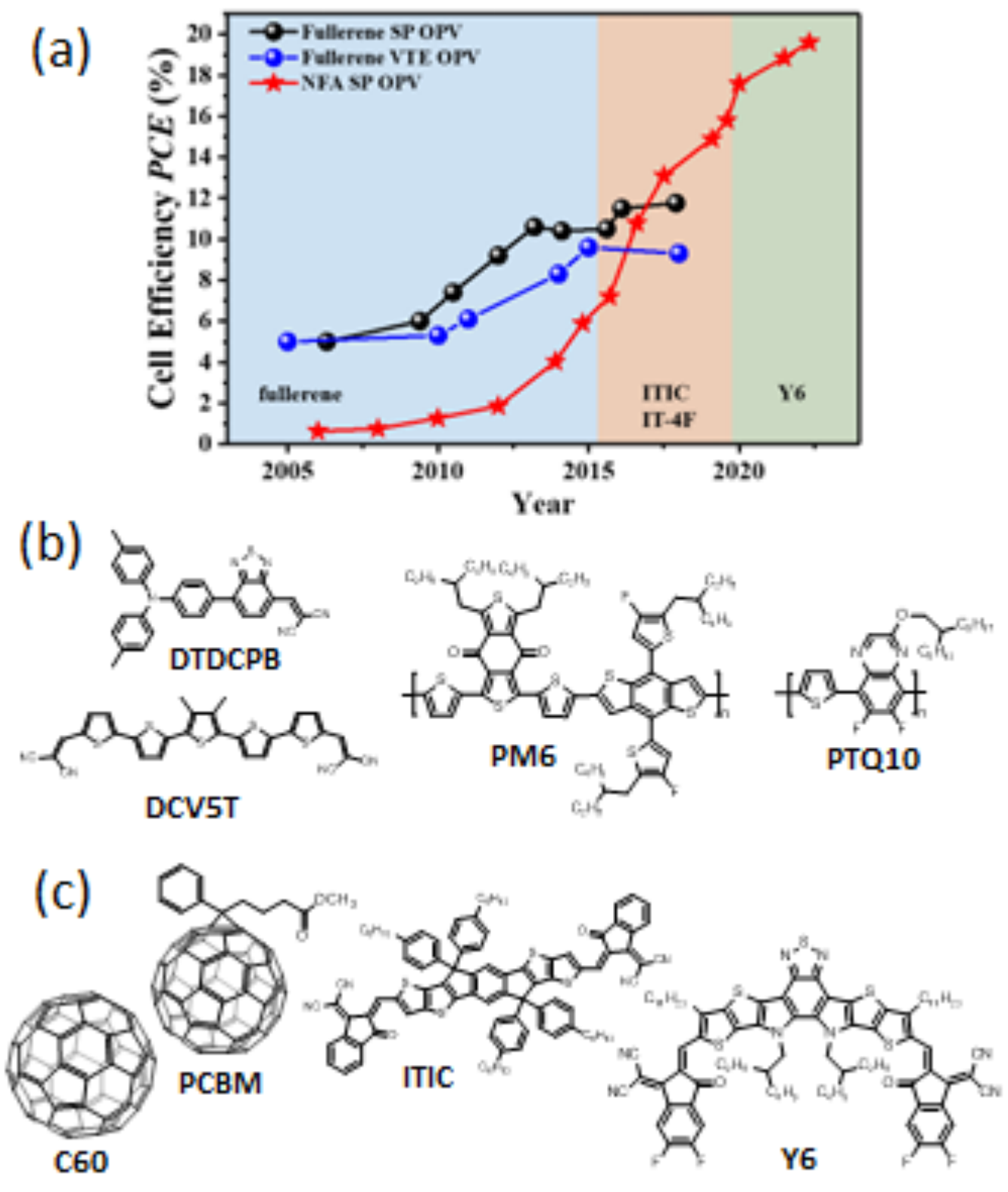

**Figure 9.** (a) Selected historic OPV single-junction cell performance. Reproduced with permission from Ref. [132]. Copyright 2021, Wiley. (b) Example donor and (c) acceptor

molecules showcasing the diversity of OPV absorber materials. Industrial OPV mostly uses undisclosed IP protected materials.

### *6.4 Concluding Remarks*

Through steady progress and disruptive advances, OPV has moved from a lab curiosity in the 1970s [146] to early-stage commercial products with fabrication capacity of several 100MWp/y online. Various technological approaches have been used, with future products expected to serve diverse growth markets. Sustained research efforts to advance the understanding of basic molecule-level processes, together with bolstering OPV's technological foundations, is expected to run side-by-side with application-oriented research. Here, research targets diverse performance metrics, providing the basis for further commercial product innovation. To unleash the potential of OPV, we encourage closer collaboration between industry – ranging from equipment manufacturers and chemical companies to module manufacturers and product developers - and academia, which requires appropriate funding structures as well as tackling barriers for collaboration. These include fostering academic work with undisclosed IP-protected materials; a situation that is common in related research on drug development and OLEDs. A better alignment between academic OPV research with the needs of industry will require a focussed effort on studying problems faced by industry and on a molecule and processing space that has relevance for practical manufacture. We see plenty of economic opportunities in the developing value chain, ranging from generating IP, upscaled molecule synthesis, to module manufacturing and installation – all of which are realistic to be achieved locally anywhere on the globe - meeting aspirations of energy security and local job generation.

**Acknowledgements**

P.K. thanks Anna Jungbluth and Irfan Habib for being a test audience. P.K. thanks EPSRC for funding for a Postdoctoral Fellowship (no. EP/V035770/1) and Linacre College for the award of a Junior Research Fellowship. We thank the EPSRC project ATIP (no. EP/TO28513/1) for financial support. A.J.G. thanks the Leverhulme Trust for an Early Career Fellowship (no. ECF-2022-445). D.G.L. thanks EPSRC for support (grant no. EP/V027131/1). T.A. acknowledges the finance support of Royal Society of Edinburgh via International Joint Project 1787.

## 7. Indoor Photovoltaics

Robert L. Z. Hoye,[1,2] Simon Aliwell,[3] Mathieu Bellanger,[3] James C. Blakesley,[4] Matthew J. Carnie,[5] James R. Durrant,[6,7] Marina Freitag,[8] Matthias Kauer,[3] George Koutsourakis,[4] Trystan Watson[7]

[1] Inorganic Chemistry Laboratory, Department of Chemistry, University of Oxford, South Parks Road, Oxford OX1 3QR, United Kingdom

[2] Department of Materials, Imperial College London, Exhibition Road, London SW7 2AZ, United Kingdom

[3] Lightricity Ltd., Edmund Halley Road, Oxford OX4 4GB, United Kingdom

[4] National Physical Laboratory, Teddington, Middlesex TW11 0LW, United Kingdom

[5] Department of Materials Science and Engineering, Swansea University, Bay Campus, Fabian Way, Crymlyn Burrows, Swansea SA1 8EN, UK

[6] Department of Chemistry and Centre for Processable Electronics, Imperial College London, 80 Wood Lane, London SW7 2AZ, United Kingdom

[7] SPECIFIC, Swansea University, Bay Campus, Swansea SA1 8EN, UK

[8] School of Natural and Environmental Science, Bedson Building, Newcastle University, Newcastle upon Tyne NE1 7RU, United Kingdom

### *7.1 Status*

Society is being revolutionized by an ecosystem of networked devices called the Internet of Things (IoT), which is embedding 'intelligence' into infrastructure and giving rise to 'smart' homes, offices, healthcare, agriculture, and many more [40]. The IoT could reduce the $CO_2$eq footprint across multiple sectors by making processes more efficient [147]. There are over 200 billion IoT devices currently [148], potentially expanding to over a trillion nodes in the near-future [149]. A critical question is how so many devices could be sustainably powered. Currently, IoT nodes are predominantly powered using primary or secondary batteries, but this will create substantial waste and strain on the supply of critical elements, as well as pose practicality challenges around changing billions of batteries each year [40]. Indoor photovoltaics (IPVs) are a highly promising alternative, given that nearly half of new IoT installations will be indoors [30], as well as the reliability and predictability of indoor light sources. Indoor lighting has a higher power density than other power sources for energy harvesters [40], with 350 μW $cm^{-2}$ available in a brightly-lit room with 1000 lux illuminance [150]. At a more common illuminance of 250 lux, IPVs with modest device area can harvest sufficient energy to power a range of communications protocols used in the IoT (figure 10a). Such IPVs can be integrated into IoT nodes (figure 10b), and either directly power the electronics (harvest-use), or store excess energy in an energy-storage device (*e.g.,* supercapacitor) to be used in the dark or at low illuminance (harvest-store-use). Such a design could lead to a perpetual power supply, enabling 'place-and-forget' IoT devices, and ameliorate the sustainability and practical challenges with the current reliance on batteries as the power source for IoT nodes.

c-Si is not well-suited for indoor lighting conditions, because of high dark currents and sub-optimal bandgaps [40], and hydrogenated amorphous silicon (a-Si:H) has instead become the commercial-standard material. But a-Si:H IPV PCEs only reach up to 21% (under specific LED conditions) [151], with most devices 4.4-9.2% efficient under most indoor lighting conditions (figure 10b) [40]. Furthermore, the device lifetime and durability requirements for IPV applications are much less stringent than for outdoor PVs. These factors open up space for new technologies to have an impact. The most common alternative technologies currently investigated for IPVs are dye-sensitized solar cells (DSSCs), organic photovoltaics (OPVs), III-V photovoltaics (commonly GaAs), and, more recently, lead-halide perovskites (LHPs). The progress in the PCEs reported from lab-based and commercial devices is shown in figure 10c. Commercial GaAs, OPV and DSSC IPVs have reported power output densities of 70–90 $\mu W\ cm^{-2}$ (GaAs) [150], [152], 15–17 $\mu W\ cm^{-2}$ (OPV)[153], 4–35 $\mu W\ cm^{-2}$ (DSSC) [152], [154], albeit all under different indoor conditions. Beyond these materials, over the past two years, Bi- and Sb-based perovskite-inspired materials (PIMs) have been discovered to have potential to reach high PCEs under indoor lighting (figure 10d). Furthermore, elemental selenium, the first photovoltaic material ever reported, was recently found to hold substantial promise for IPVs [155]. A rich materials space is therefore developing, offering multiple routes to achieving efficient and durable energy harvesters that can more sustainably power the billions of IoT devices.

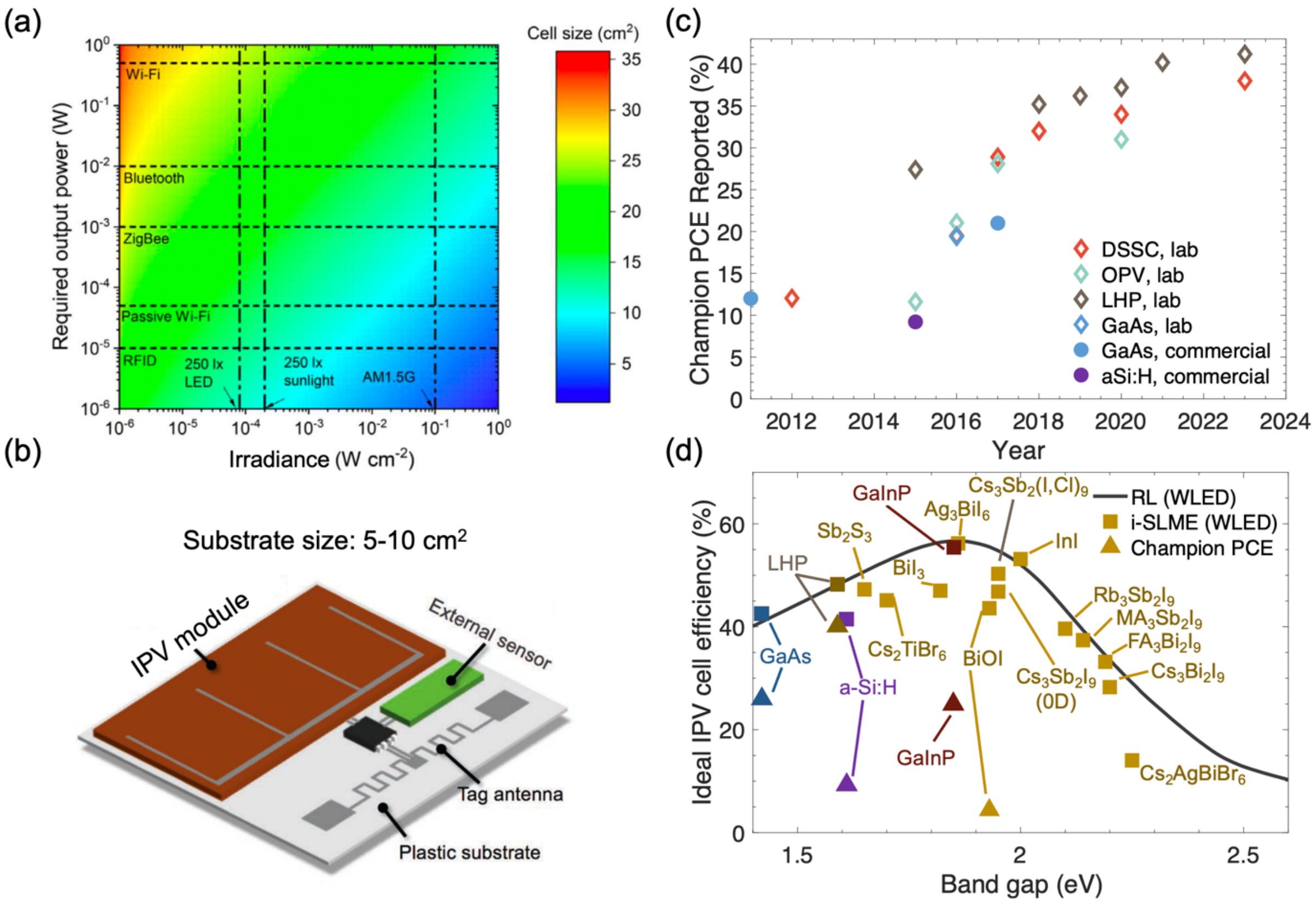

**Figure 10.** (a) Required output of common IoT communications protocols, and the power available from IPVs, depending on the light source and cell area. Reproduced under the terms of the CC-BY license from Ref. [156]. Copyright 2022, The Authors. (b) Illustration of integrating an IPV device with a wireless sensor. Reproduced with permission from Ref. [157]. Copyright 2019, Wiley. (c) Plot of champion efficiency over time for lab-based and commercial dye-sensitized solar cells (DSSCs), organic photovoltaics (OPVs), lead-halide perovskite photovoltaics (LHPs), and photovoltaics based on III-V compounds, along with a reference point for a-Si:H. Note: All devices measured under indoor lighting, but test conditions may vary between different groups, limiting the ability to make a direct comparison between different reported values. Refer to Table S1, SI for details. (d) Plot of the spectroscopic limited maximum efficiency (SLME) for a-Si:H, DSSCs, OPVs, LHPs, III-V and selected perovskite-inspired materials (PIMs) under 1000 lux white light-emitting diode illumination.

### *7.2 Current and Future Challenges*

As the IPV field grows in importance, with many new classes of materials appearing, several critical challenges will need to be addressed (figure 11a). Chief among them is adopting a rigorous set of protocols for device performance measurement and reporting, similar to those that are now commonplace in the outdoor PV field. Until very recently, there were no standards for IPV testing, with no specifications on how measurements should be made. Kirchartz and co-workers recently showed how a slight change in the spectrum of the LED light source (figure 11b) changes the PCE of the same OPV device from 11.67% to 12.93% (figure 11c) [158]. Furthermore, stray ambient light in the testing box, or shadowing of the diffuse light due to an aperture that is too thick, can lead to unintended errors in the reported performance [159], [160]. Developing standard indoor test conditions and requirements for measurement and reporting will be critical to ensure fair comparisons between groups, and support the growth of this area of research. A set of standards for indoor PV measurements has just been published: IEC TS 62607-7-2:2023 [161]. These standards will support the development of the necessary products and market for indoor PV testing, such as calibrated standard light sources, calibrated reference cells, various measurement sensors and even whole indoor PV testing systems.

Part of the challenge with IPVs is the wide range of conditions experienced by the devices, which vary depending on the application. For example, it is common in the academic literature to report PCEs over a narrow range of illuminances (often 1000 lux). But IoT nodes can experience illuminances from <50 lux (in corners) to >1000 lux (*e.g.*, near windows), and the performance of devices can change widely over this illuminance range [42], [160], [162]. This demonstrates that the linearity of indoor PV products across the relevant illuminance range is a crucial parameter. Furthermore, the stress factors (*e.g.*, thermal, moisture, dust) experienced by IPVs can widely vary by application. Developing protocols for accelerated degradation testing will be important for achieving commercial IPVs that are durable over the lifetime of the IoT node, and to support the development of effective encapsulation materials. In addition, it should be ensured that IPV devices that could be exposed to direct sunlight (taken outdoors or placed next to a window) are safe for the users, even if such conditions could degrade them, and meet all existing product safety standards for consumer electronics.

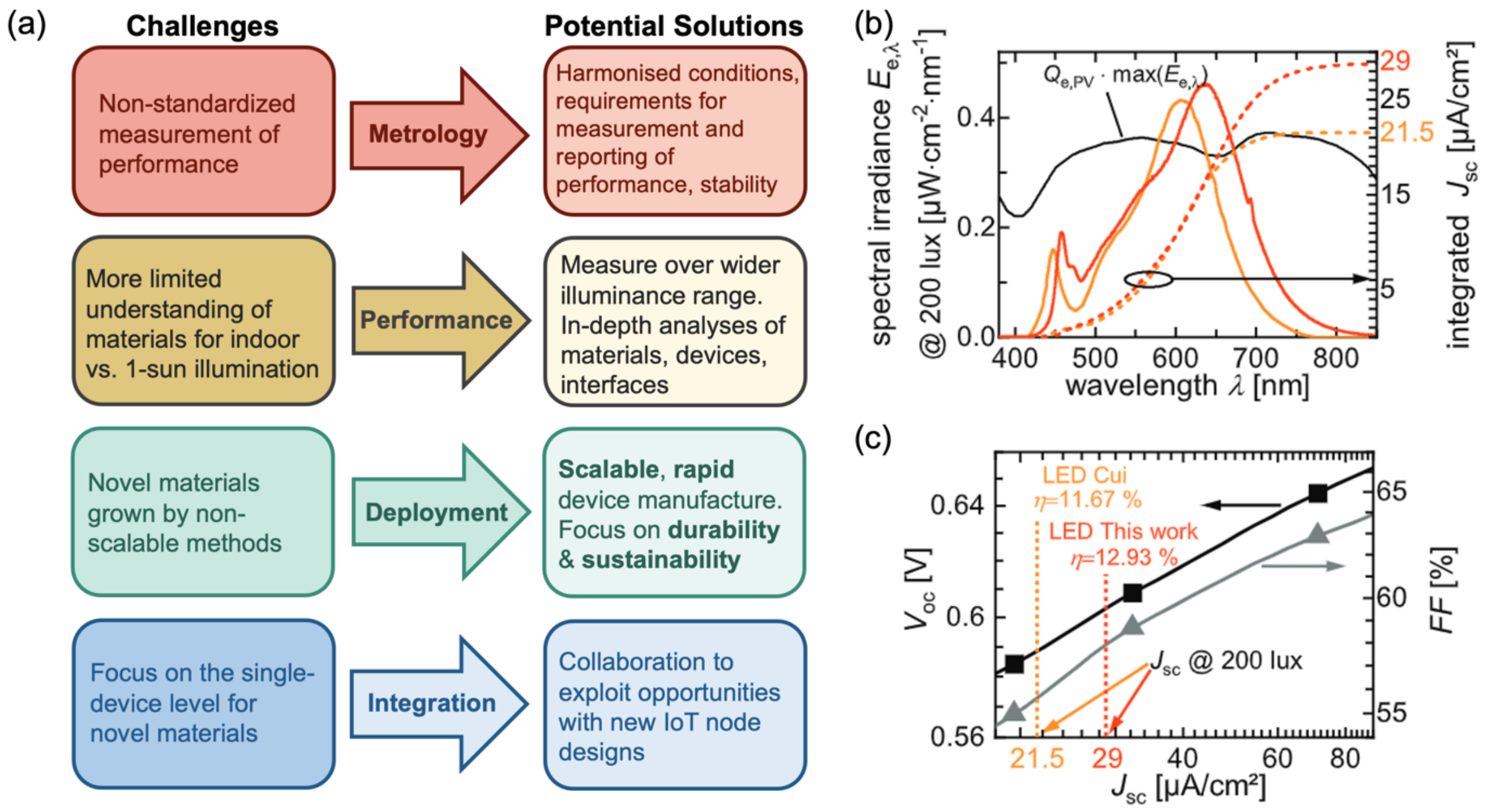

**Figure 11** (a) Key challenges and potential solutions for IPVs. (b,c) Illustration of how much efficiencies can change depending on the light source, emphasizing the need for standardization and methods to compare devices made between different groups. This example is with an IPV based on the organic blend PBDB-TF-T1:Y12. Part (b) shows the IPV external quantum efficiency, along with the LED light sources used by two different groups, and the corresponding short-circuit current densities ($J_{SC}$s) that would be obtained from the same device. Part (c) shows a plot of the open circuit voltage ($V_{OC}$) and fill factor (FF) against $J_{SC}$, and the interpolation of $V_{OC}$ and FF from the integrated $J_{SC}$ value, thus showing how the power conversion efficiency ($\eta$) changes depending on the light source. Reproduced under the terms of the CC-BY license from Ref. [158]. Copyright 2021, The Authors.

An important advantage of the novel materials investigated for IPVs is their widely tunable optoelectronic properties based on the molecular structure (OPV, DSSCs) and composition (LHPs, PIMs). Whilst most efforts have been to optimize these materials for 1-sun illumination, it will be critical to focus on tailoring these materials to indoor light spectra, especially to achieve high operational voltages that linearly vary with illuminance. This is to minimize the number of cells that need to be connected in series to deliver sufficient voltage to operate IoT electronics, and ensure reliability in their performance over a wide range of illumination levels. On the other hand, requirements for sheet resistance and cell size might be less restrictive than for outdoor applications, since lower current densities are expected at such low irradiances as indoor conditions. These factors can result in optimal layer thicknesses being larger for IPVs compared to outdoor PVs. In-depth understanding of how defects in the bulk and interfaces in these IPV devices affect charge-carrier recombination and transport under weak light is also needed. For example, for outdoor PV applications of emerging technologies (such as LHPs), it is common to use light-soaking to saturate interface traps, leading to an initial increase in PCE. The substantially lower power densities in indoor lighting may be insufficient to produce similar burn-in effects, and the ratio of photo-to-dark current will be lower, which could lead to increased trapping rates. As such, controlling trap densities can be particularly important for IPVs.

The scalability and manufacturability of new materials also needs to be considered. Whilst lab-based studies commonly focus on <1 $cm^2$ area devices, device areas on the order of tens of $cm^2$ would be needed (figure 10a). Methods to manufacture materials rapidly over large area and match the performance of lab-based prototypes, as well as module designs tailored for low irradiance illumination, will be critical to realize hundreds of thousands to millions of devices per year that can have an impact on the IoT ecosystem [37], [163]. At the same time, it will be important to achieve highly-efficient IPVs that are reliably made on the $mm^2$ area level or smaller to power miniature devices, such as smart dust.

### *7.3 Advances in Science and Technology to Meet Challenges*

The UK National Physical Laboratory, along with international colleagues, have been working to develop standardized conditions for IPV testing, which include a typical white LED spectrum, specific illuminance levels and specific recommended measurement procedures. If these protocols are widely adopted in the future, and commercial testing equipment following these protocols becomes readily available, then reliable comparisons could be made between labs. Furthermore, the independent, accredited validation of published results becomes possible.

Beyond metrology, the sustainability and safety of IPV technology needs to be considered through lifecycle analyses. In this regard, the toxicity of readily-accessible and bioaccumulative lead from LHPs may not be appealing for IPVs used in consumer electronics. It is also important to consider the toxicity of solvents and catalysts used in the preparation of materials, especially for OPV and DSSCs [40]. Furthermore, strategies to recycle IPVs, and reuse valuable materials at the end-of-life, will be important, especially if these devices are to be deployed at the billion-device level or beyond. Developing devices that are not reliant on scarce elements (*e.g.*, In in indium tin oxide) will also be important. Recent analyses have shown that PVs are not as sensitive to the series resistance under indoor lighting as under 1-sun, opening up the use of transparent conducting electrodes with higher sheet resistance, but based on Earth-abundant elements [156].

The solar resource for outdoor PV systems is well modelled and understood, which allows accurate energy rating, yield modelling, performance assessment and long term forecasting of solar PV generation. Such tools and data are not available for the indoor light resource. There is currently a need for similar studies for indoor environments, to better understand the indoor light resource, to be able to conduct yield estimations and shape requirements for indoor PV materials and devices. Indoor light is highly variable, depending on the specific environment of the application, so potentially an application targeted approach of modelling the indoor light resource and a series of typical conditions might prove useful for modelling, energy rating and yield estimations.

Finally, the current development of novel materials for IPVs has focused almost exclusively on the IPV itself. Stronger collaborations with groups developing IoT nodes will be important to tailor the materials to the specific needs to the IoT, as well as to exploit new opportunities, such as on-node machine learning [42], [164].

### *7.4 Concluding Remarks*

The IPVs field has recently been reinvigorated with a wide range of new materials classes that offer the opportunity to achieve high-performance, durable, scalable and cost-effective devices that can be used to power IoT nodes. But rationally developing these materials requires the harmonization of performance measurements, as well as accelerated degradation testing. As the IPVs field heats up, dedicated efforts to develop materials with tailored optoelectronic properties for indoor light spectra will be necessary, especially to achieve a linear variation in the operational voltage with illuminance. Understanding and controlling the role of traps in the bulk and interface under diffuse lighting will also be necessary. As manufacturing processes are developed for large-scale deployment, it will be essential to consider the lifecycle of the devices, with processes minimizing environmental impact and embedded energy, as well as approaches to recycle devices at end-of-life. Stronger cross-disciplinary collaboration will be needed to ensure greater synergies in the development of new IPV technology with improvements in IoT node electronics to realize new opportunities and expand the reach of the IoT. Addressing these challenges and seizing the opportunities opened up by emerging IPV technology can deliver a more sustainable IoT ecosystem and support its rapid expansion.

**Acknowledgements**

All co-authors arranged alphabetically by surname. R.L.Z.H., M.B., M.K. and S.A. acknowledge funding from the Henry Royce Institute Industrial Collaborative Programme scheme, funded by EPSRC (no. EP/X527257/1). J.C.B and G.K acknowledge funding from the Department for Science Innovation and Technology through the National Measurement System (NMS) programme. J.B, M.C., J.R.D and TW acknowledge funding from the EPSRC project ATIP (no. EP/T028513/1). M.C. also acknowledges funding from EPSRC (no. EP/W025396/1).

## 8. Indoor Dye-Sensitized Photovoltaic Cells: Advances, Challenges, and Prospects for IoT Applications

Elizabeth A. Gibson[1] and Marina Freitag[1]

[1]School of Natural and Environmental Sciences, Bedson Building, Newcastle University, NE1 7RU, Newcastle upon Tyne, UK

### *8.1 Status*

Dye-sensitized photovoltaic cells (DSCs) consist of a mesoporous semiconducting layer, usually titanium dioxide ($TiO_2$), functioning as a working electrode (WE), with sensitizer molecules adsorbed onto it. The counter electrode (CE) faces the sensitizer, and a redox mediator is placed between the CE and WE. Upon light absorption, photo-induced electron transfer occurs from the sensitizer to the $TiO_2$. The redox mediator regenerates the dye, facilitating the transfer of positive charges from the WE to the CE. DSCs primarily absorb in the 400 to 650 nm visible wavelength region, outperforming GaAs solar cells under diffuse light conditions while being inexpensive and eco-friendly [165].

DSCs can maintain high photovoltage even under ambient light illumination, attributed to the tunable energy levels in Cu(II/I) electrolyte systems, reduced non-radiative recombination, and fast charge separation processes in organic dyes. Molecular engineering of the dyes and their combination (co-sensitizers) for improved matching of absorbance with the emission of artificial light sources has significantly pushed the efficiency of DSCs for indoor photovoltaics (IPVs, figure 12). A PCE of 28.9% was observed under 1000 lux fluorescent light tube using $[Cu(tmby)_2]^{2+/1+}$ redox coupled with $TiO_2$ films co-sensitized with dye D35 ((*E*)-3-(5-(4-(bis(2′,4′-dibutoxy-[1,1′-biphenyl]-4-yl)amino)phenyl)thiophen-2-yl)-2-cyanoacrylic acid) and XY1 ((*E*)-3-(4-(6-(7-(4-(bis(2',4'-bis((2-ethylhexyl)oxy)-[1,1'-biphenyl]-4-yl)amino)phenyl)benzo[c][1,2,5]thiadiazol-4-yl)-4,4-bis(2-ethylhexyl)-4H-cyclopenta[2,1-b:3,4-b']dithiophen-2-yl)phenyl)-2-cyanoacrylic acid) [150]. The continued development of panchromatic rigid-structure dyes, alternative hole transport materials, and design flexibility has enabled improved PCEs, currently reaching 13% under AM 1.5G illumination and 34% under indoor lighting (1000 lux fluorescent lighting) [42].

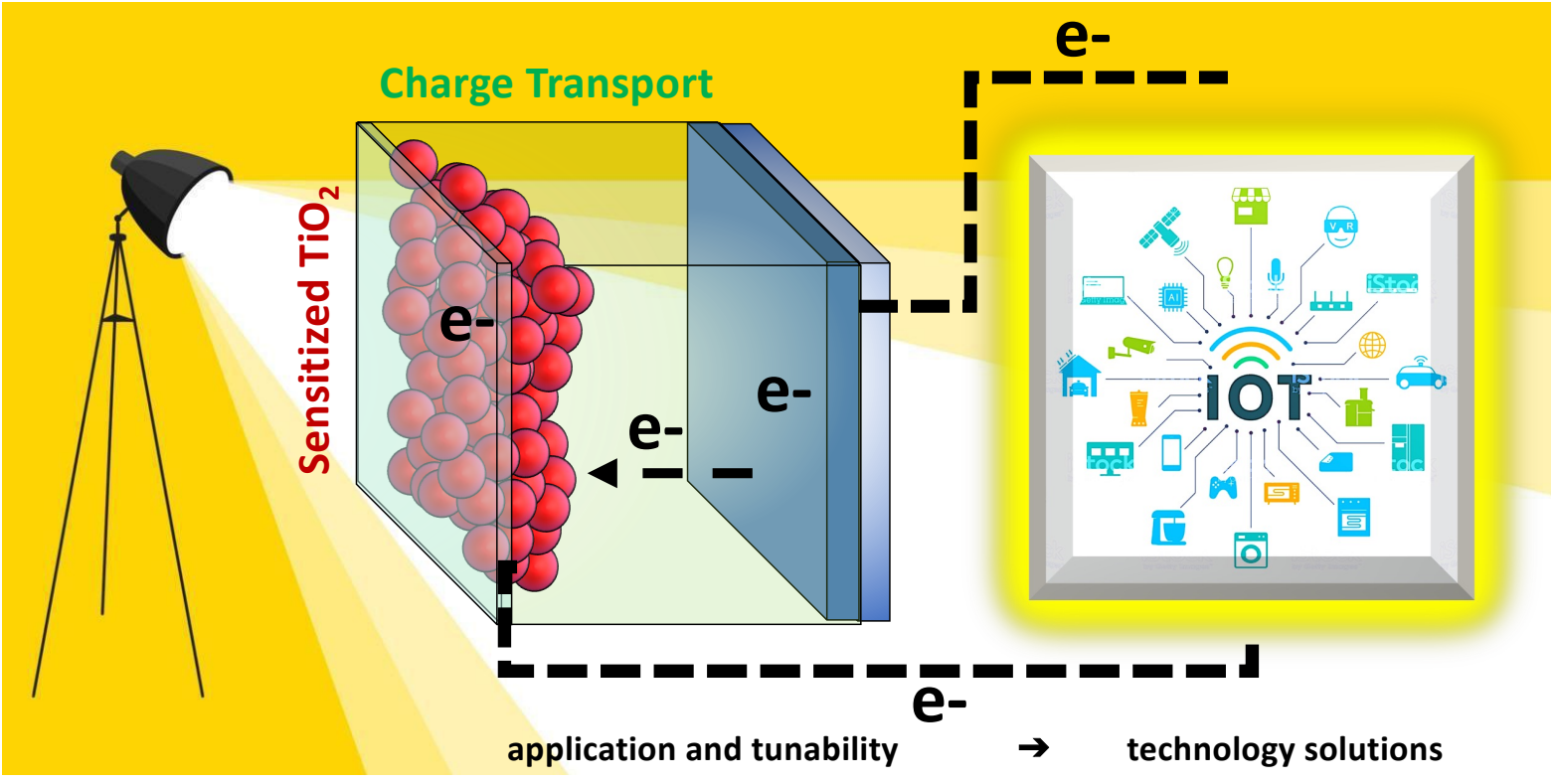

**Figure 12.** Illustration of a DSC-based power source for IoT applications, utilizing dye-sensitized materials to convert indoor light into electrical energy.

Recently, DSCs have made remarkable progress in harvesting energy from artificial light sources, making them suitable for various low-power indoor devices. In 2020, DSCs successfully powered battery-free IoT devices capable of machine learning under ambient light conditions. As the number of IoT devices rapidly increases (see Section 7), there is an urgent need for sustainable and efficient energy sources and management practices in ambient environments. In response, a high-efficiency ambient photovoltaic based on sustainable non-toxic materials was developed, featuring an unprecedented PCE of 38% and 1.0 V open-circuit voltage at 1000 lux (fluorescent lamp). This development utilized a long short-term memory (LSTM) based energy management system with on-device prediction on IoT sensors solely powered by ambient light harvesters [166]. By harvesting ambient light combined with artificial intelligence, fully autonomous self-powered sensor devices can be created for industry, healthcare, homes, and smart cities.

### *8.2 Current and Future Challenges*

To enable the widespread implementation of self-powered wireless devices, indoor photovoltaics (IPVs), particularly DSCs, must overcome several challenges to achieve power conversion efficiencies (PCEs) closer to the maximum theoretical value of 52%. The integration of DSCs as a sustainable power source for IoT devices necessitates addressing these challenges.

High PCEs in indoor DSCs can be achieved using co-sensitized systems employing dyes that absorb at 550-600 nm and a co-sensitizer absorbing in the blue region, providing an excellent match to ambient light spectra. This approach reduces electron recombination rates. However, further improvements require interfacial engineering, high redox potential redox mediators, and the development of large-scale solid-state devices. Solid-amorphous copper-based hole transport materials (HTMs) show promise, but scalable deposition methods are needed.

The IPV community faces challenges due to diverse commercial light sources and fluctuations in illuminance and spectra (see Section 7). To address this, efficient IPVs must have ideal absorber materials with photoresponse spectra matching indoor light, high charge carrier mobility, high external quantum efficiency (EQE), and low trap-aided charge recombination.

The Shockley-Queisser efficiency limit for IPVs under 1.5 W $m^{-2}$ (500 lx) from a White-LED lamp is around 52%.

Addressing these challenges requires interdisciplinary collaboration between chemistry, engineering, and computer science. Molecular engineering and novel semiconductor materials can lead to increased dye loading and reduced interfacial electron recombination. Alternative charge carrier materials, solid-state devices, and backside illumination techniques can improve DSC performance and long-term stability. Innovative energy storage solutions, such as supercapacitors or non-conventional storage approaches, enable constant power delivery even without indoor light. Integration of artificial intelligence and machine learning techniques can create efficient energy management systems that adapt to changing conditions and optimize power usage.

Indoor photovoltaics (IPVs) are potential power sources for IoT devices, but DSC efficiency under one sun conditions has not improved significantly over the past decade. The maximum feasible efficiency for DSCs under one sun conditions is 23.78% at a bandgap of 1.34 eV, while under indoor light conditions, efficiency ranges from 35-38% [166]. DSCs are suitable for powering IoT devices in smart homes, with a maximum power intensity of >100 μW $cm^{-2}$ under indoor conditions, as shown in figure 13.

To improve DSC performance, researchers should focus on reducing voltage loss by optimizing dye and electrolyte structures, increasing the molar extinction coefficients of dyes, and developing environmentally friendly solid-state or quasi-solid state DSCs for indoor applications. By addressing these challenges, we can pave the way for efficient, self-powered wireless devices, benefiting buildings, industries, and households.

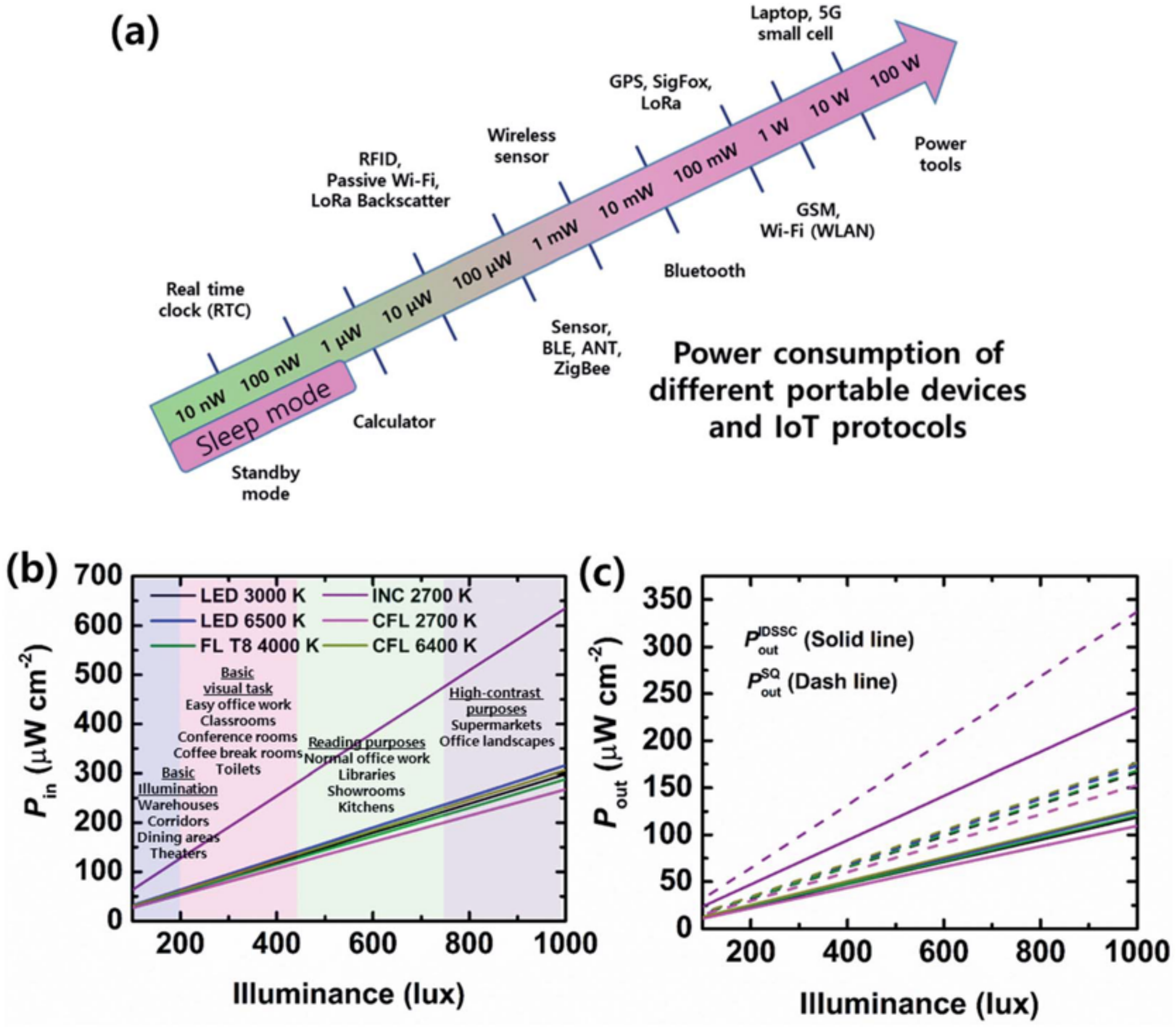

**Figure 13.** (a) Power consumption comparison for various portable devices and IoT protocols(b) Incident power intensity for different light sources across diverse lighting consumption zones. (c) Output power intensity for ideal SQ photovoltaic cells and IDSCs under a variety of light sources. Reproduced with permission from Ref. [167]. Copyright 2021, The Royal Society of Chemistry.

### *8.3 Advances in Science and Technology to Meet Challenges*

Achieving significant progress in indoor applications of DSCs necessitates interdisciplinary collaboration, encompassing chemistry, materials science, engineering, and computer science. Addressing the limitations in existing DSCs requires the development of new materials, strategies, and device architectures [167], [168]:

1. New sensitizers and co-sensitizers: Designing novel organic and inorganic dyes with extended absorption spectra, high molar extinction coefficients, and better molecular orbital energy alignment with the semiconductor's conduction band will improve energy conversion efficiency. Additionally, the synergistic combination of dyes as co-sensitizers could increase the absorption range, reduce charge recombination, and enhance stability [169].
2. Advanced semiconductors: Exploring alternative semiconducting materials with higher conduction band energies and improved surface properties, such as $Zn_2SnO_4$, $SrTiO_3$, and $BaTiO_3$, will aid in boosting $V_{OC}$ and reducing electron recombination. Novel synthesis methods and surface treatments can further optimize dye loading and charge transport [170].
3. Solid-state hole transport materials (HTMs): Replacing liquid electrolytes with solid-state HTMs can mitigate issues like leakage, solvent evaporation, and corrosion. Conducting polymers, metal-organic frameworks, and perovskite-based materials offer potential avenues for solid-state HTMs with efficient charge transport, enhanced stability, and reduced recombination losses.
4. Innovative device architectures: Developing new device structures, such as backside illumination designs and carbon-based composites at counter electrodes, can facilitate solid-state monolithic devices with improved performance and stability.
5. Machine learning and data analysis: Implementing machine learning algorithms for device optimization and data analysis can lead to the discovery of new materials and strategies, accelerating the development of high-performance DSCs for indoor applications.
6. Energy management and storage: Investigating efficient power converters and energy storage solutions, such as supercapacitors and novel energy storage devices, can ensure constant power delivery to indoor electronics during periods of insufficient lighting [42].

In summary, overcoming current challenges in DSCs for indoor applications demands the development of novel materials, a deeper understanding of dye-charge transport material interactions, and the implementation of solid-state components and device architectures [171]. Coupling these advances with efficient energy management and storage solutions will result in the next generation of DSCs, revolutionizing indoor photovoltaics and powering a wide array of electronic devices [42], [168].

### *8.4. Concluding Remarks*

In conclusion, the synergy between DSC and IoT development is crucial for realizing self-sustaining, energy efficient IoT devices. This involves boosting DSC performance by refining dye and electrolyte structures, elevating molar extinction coefficients, and introducing eco-friendly solid-state or quasi-solid-state electrolytes. Concurrently, IoT devices should prioritize energy-efficient designs, adopting low-power wireless communication protocols like Zigbee and LoRaWAN, and implementing adaptive sensing algorithms. Incorporating advanced energy management systems with machine learning and edge computing strategies will optimize power consumption. Moreover, efficient system integration should facilitate seamless integration of DSCs, energy storage components like supercapacitors, and energy-conscious hardware like low-power MCUs. By concentrating on these technical aspects, researchers can usher in a new era of smart, interconnected systems that excel in both indoor and outdoor environments, revolutionizing the IoT domain.

## Acknowledgements

M.F. acknowledges the support by the Royal Society through the University Research Fellowship (no. URF\R1\191286), Research Grant 2021 (no. RGS\R1\211321), and EPSRC New Investigator Award (no. EP/V035819/1).

## 9. Space Photovoltaics

Jihoo Lim[1,2], Craig Underwood[3], Dan Lamb[4], Louise C. Hirst[5,6], and Jae Sung Yun[1]

[1] School of Computer Science and Electronic Engineering, Advanced Technology Institute (ATI), University of Surrey, Guildford GU2 7XH, United Kingdom
[2] Australian Centre for Advanced Photovoltaics (ACAP), School of Photovoltaic and Renewable Energy Engineering, University of New South Wales, Sydney, NSW 2052, Australia
[3] Surrey Space Centre, School of Computer Science and Electronic Engineering, University of Surrey, Guildford, Surrey, GU2 7XH, United Kingdom
[4] Oxide and Chalcogenide Facility, Centre for Integrative Semiconductor Materials, Faculty of Science and Engineering, Swansea, University, SA1 8EN United Kingdom
[5] Department of Physics, University of Cambridge, Cambridge, CB3 0HE, United Kingdom
[6] Department of Materials Science and Metallurgy, 27 Charles Babbage Road, Cambridge CB3 0FS, UK

### *9.1 Status*

Solar cells for space applications have served as a primary source of power for thousands of satellites. For over sixty years, space solar cells have been utilized in spacecraft and satellites starting in 1958, with the launch of the first solar powered satellite Vanguard 1 employing single junction silicon solar cells (SSCs) with efficiency around 7-8% (see figure 14(a) inset) [172]. In the 1960s, the efficiency of such cells was enhanced to 12% and cadmium sulfide (CdS) solar cells were also explored due to their low mass and flexibility. However these cells still suffered from low efficiency and instability [173], thus during the 1970s, SSCs remained the preferred material for photovoltaic technology in space [174]. Demand for more powerful satellite payloads led to the development of higher efficiency alternatives, and in recent decades III-V triple junction devices have become the industry standard, powering satellites delivering spaced based services, as well as notable space exploration missions such as the James Webb Space telescope with its record breaking 60 $m^2$ panel. A notable exception is the International Space Station (ISS), which has been running on silicon arrays since its first deployment in 1998 and was only recently began adding III-V multijunction arrays. In addition to the powering the satellites, Space Based Solar Power (SBSP), first studied seriously in the 1970s by NASA, has re-emerged as a potential solution to the problem of providing terrestrial baseload electrical power from carbon-free energy sources under "net zero" [175]. It involves building large solar arrays in space and transmitting the energy to terrestrial stations via microwaves or lasers. These space arrays offer continuous illumination, making them more efficient than ground-based solar farms. However, the main challenge lies in the enormous size of the space infrastructure required, thus, highest possible specific energy PV cells are ideal for this purpose.

Industry standard III-V multijunction solar cells are fabricated using Metal Organic Vapor Phase Epitaxy (MOVPE), to achieve near pristine single crystal semiconductor layers at the wafer scale. Producing devices with this technique costs ~\$40 $W_p^{-1}$ [176], exceeding that of a silicon equivalent by two orders of magnitude. Historically, the additional cost of the higher efficiency III-V devices has been small in the context of over mission cost. However, reductions in the cost of launch, down to US\$1400 $kg^{-1}$ on recent Falcon Heavy demonstrations, over an order of magnitude lower than that of the Space Shuttle missions of the preceding decade, have created renewed interest in potentially viable cheaper power system options using thin film solar cells such as cadmium telluride (CdTe), copper indium gallium diselenide (CIGS), and perovskite solar cells for space applications are required to be considered. Apart from their cost, thin film solar cells have the potential to achieve high specific power (W $g^{-1}$) in comparison to MJSCs. Figure 14(b) provides an overview of the efficiency, mass density (mg $cm^{-2}$), and specific power potential of various solar cell technologies, with a focus on approximately 70% SQ limiting efficiency. The chart indicates that multiple junction perovskites, denoted as 2J or 3J solar cells, hold promising potential in terms of specific power.

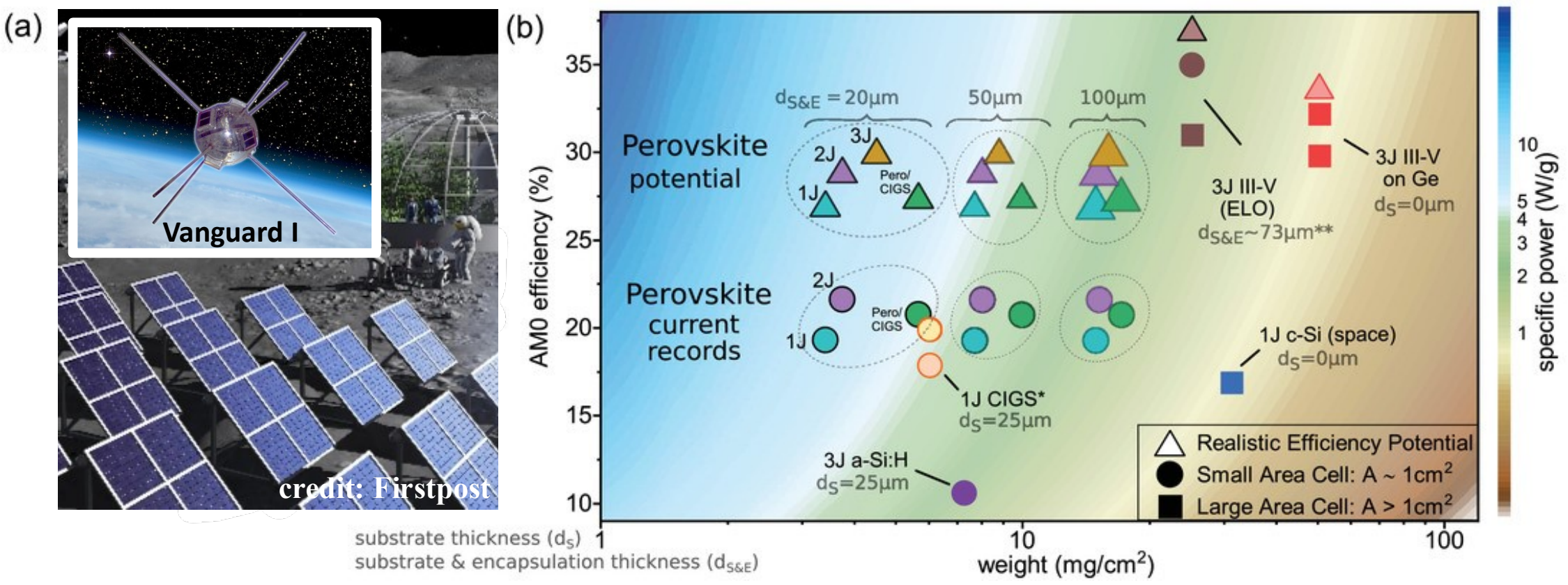

**Figure 14.** (a) Illustration of PV power generation on a planet in space. Inset is an image of Vanguard 1, the earliest satellite powered by solar energy (credit: National Space Science Data Center). (b) An overview of the efficiency, mass density, and specific power potential of diverse solar cell technologies. Reproduced under the terms of the CC-BY license from Ref. [177]. Copyright 2021, The Authors, published by Wiley.

### *9.2 Current and Future Challenges*

The current industry workhorse technology is a III-V triple solar cell on a germanium substrate (InGaP/InGaAs/Ge). These devices can supply efficiencies in excess of 30% under AM0 illumination (extra-terrestrial spectrum). Their efficiency is limited by subcell current matching, with relative subcell thicknesses engineered to balance high beginning of life performance with extended on-orbit lifetime following exposure to damaging particle radiation. Key challenges for this technology type are:

1) Lowering the cost of manufacture in line with cost of launch reductions to meet demand for large area applications such as space based solar power. A recent NREL technoeconomic analysis estimates that fabrication costs could be reduced to US\$0.4 $W^{-1}$ through a number of measures, including industry scale adoption of low cost epitaxy techniques, such as Hydride Vapour Phase Epitaxy (HVPE) [176], [178];
2) Further increasing the AM0 PCE by increasing the number of junctions employed, through alternative device fabrication approaches such as metamorphic growth, wafer bonding and mechanical stacking of subcells.

Hence, low-cost thin film solar cells have been considered as viable alternatives. CdTe thin film single junction solar cells have been investigated for space applications. Lamb *et al.* have published extensively on CdTe deposited on 100 μm thick cover glass and reported the first *I-V* data from a low-earth-orbit demonstration experiment [179]. No delamination effects have been observed after 6 yr in space. Depositing directly on cover glass has multiple benefits: a bend radius of 50 mm for the CdTe device structure offering stowage volume reduction and supporting role out of new deployment techniques, increased specific power by replacing the conventional substrate with the lightweight radiation-hard cover glass and finally, cost reduction by eliminating the need for both a substrate and cover glass [180]. CdTe has been shown to be up to 2 orders of magnitude more radiation tolerant than the MJSC, which could increase end-of-life efficiency and be crucial for certain mission environments [181].

Recently, there has been a notable rise in interest in using lead-halide perovskite solar cells (PSCs) for space applications due to their exceptional radiation stability and the highest specific power of 29.4 $W_p$ $g^{-1}$ of all PV technologies [182]. PSCs have demonstrated superior tolerance to electron radiation, and proton radiation. However, perovskite films can be decomposed under ultra-high vacuum conditions with illumination [183]. The requirements for solar cells used in space applications are not only confined to radiation but also include thermal stability, as space solar cells are subjected to severe cyclic temperature fluctuations. Of utmost significance is that PSCs have lower thermal stability if manufactured at less than 150 °C, which is affected by the organic charge transport layers used, in addition to the intrinsic properties of the lead-halide perovskite. Phase transitions can also occur in the perovskite absorber over a wide temperature range. In addition to the materials themselves, the overall stability of a device is significantly influenced by its interfaces. Therefore, it is of utmost importance to address and mitigate any interfacial defects.

### *9.3 Advances in Science and Technology to Meet Challenges*

To evaluate all relevant effects of the environmental parameters simultaneously, it is crucial to comprehensively monitor the performance of different kinds of solar cells under an in-flight test. In the past and present, silicon solar cells have been widely utilized for most space solar applications. However, studies have shown that these cells can degrade due to radiation exposure in a relatively short time [184]. To address this, the use of GaAs/Si solar cells with higher efficiency and specific power was investigated. However, the performance still suffered from radiation damage, as demonstrated in the Engineering Test Satellite (ETS-VI) [185]. CIGS solar cells have shown exceptional radiation tolerance as demonstrated by the MDS-1 Tsubasa satellite, which operated in a geostationary transfer orbit with a severe radiation environment. Nonetheless, there are still some issues that need to be addressed, such as the damage that can occur to the electrical interconnection affected by the harsh space environment, as well as the need for a protective layer to ensure long-term stability. Figure 15(a) shows the daily fluence rate for protons and electrons of the MDA-1 satellite after passing through 100 µm-and 500 µm-thick cover glasses [186]. During the in-flight test, the CIGS cells with 9.2% efficiency and with 500 µm thick cover glasses showed almost no degradation (figure 15(b)) whereas the 24.5% efficient InGaP/GaAs 2J cells demonstrated some degradation of over 10% in maximum power (figure 15(c)) [187]. Furthermore, an inflight-test study of CdTe solar cells on AlSat-1N CubeSat over a period of up to 6 yr indicated no delamination and no deterioration in short circuit current or series resistance [179]. Nevertheless, there was a decrease in shunt resistance, which resulted in a reduced fill factor and open-circuit voltage as shown in figure 15(d). This was ascribed to the diffusion of gold from the back contact into the CdTe layer forming micro-shunts along the grain boundaries – pre-flight proton and electron irradiation tests showed that the technology was two orders of magnitude more robust to radiation damage than silicon or conventional III-V cells. PSCs have undergone rather limited real space environment tests compared to other PV technologies. There has been a short suborbital rocket flight test (MAPHEUS-8) [188]. The device performance of the PSCs was measured for approximately 6 min, at an altitude of 239 km. While the results indicate that PSCs can be a viable option for space applications, it is still just the initial phase. Further testing and experiments in the context of space are necessary to identify any key limitations. To improve stability, researchers should also investigate alternative materials such as all-inorganic and low-dimensional perovskite.

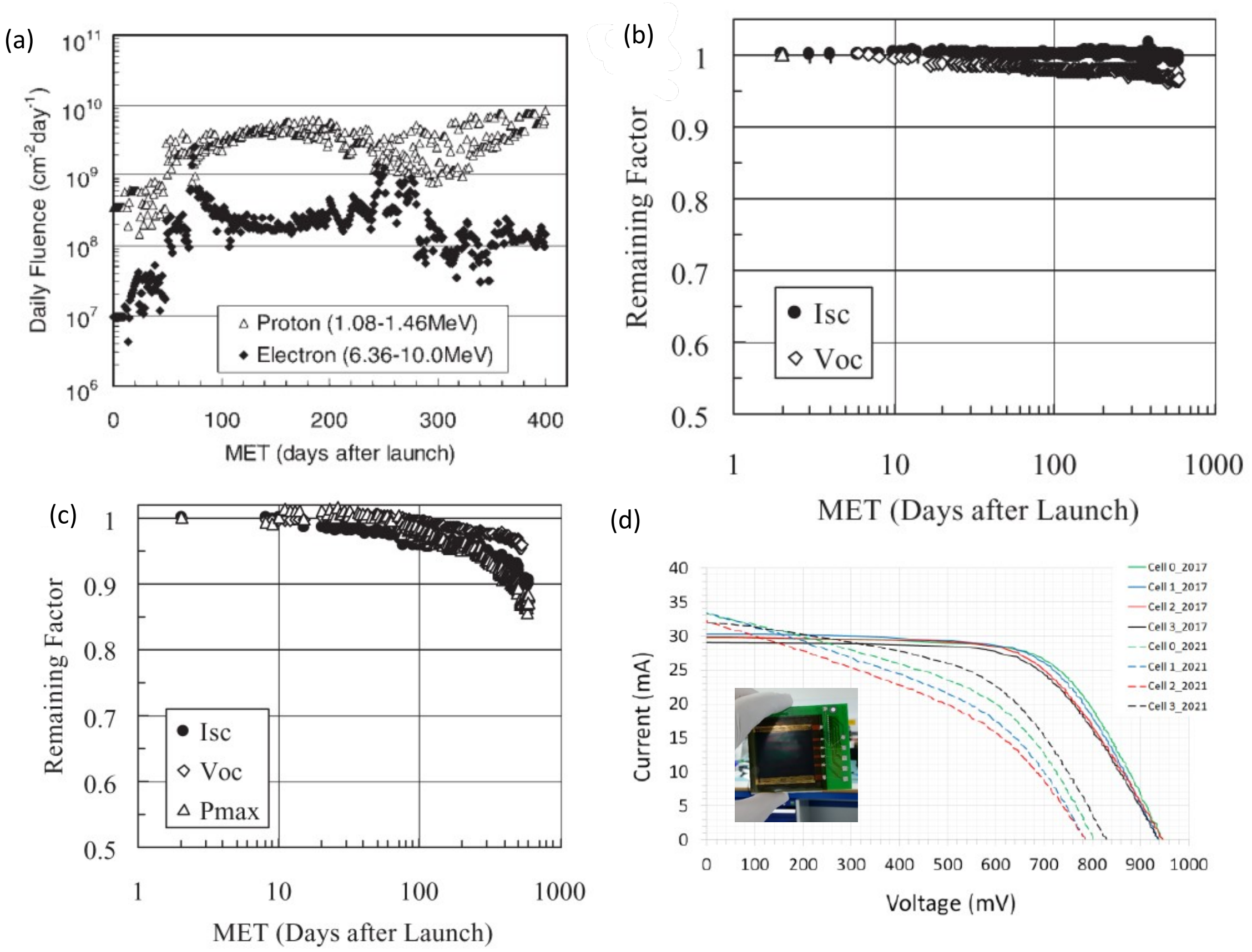

**Figure 15.** (a) Differential incident proton dose spectra on MDS-1 mission and the transmitted spectra after passing through 100 µm- and 500 µm-thick cover glasses. Reproduced under the terms of the CC-BY license from Ref. [186] Copyright 2004, The Authors, published by Wiley. (b) Degradation trend of $I_{sc}$ and $V_{oc}$ of the CIGS cell with 500 µm thick cover glass on MDS-1 expressed as a remaining factor. Reproduced under the terms of the CC-BY license from Ref. [187]. Copyright 2005, The Authors, published by Wiley. (c) Degradation trend of $I_{sc}$, $V_{oc}$ and $P_{max}$ of the InGaP/GaAs 2J cells with 500 µm thick cover glass on MDS-1 expressed as a remaining factor. Reproduced under the terms of the CC-BY license from Ref. [187]. Copyright 2005, The Authors, published by Wiley. (d) AlSat-1N TFSC payload in-orbit I-V results for 5th January 2017 compared with those from the 8$^{th}$ of February 2021. Inset shows AlSat-1N TFSC payload external board (front surface showing the contacts to the four 1 cm$^2$ rectangular cells and two gold bus bars through the 100 µm-thick space glass superstrate).

In addition to powering satellites, Space Based Solar Power (SBSP), first studied seriously in the 1970s by NASA, has re-emerged as a potential solution to the problem of providing terrestrial baseload electrical power from carbon-free energy sources under “net zero” [175].

SBSP involves constructing kilometric-scale solar arrays in space, coupled to a means of transmitting the energy generated to terrestrial receiving stations *via* microwaves or laser, where it is converted back into electrical power, and fed into the national energy grid. The advantage of placing the arrays in orbit, rather than on the ground, it that they can be illuminated permanently, 24 h a day and thus, even given the transmission losses, can be a more effective means of using solar cells to generate terrestrial electricity than building solar farms.

The challenge is due to the vast size of the space infrastructure required: for 2 $GW_p$ delivered to the grid, a typical SBSP satellite would require a microwave antenna 1.6 km in diameter and would have a mass of ~2000 tonnes. The highest possible specific energy PV cells would be required. In this context, solar concentrated thin-film PV cell technologies look very attractive – especially as many of these also exhibit excellent radiation tolerance, and therefore would be expected to have a long operational lifetime in orbit (approaching 100 years). While it sounds like science fiction, a recent study [189] has shown that such stations are practical propositions and should have an LCOE which is significantly less than that of a nuclear plant, and only ~25% more than wind energy.

### *9.4 Conclusions*

There have been significant advances in space solar cells with the development of new materials and designs aimed at enhancing their efficiency, reliability, and performance in space. Multi-junction GaAs based solar cells have demonstrated the best choice for space applications. However, given the growing demand for space-based services and exploration, it is crucial to develop cheaper, more efficient, and reliable space solar cells to meet the increasing power demands of spacecraft and satellites. Lightweight thin film solar cells, such as CIGS, CdTe, ultrathin GaAs, and halide perovskite solar cells, are emerging as future technologies with high radiation tolerance. In-flight tests of CIGS and CdTe have proven their strong radiation tolerance, but their output power efficiency falls far behind that of multi-junction GaAs-based solar cells. Multijunction perovskite solar cells can be a potential candidate, however, a thorough examination, such as inflight tests, are required.

The development of more efficient and reliable space solar cells is crucial for the success of future space missions, given the increasing demand for space-based services and exploration, and SBSP applications. Advances in solar cell technology can enhance the power capabilities of spacecraft and satellites, allowing for further advancements in space exploration and discovery.

**Acknowledgements**

D.L. would like to thank financial support from the European Regional Development Fund (ERDF) for funding the 2nd Solar Photovoltaic Academic Research Consortium (SPARC II). J.S.Y. gratefully acknowledges the support by the National Research Foundation of Korea (NRF) funded by the Ministry of Science and ICT (no. RS-2023-00257494).

## 10. Agrivoltaics – The Next Frontier in Agriculture

Nicola Gasparini,[1] Hamish Beath,[2,3] James R. Durrant,[1,4] Jenny Nelson,[2,3] Paloma Ortega-Arriaga,[2,3] James W. Ryan,[5] Benedict Winchester[2,3]

[1]Department of Chemistry & Centre for Processable Electronics, Imperial College London, London, W12 0BZ, UK

[2]Department of Physics & Centre for Processable Electronics, Imperial College London, London, W12 0BZ, UK

[3] Grantham Institute – Climate Change and the Environment, Imperial College London, London, SW7 2AZ, UK

[4] SPECIFIC, Swansea University, Bay Campus, Swansea SA1 8EN, UK

[5] Department of Chemistry, Swansea University, Swansea SA2 8PP, UK

### *10.1 Status*

Global demand for energy, food, and water has been increasing alongside population and economic growth, rapid urbanization, and changing diets, putting pressure on land use. PV technologies are crucial for climate change mitigation, but the required deployment of this technology to meet emissions reductions targets needs considerable land resources, especially for large-scale ground-mounted systems. At the same time, agricultural production must increase substantially to guarantee food security to an expected global population of 9.7 billion people by 2050 [190]. Agriculture is the largest consumer of the world's freshwater resources, and more than one-quarter of the world's energy is expended on food production and supply. However, only 38% of the planet's 13 billion hectares of land are available for farming, and expanding agricultural land would require deforestation, which negatively impacts biodiversity and greenhouse gas removal [191]. Therefore, solutions that address the competing demands for energy and food are necessary. Pairing PVs with agriculture can more efficiently use additional farmland, potentially reducing land-use conflicts. Hence, agrivoltaics is a promising area for the water-food-energy nexus and is a potential solution to the climate crisis.

Agrivoltaics is one application of PV integration into the human environment, alongside building-, floating- and vehicle-integrated systems [192], [193]. The applications of integrating PVs in agriculture can be generally classified as crop and food production, livestock production, ecosystem services provision via vegetation management, and solar greenhouses, as depicted in figure 16 [194]. The resulting agrivoltaic systems typically consist of common crystalline silicon PV modules mounted in different configurations to traditional solar farms (greater spacing, raised mounting, different orientations including vertical) all of which allow more light to pass through the installation (figure 16); or they may involve new PV technologies that are semi-transparent or flexible, and can be integrated on curved surfaces. Although implementing PVs with crops effectively means reducing the amount of solar radiation that plants receive, this can be beneficial given that plants use only a fraction of the radiation they absorb and often perform better under lower intensity (see for example Ref. [195]) and that reductions in temperature or in moisture loss due to shading can improve yields. Some of the benefits of agrivoltaics thus include alleviating space constraints for large-scale solar projects, improving agricultural yield by providing shade or reducing moisture loss, increasing crop resilience by providing protection against harsh weather conditions, and raising module power output through decreases in panel temperature [192], [196], [197]. In off-grid contexts, agrivoltaics may help to conserve water, improve crop yield by providing local power for irrigation, offer an additional source of income to farmers from the sale of surplus power, and improve access to electricity for surrounding communities [198], [199].

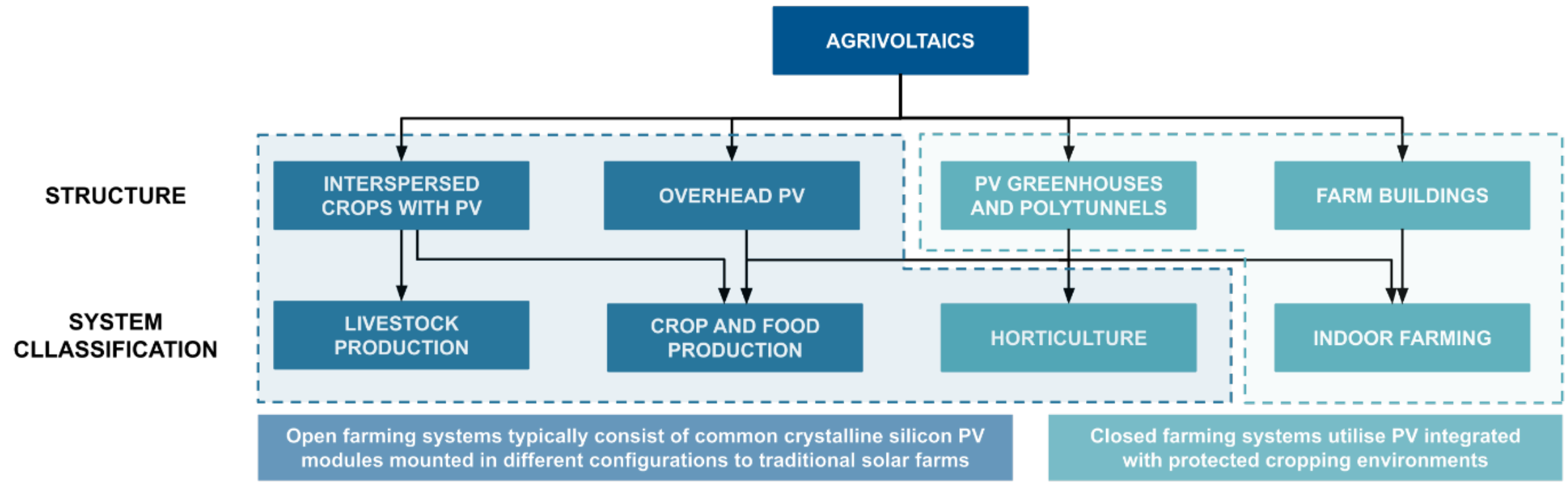

**Figure 16.** Examples of the applications of agrivoltaics. Based on ref. [198].

Previous studies on agrivoltaics have focused on plant cultivation, aquaponics, and livestock production, demonstrating the technical and economic feasibility of agrivoltaic systems in different applications, locations, and climates [200]. In 2021, the global installed capacity of agrivoltaics was approximately 14 gigawatt-peak ($GW_p$), with agrivoltaics found in almost all regions of the world, but concentrated in Japan, China, France, the United States, and South Korea due to government funding programs in these countries [192], [201]. Despite this, there are challenges to the implementation of agrivoltaics including, but not limited to, a lack of research and understanding on impacts and best practice, cost challenges, regulatory frameworks and, social acceptance [192], [194].

### *10.2 Current and Future Challenges*

Current and future challenges to increased deployment of agrivoltaic systems include alternative PV module types that may allow better integration with some kinds of agriculture, insufficient experience or knowledge of current and future impacts, cost challenges associated with different agrivoltaic types, and additionally, regulatory hurdles.

Although most agrivoltaics systems use mature, c-Si modules, there is growing interest in alternative technologies such as bifacial silicon (or other) modules, semitransparent modules that allow part of the spectrum to pass, modules incorporating partial shading, flexible *e.g.*, CIGS modules that could be incorporated into curved roofs such as polytunnels, and lightweight membrane modules that could be extended over fields with minimal mounting.

Whilst for agrivoltaics that use c-Si panels the PV technology is mature, the designs of the agrivoltaics *systems* being implemented are in an early stage of maturity [194]. As a result, knowledge and understanding of impacts and best practices are still being developed [202], [203]. Research highlights the considerable uncertainty regarding the impacts on different crops and soil types [196], [204]. Crop-yield changes and crop suitability for agrivoltaic systems also vary depending on climatic zone: in hotter climates, shading from stilted agrivoltaics may help to improve crop yields, but not in cooler climates [192]. Possible challenges associated with agrivoltaics include excess soil moisture or drainage problems, reduced crop yields due to overshading, and reduced lifetimes of solar infrastructure [192], [196]. New research and pilot projects are needed to examine the impacts across different regions to ensure that the benefits of agrivoltaics are maximized and the limitations better characterized [194], [202], [205]. Research should also consider the influence on crops or PV systems of future climate change, and consider how agrivoltaics can be most resilient to increased extreme weather events, and even how agrivoltaics may be used in planning for adaptation.

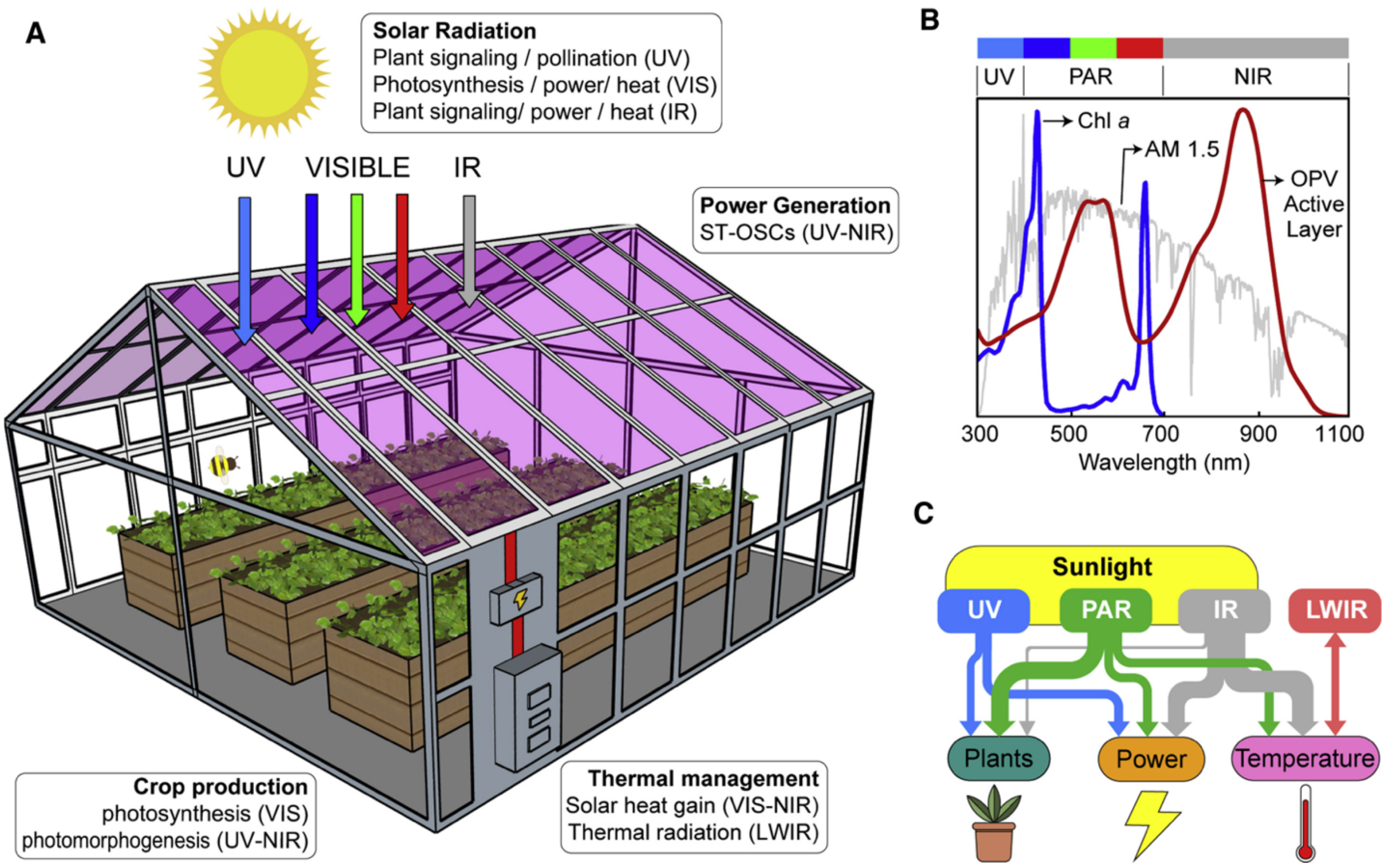

**Figure 17.** Agrivoltaics based on organic photovoltaics: (A) an overview of an organic-solar-cell-integrated greenhouse indicating spectral use of sunlight, (B) the absorption spectra of chlorophyll a and active layer of an organic solar cell, indicating complimentary absorption, (C) and the schematic flowchart indicating distribution of sunlight toward management of plant growth, electricity generation, and greenhouse temperature management. Adapted under the terms of the Creative Commons Attribution (CC-BY-NC-ND) license from Ref. [206]. Copyright 2021, The Authors.

An additional set of challenges for agrivoltaics are financial, with the nature of these varying depending on the type of PV technology in use, the agricultural activity, the project size, and the location [203]. For novel PV technologies, such as semi-transparent panels, there is a considerable financial challenge due to their relative nascency when compared to c-Si PVs. Further research and manufacturing scale-up is needed to improve the attractiveness of these technologies in agrivoltaic contexts (see section 10.3 below).

For agrivoltaics utilizing c-Si panels arranged on novel mounting and racking systems to permit the co-location with agriculture, there are higher costs associated with their development, installation, and operation than for conventional installations [207]. During project development, greater costs are incurred due to more complex designs, and a need to coordinate across additional stakeholders compared to conventional solar plants, such as farmers [207]. Installation costs for agrivoltaics are often greater due to increased raw material costs, stronger foundations, and specialist equipment needed to install arrays [194], [207]. For raised agrivoltaics, the biggest differences in cost when compared to conventional ground-mounted PVs are in the substructure (up to five times greater cost) and the surface preparation and installation costs (up to 2.8 times greater cost) [203]. Operation and maintenance costs are frequently higher for agrivoltaics installations due to the need for the use of specialist equipment and approaches [194]. These cost premiums act as a potential impediment to future deployment of agrivoltaics. Increased deployment and dissemination of best practice, and the creation of open-source, adaptable racking and mounting, will help to bring costs closer to conventional ground-mounted PVs [205], [207].

The final set of challenges to further deployment are regulatory in nature. There may be land classification or zoning restrictions on the deployment of conventional ground-mounted PVs [200]. Planning regulations designed for conventional ground-mounted PVs may not accommodate the permitting of agrivoltaic plants with novel designs such as raised panels. Local and national policies that accommodate the permitting of agrivoltaic plants and dual use of land provides an enabling environment for projects [208]. Without legal frameworks to accommodate the dual-use of PVs with crops, agrivoltaics may be restricted to the same land classifications as conventional ground-mounted PVs, and would therefore be prevented, or severely limited, in their deployment [192]. Additionally, consideration for subsidy payments is needed. In the instance that subsidies exist for farmers (e.g., EU Common Agricultural Policy) and there is no provision for dual land use, farmers may be prevented from receiving subsidy payments, making agrivoltaics unattractive for farmers [202].

### *10.3 Advances in Science and Technology to Meet Challenges*

Most barriers to the wider adoption of agrivoltaics are around the implementation of existing technologies, whether in the uncertainties surrounding the best implementations to synergistically co-locate crops with PV panels, or whether political and legislative in nature. However, there remain some technical challenges which, if overcome, could increase the attractiveness of agrivoltaics: PV technologies with greater module transparency; improvements to, or investigations around, crop use and selection, including modelling of whole agrivoltaic energy systems, as well as investigations into the long-term impact on the land and local ecosystems.

As discussed in Section 2.1, c-Si solar cells dominate the PV market globally. Agrivoltaic installations that utilize silicon modules generate patches of deep shade with adverse effects on crop yield [205]. Semi-transparent panels, such as spectrally-tunable organic solar cells, could allow light in the photosynthetically active region of the spectrum (400–700 nm) to pass through to the crops below whilst capturing lower energy infrared photons and converting them to electricity. Systems using such spectral-sharing PV modules may have potential to improve crop performance whilst generating power. Organic solar cells (Section 6) and dye-sensitized solar cells (Section 8), provide alternative solutions which may allow more light through for plant growth whilst avoiding toxic compounds. Their lower module efficiency and higher current cost compared to silicon-based PV, however, present a barrier to uptake [205], [209].

For silicon-based agrivoltaics, further work is needed on the performance trade-offs of transparent-backed silicon cells and greater gaps between cells [210]. Modelling is needed to assess the performance of technological solutions, both at a device level, and at a system level, considering, not only the performance of the energy system, but also the impact on the productivity of the agricultural solution. More investigations around novel solutions are hence needed to produce commercially viable technologies.

Agrivoltaics at its core aims to split the solar spectrum, either by amplitude (through partial-shading solutions such as interspaced overhead PV panels) or by wavelength (*e.g.*, semi-transparent cells in PV greenhouses or opaque buildings). There remain issues specific to wavelength-based splitting due to the impact of both altering and reducing the solar spectrum incident on crops being unknown. Biomass reductions have been reported in protected (closed) agrivoltaic setups, similar to that shown in figure 17 [206], with reductions in both root and above-ground biomass varying significantly depending on the choice of solar technology [211], [212]. An optimized synergistic utilization of the solar spectrum has the potential to improve the water uptake and $CO_2$-absorption capacity of plants compared with conventional PVs [196], [213], whilst inefficient use can increase water demands and reduce $CO_2$-capture potential. Thus, solutions which better utilize those parts of the solar spectrum which are not harnessed fully by plants need further technological investigation.

As discussed, as well as the need for a more thorough investigation into which plants are most suited to agrivoltaic installations [196], there exists scope to develop plants which are more capable of effectively utilizing shade, either by selective breeding or through genetic engineering. In this way, the heavy-shading conditions effected by c-Si PV panels can be better suited for use in agrivoltaic installations [210].

Further investigation is needed into the use of building-integrated agrivoltaics for not only crop growth, but also other parts of the agricultural sector. Commercial algal and bacterial production is a market where there exists scope to utilize part of the solar spectrum for PV energy production and part for commercial plant growth [214]. The drying and dehydrating of agricultural products further down the value chain also presents an opportunity where novel technological solutions utilizing semi-transparent agrivoltaics can present synergistic commercial benefits [215].

### *10.4 Concluding Remarks*

Agrivoltaic systems of different types have great potential for more efficient land and water use, and for more sustainable energy and food supplies. However, not enough is known about which crops and solar technologies to employ and how systems should best be installed to maximise their benefits such as potential synergistic crop yield improvements. Work to address these knowledge gaps is needed to increase the installed capacity of agrivoltaics. Additionally, research to improve the efficiency, longevity, and cost profile of novel PV technologies, such as semi-transparent panels, could significantly broaden the range of viable agrivoltaic applications. Legislative frameworks must also be adapted to accommodate the dual- use of land. If these barriers can be overcome, agrivoltaics can play an important role in the transition to more sustainable food and energy systems.

**Acknowledgements**

Authors, apart from first author, arranged alphabetically by surname. H.B., B.W. and J.N. acknowledge funding from EPSRC (no. EP/X52556X/1). J.N acknowledges The European Research Council (no. 742708 (CAPACITY)) and thanks the Royal Society for the award of a Research Professorship. N.G, J.D, J.R and J.N acknowledge funding from EPSRC, UK (no. EP/T028513/1 (ATIP)). B.W. acknowledges funding received from NERC, UK (no. NE/2451429). P.O thanks the Mexican National Council for Science and Technology (CONACYT) and the Mexican Ministry of Energy (SENER).

## 11. Characterization Techniques and Standards for Photovoltaic Materials and Device Development

George Koutsourakis,[1] James C. Blakesley[1], Stuart A. Boden[2], Fernando A. Castro[1], Jeff Kettle[3]

[1] National Physical Laboratory, Hampton Rd, Teddington, UK,

[2]School of Electronics and Computer Science, University of Southampton, Southampton, UK

[3]James Watt School of Engineering, University of Glasgow, Glasgow, UK.

### *11.1 Current status*

The continuous progress in PV materials and devices is based on different iterative development steps: discovery and refinement of materials; optimization of fabrication and upscaling methods; development of reliable and trustworthy PV devices along with measurement methods for validation of their performance. Characterization and measurement techniques play an important role in all of the above. To accelerate innovation and facilitate technology uptake, it is critical that measurements provide accurate, robust and reproducible results. However, different PV materials and structures create different needs, requiring matching innovations in characterization techniques. Research in measurement and validation of methods leads to best practice, better reproducibility in research, and provides the technical basis upon which international standards can be established to support industrialization and world trade. The process for developing new standards is illustrated in figure 18.

The International Electrotechnical Commission (IEC) publishes international standards for all electrical, electronic and related technologies, which address product development, performance, compatibility and environmental safety. Standardization activities have played a major role in providing quality control in the PV industry, from the IEC 60904 series, which underpins the standardized power rating of PV products, to the IEC 61215 series which has built confidence in the reliability of products. Such standards are constantly revised, and new standards or relevant documents (technical specifications and reports) are created to accommodate for advances in specific PV technologies and serve new technologies. A large part of this progress is based on continuous improvements in characterization techniques, and the development of new measurement methods for PV materials and devices.

Imaging and spatial characterization techniques contribute towards improving performance and stability, quality control and tackling upscaling challenges throughout the development steps of PV technologies. Electroluminescence (EL) imaging is now an established technique for PV modules, which is widely used for PVs field condition monitoring [216], as is large area photoluminescence (PL) imaging [217]. Current mapping methods for solar cells have been improving as well, now providing megapixel resolution at realistic measurement speeds [218], [219]. All the above techniques have been vital to support the deployment of PV products in the field, supplement and provide additionality to standardized tests, and have shown that they will continue to do so with future PV technologies [220]. Therefore, the development of best practice guidelines, harmonized measurement protocols and standards is crucial to ensure reproducibility of results.

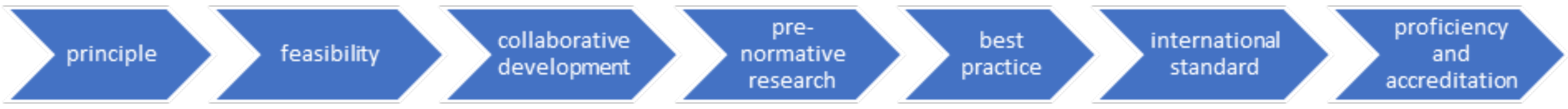

**Figure 18.** Different steps of a characterization or testing method, from proof of concept up to standardization. Not all measurement techniques reach the standardization step.

Materials characterization techniques, from Raman and PL spectroscopy to X-ray based methods, are increasingly used at the nano- and micro-scale to support development of new and emerging materials. Atomic Force Microscopy (AFM) and its different modes have provided insights into perovskite PV materials in recent years, towards investigating charge dynamics and hysteresis effects [221] or improving stability [222]. Scanning electron microscopy (SEM), transmission electron microscopy (TEM) and related methods (Electron-Beam-Induced Current (EBIC), Cathodoluminescence) are paramount for studies and development of new PV materials and film structures, but it is critical to first determine the measurement conditions to avoid beam damage [223]. In addition, nano and microscale characterization techniques also play a crucial role in providing the necessary background information for the revision or development of new reliability standards [224].

### *11.2 Current and Future Challenges*

With PV installations exceeding 1 $TW_p$ scale in 2022, there is a need to devise strategies to continuously improve the operational lifetime of PV systems, and thus, to reduce the electricity cost and the burden of the industry on global material resources. The latest IEC 61215-1:2021 and IEC 61215-2:2021 documents describe the test requirements and procedures for the design qualification of PV products for long-term operation in the field. Although such tests are necessary and provide confidence in the PV products' quality, it is difficult to stimulate the specific failure modes that will occur in the varying international climatic conditions in the field. In addition, such pass/fail tests provide limited insights into the failure factors of different technologies, while new technologies will have their own failure mechanisms which might not be stimulated through current qualification tests. The so-called agnostic stress tests use combined accelerated and sequential testing and could be a solution to trigger failure modes that are observed in the field, but not during qualification tests [225]. Capabilities for multi-method *in-situ* characterization during accelerated tests could provide insights into the origin and mechanisms behinds failures, supporting R&D and accelerating innovation [226].

Applying specifically designed accelerated stress tests is an ideal way of looking into specific failure mechanisms of new PV technologies such as perovskite or organic PV devices. Such tests can also evaluate different encapsulation schemes and materials. Nevertheless, consistency in reporting the test procedures and resulting data is necessary across the scientific community when publishing results. Consensus regarding the ISOS protocols, originally designed for organic PVs, has been extended to perovskite PV testing [102], and provide good guidance that should be followed when publishing results. Such requirements should be safeguarded by the scientific community in this area, including reviewers and editors of publishing journals in this field.

While the IEC 60904 series underpins *I-V* testing for PV devices, new PV technologies create additional complexities and ambiguity when acquiring *I-V* curves at standard test conditions (STC). In the case of bifacial PV devices, where the recent technical specification IEC TS 60904-1-2:2019 defined testing procedures for bifacial modules, there are still challenges when measuring individual cells, some of which the latest IEC TS 63202-3:2023 has tried to overcome for standard crystalline silicon cells, but does not yet accommodate for semi-transparent devices, non-silicon technologies or tandem devices. In these latter cases appropriate measurement protocols or guidelines are not fully set, making commercialization and market penetration of such products difficult.

Another challenge is the power rating of highly metastable PV devices such as perovskite solar cells (figure 19). Metastable devices often require a long time (seconds to hours) under constant illumination and voltage to reach steady state, while it is not always straightforward to distinguish metastable behavior from unstable (irreversible) behavior. Although several methods have been reported recently that allow acquisition of a stable Maximum Power ($P_{max}$) value, there has been no consensus on a standardized method for power rating of metastable devices, with the IEC TR 63228:2019 document being an advisory aid towards such measurements. This also creates challenges in outdoor monitoring of metastable devices, especially regarding appropriate maximum power point tracking. Measurements of spectral responsivity, required for spectral mismatch corrections, also suffer from similar issues when measuring metastable devices.

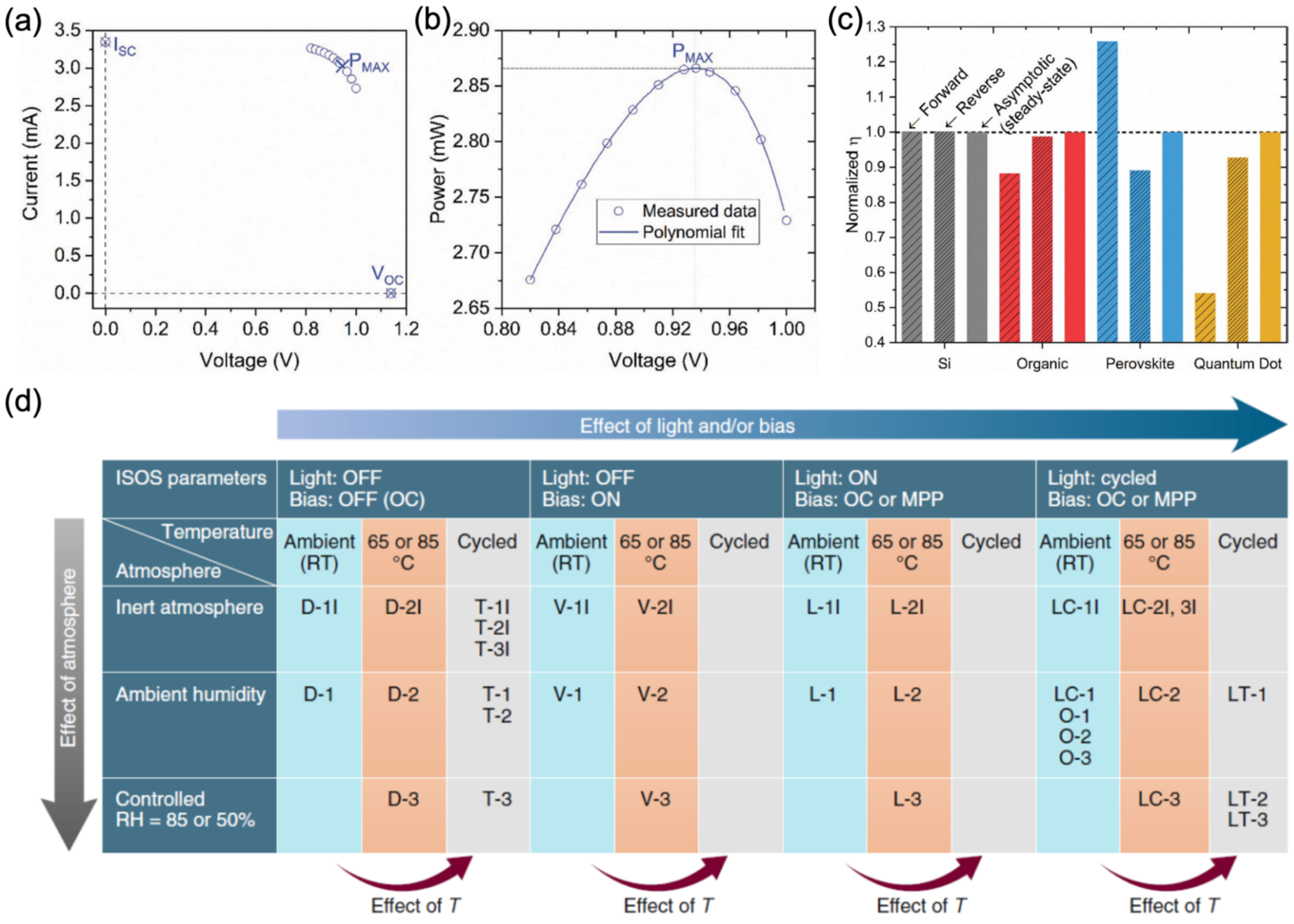

**Figure 19.** (a) Illustration of an asymptotic $P_{max}$ scan curve from a lead-halide perovskite solar cell. (b) Power versus voltage with polynomial fitting to obtain $P_{max}$ from the left asymptotic I–V curve. (c) Performance comparison between fast *I–V* scans (forward and reverse) and steady-state asymptotic scan on three types of cells: Si, organic, perovskite, and quantum dot. All efficiencies are normalized to the steady-state efficiencies of each device. Parts (a) – (c) reproduced with permission from Ref. [227]. Copyright 2021, Wiley. (d) Stress factors affecting solar cells in different ISOS protocols. Rows correspond to different atmospheres, columns correspond to different combinations of light and electrical bias, and sub-columns depict different temperature regimes. Arrows alongside the table guide the eye for comparing protocols to identify the effects of atmosphere, temperature, light or electrical bias. Reproduced under the terms of the CC-BY license from Ref. [102]. Copyright 2020, The Authors.

With emerging PV technologies such as lead-halide perovskite or quantum dots solar cells relying on material engineering and investigations in the nanoscale, a simultaneous and continuous improvement in nanoscale characterisation methods is also necessary. For new materials and early-stage nano-scale structures development, high precision techniques realized under vacuum or inert atmospheres are necessary for repeatability and reproducibility of measurements, in order to produce useful and meaningful results. Facilities for nanoscale characterization at controlled environments (controlled oxygen, moisture, heat and illumination) will play a key part in decoupling the different factors that contribute to material degradation during initial studies, before moving to the scale up and commercialization phase.

### *11.3 Advances in Science and Technology to Meet Challenges*

Non-destructive characterization techniques for manufacturing and scale up are invaluable tools in understanding losses when scaling up PV technologies, from small cells to mini-modules and eventually to large commercial PV products. The field of PV characterization can benefit from recent advances in imaging technology, such as the increase in resolution and spectral range (Short-wave Infrared imaging) of cameras, camera based hyperspectral imaging developments [228], applications of machine learning methods [229] and digital light processing capabilities [230]. Combined with appropriate advances in modelling, such advances will aid towards the development of robust quantitative rather than qualitative imaging methods, which can also be used for contactless inline measurements [231]. Multimodal characterization is also important for linking together different properties (*e.g.*, structure and luminescence) under realistic operating conditions.

Collaborations and data sharing will be key to the advancement of PV material and device technologies, especially regarding stress tests and characterization of new materials, device architectures and encapsulation schemes. An example of such an activity is the Perovskite Database [232], an open-access database for perovskite solar cell device data, populated with data for devices described in peer-reviewed literature. Such activities can help reduce duplicate research, allow faster dissemination of recent work, while with enough data machine learning methods can be utilized to accelerate future development. In parallel, it is crucial to ensure the quality of the data based on aspects such as reproducibility and uncertainty of the methods used, to ensure the results are not biased towards incorrect solutions or misleading conclusions.

Similarly, international intercomparison activities are always necessary, to validate new methods, to evaluate recent standardized methods [233], or investigate required revisions of protocols and standards for new technologies [234]. In parallel, development and availability of facilities for outdoor testing and monitoring of emerging PV devices will be important in providing the necessary data towards development of energy yield models, also connecting outdoor performance and failure mechanisms observed outdoors with the development of suitable power rating evaluation and specific stress tests in the lab.

An important aspect necessary for any technological advancement is the availability of skilled scientists and engineers to move forward the advancements in PV metrology and characterization. From the nanoscale up to PV systems level, most characterization methods require specific technical expertise, which means that skills, training and knowledge transfer will play a pivotal role in future PV metrology development.

### *11.4 Concluding Remarks*

Characterization methods and testing standards underpin PV technology development, from the discovery of new materials up to the deployment of PV systems in the field. Best practice documents, harmonized protocols or standards where applicable, ensure the uptake of high quality measurements, especially when all information of interest is logged when publishing results. Material innovation in PV technology is complex also due to interactions within a PV module, where the advantage of one material may be outweighed by its interaction with another component. New applications of PV technologies including indoor PVs, vehicle integrated PVs, high altitude and new space applications, will be creating new measurement challenges in the near future.

Research in PV materials can be accelerated by establishing multi-user accessible facilities for collaborative studies such as material and device degradation testing, outdoor testing and monitoring of PV devices or high cost nanoscale facilities. Such concepts will also allow the realisation of intercomparisons between researchers, which can help validate measurement results and develop new measurement protocols or standards. Better dissemination and access to best practice or standards documents is also crucial, since in many cases standards documents or best practise guides exist for specific measurements, but are not followed by the PV community, in most cases due to lack of awareness.

**Acknowledgements**

G.K, J.C.B and F.A.C acknowledge funding from the Department for Science Innovation and Technology through the National Measurement System (NMS) programme. JK acknowledges the support of the EPSRC through EP/W018632/1 and EP/W019248/1.

## 12. Computational Materials Discovery and Design for Photovoltaics

Alex M. Ganose[1], Marina R. Filip[2], Jarvist M. Frost[1], Pooja Goddard[3], M. Saiful Islam[4], Keith P. McKenna[5], David. O. Scanlon[6,7], Aron Walsh[8], Lucy D. Whalley[9]

[1] Department of Chemistry, Imperial College London, 80 Wood Lane, London W12 0BZ, UK

[2] Department of Physics, University of Oxford, Clarendon Laboratory, Parks Road, Oxford OX1 3PU, UK

[3] School of Science, Loughborough University, LE11 3TU, UK

[4] Department of Materials, University of Oxford, Oxford, OX1 3PH, UK

[5] School of Physics, Engineering and Technology, University of York, Heslington, York YO10 5DD, UK

[6] Department of Chemistry, University College London, 20 Gordon Street, London, UK

[7] Thomas Young Centre, University College London, Gower St, London, UK

[8] Department of Materials, Imperial College London, London SW7 2AZ, UK

[9] Department of Maths, Physics and Electrical Engineering, Northumbria University, Ellison Place, Newcastle Upon Tyne, NE1 8ST, UK

### *12.1 Status*

Theory and computation have long played a vital role in the design and optimization of photovoltaic materials and devices. In the 1950s, models of trap-assisted recombination provided insight into the origins of voltage losses in semiconductors [235]. The 1960s saw Green’s functions approaches applied to identify the microscopic origin of defect traps in silicon [236]. The first computational screening of emerging absorbers GaAs, CdTe, and $ZnSiAs_2$ based on their optoelectronic properties and minority carrier lifetimes was performed in 1979 [237]. Today, computational modelling is used at every scale of photovoltaic research and development, from nanoscale high-throughput materials discovery [238], micron-scale ionic-diffusion modelling [239], to production line diagnosis of cell imperfections using deep machine learning (figure 20) [240].

Much of the success of modelling comes from the insight gained into the microscopic processes that underpin the properties of PV materials and that control charge-carrier generation, transport, and collection. Heterojunction interfaces pose a challenge for conventional experimental characterization. Computation has filled the gap with insights into interfacial phase formation, energy barriers to electron transport, ion migration, and degradation mechanisms, enabling reduced contact losses and optimized device architectures. For example, the record 22.1% efficiencies of CdTe cells were enabled by theoretical insights into the band alignment of CdS and MgZnO buffer layers [241]. Similarly, recent advances in first principles defect modelling now allow the quantitative prediction of charge-carrier capture cross-sections and recombination rates [242]. This knowledge has informed the identification of ideal growth conditions and passivation strategies to minimize deep trap formation and resulting voltage losses [243].

The wide design space of diversified photovoltaic technologies (including outdoor, indoor, and multijunction PVs) is well suited to computational materials discovery. Computational materials databases have facilitated the extensive screening of known inorganic compounds as novel photovoltaic absorbers. The full compositional spaces of entire materials classes (including metal halide perovskites [244], double perovskites [245], and chalcopyrites [246]) has been explored using simple performance metrics (band gap, effective masses, optical absorption). Similar high-throughput searches have been applied to organic photovoltaics. The Harvard Clean Energy Project screened over 2.3 million donor materials using density functional theory [247], while more recently, Lopez *et al.* evaluated a library of 51,000 non-fullerene acceptors through theoretical predictions of device efficiency [248]. However, it is difficult to point to a new photovoltaic technology that has been found by these computational sifts, rather than by empirical exploration. The existence of these large open databases of calculations is of incredible utility for machine-learning approaches.

Photovoltaic technologies require more than an active absorber layer. Materials discovery for transparent conductors is one active area, with computation enabling improved dopants in existing technologies for higher conductivity and increased transparency [249], the realization of new *n*-type oxides [249] and *p*-type TaIrGe [250], and the widescale screening of known materials using diverse design criteria, including forbidden optical transitions [251] and hole transport metrics [252]. Similar methods have been applied to search for new back contacts, with recent work identifying five candidate materials for CdTe cells based on band alignments and dopability [253].

### *12.2 Current and Future Challenges*

Theory and simulation can provide diagnostic insights into the limiting factors of current materials and identify new materials worthy of further study. A critical issue for computational materials discovery is that the most accessible design metrics (such as band gap and optical absorption), which are direct quantum observables of an idealized crystal or molecule, are not sufficient to guarantee high device performance. An operating photovoltaic device relies on a careful balance of phenomenological device physics properties (such as charge-carrier transport and recombination), and only produces power when it is out of equilibrium. To start to understand this device physics, atomistic modelling must consider defects, ion transport, secondary phases, interfaces, and finite temperature. This is exemplified by $Cu_2ZnSnS_4$ which, despite its ideal band gap of 1.5 eV, suffers from deep recombination centres [254] and only recently achieved efficiencies of 13.0% (well below its maximum theoretical efficiency greater than 30%) in the form of the double solid-solution $(Cu,Ag)_2ZnSn(S,Se)_4$ [255].

Screening the defects of many materials is challenging due to the considerable computational expense. Dahliah *et al.* considered 7000 copper-based materials as defect-tolerant absorbers, accounting for trap-assisted recombination through an automated configuration-coordinate workflow [256]. This work identified 8 novel materials with high predicted carrier lifetimes, highlighting the promise of the approach in future searches.

Calculating quantum-mechanical properties that go beyond the Born-Oppenheimer approximation, such as charge-carrier transport and optical absorption, is a long-standing challenge in solid-state physics and is critical for the understanding of photovoltaic performance under realistic operating conditions. Recent advances in modelling electron-phonon interactions mean it is now possible to simulate carrier mobilities, indirect optical transitions, and the role of ionic vibrations in electron-hole interactions, albeit at significant computational cost. While simplified models of phonon-assisted absorption have been used to screen indirect gap semiconductors for use as thin-film absorbers [257], deploying these techniques as part of a standard evaluation procedure is still a major challenge.

Another challenge the field faces is modelling microstructure. Identifying and simulating all possible grain boundaries in a material is not practical and typically only a subset is chosen as part of a statistical ensemble. However, a small fraction of grain boundaries can be responsible for limiting device performance and thus there is a need for a greater understanding of structure-property relationships to identify relevant key features and to enable their passivation. Furthermore, while computational methods can now predict thermodynamic phase preference with high accuracy, progress in understanding the impact of crystal kinetics has been slow. *Ab initio* molecular dynamics is a valuable tool to probe secondary phase and intergrowth formation, but the time and length scales involved are often prohibitive. The microstructure of films is also highly sensitive to the deposition method, and subtle experimental conditions. Classical molecular dynamics have revealed the origins of void formation in sputtered CdTe films and identified mitigation strategies [258]. Such approaches have proved powerful when combined with experimental observations but cannot yet predict optimized synthesis routes a priori.

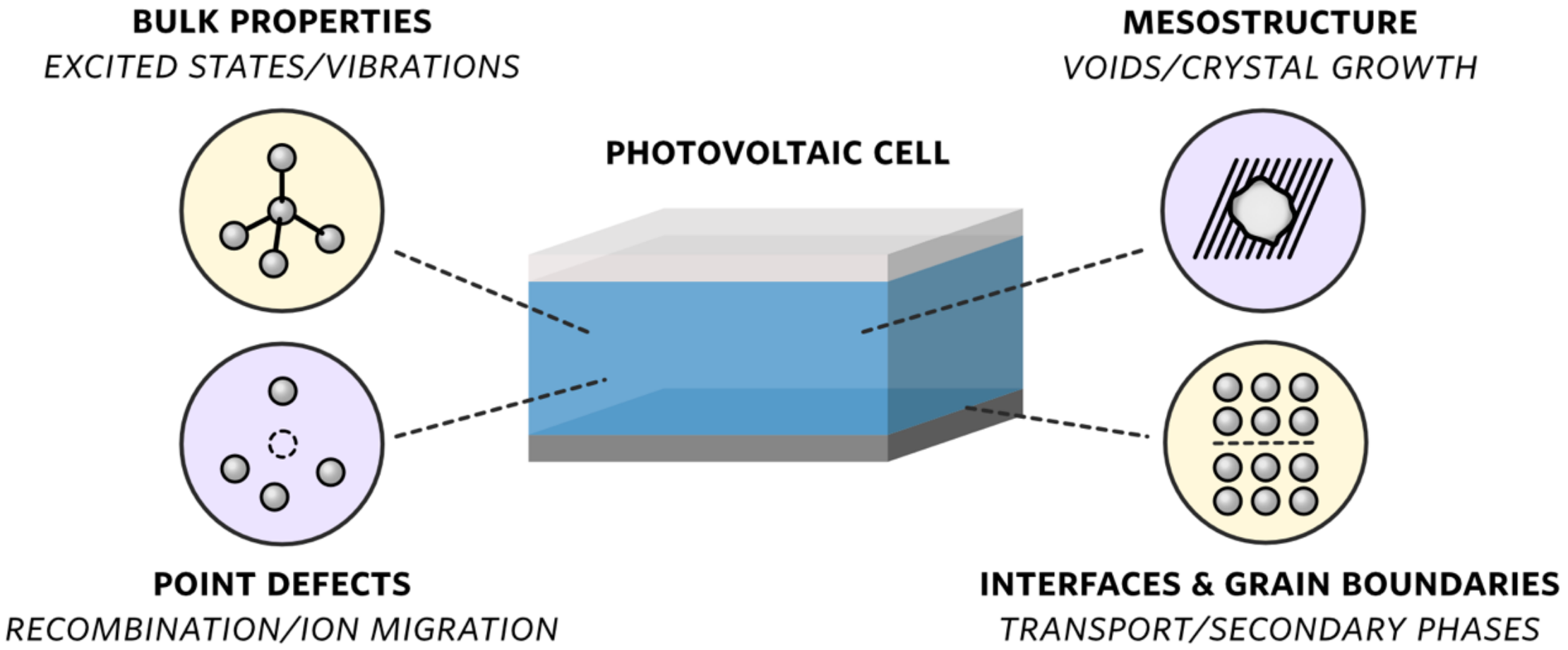

**Figure 20.** Current challenges in the modelling of photovoltaic materials.

Expanding the search for novel photovoltaics to new chemistries and structures is both a challenge and an opportunity. To date, almost all computational materials discovery attempts have involved screening databases of known inorganic phases (or elemental substitutions of known phases). It is likely that there are many undiscovered materials with properties suitable for photovoltaics that warrant investigation. One route to realizing new phases is through crystal structure prediction. Such methods enable technoeconomic factors to be considered by restricting the search to designer compositions with earth-abundant and non-toxic elements. This approach has recently been applied to discover a new candidate absorber $CuBiI_4$ [259].

### *12.3. Advances in Computational Methodology to Meet Challenges*

Progress in the computational modelling of photovoltaics has been driven by various factors, including theoretical method development; the implementation of these methods in new software packages; and increased computational resources. To achieve greater predictive accuracy across more diverse structure types, there is a need for fundamental method development that tackles higher-order phenomena. At realistic operating conditions, defects are mobile and coupled to other processes such as light, heat, and internal electric fields. Recent calculations have highlighted the diverse behavior of trap levels at finite temperatures due to the competing effects of volume-mediated free-energy corrections and band gap renormalization [260]. While these methods can now be applied to simple cubic semiconductors, the systematic screening of mechanically soft or low-dimensional absorbers will require new theoretical advances.

The complexity of processes in photovoltaic materials demands sophisticated treatment that is too computationally costly to apply across large numbers of materials. The advent of large computational datasets has enabled the use of machine-learning approaches to accelerate materials discovery. Machine-learned interatomic potentials can predict energies and atomic forces with accuracy comparable to first principles methods but with several orders of magnitude less computational expense, and without the investment of expert researcher time in deriving empirical potentials. As such, these methods can simulate much larger systems than is possible with *ab initio* approaches, while still being part of an automated workflow. For example, they have been used to investigate long-range intermolecular correlations in hybrid lead halide perovskites revealing impacts on halide migration [261]. Liu *et al.* modelled grain boundaries in $CsPbBr_3$, revealing spontaneous gliding and formation of mid-gap states [262]. Furthermore, graph networks have been applied to predict the formation energies of point defects in semiconductors [263]. However, at present, the relatively large errors around 0.3 eV remain unsuitable for quantitative predictions. A further outstanding challenge is the inclusion of charge transport and redox processes in machine-learned approaches.

An emerging opportunity is the use of generative models for crystal structure prediction. Generative approaches based on machine learning offer several advantages over traditional structure prediction including the ability to drive predictions towards regions of phase space with tailored properties (e.g., band gap and effective masses). While these methods are still in their infancy and have not yet been applied to the prediction of novel photovoltaics, successes in generative molecular design for drug discovery highlight their potential.

Computational studies rely on suitable infrastructure on which to perform calculations. Indeed, the development of computational materials science has been assisted by the sustained strategic investment in academic supercomputing facilities. Likewise, research software engineers are required to implement and maintain research codes that run on increasingly diverse platforms (GPU acceleration, etc.) and massively parallel computers. The continued and diverse investment in exascale high-performance computing centers and communities of practice is essential to enable the field to progress towards the simulation of photovoltaic devices under realistic operating conditions. Longer-term investment in quantum computing platforms and algorithms should also be prioritized due to their potential to obtain insights beyond classical computers.

### *12.4 Concluding Remarks*

The continued development of computational methods to predict the properties of materials properties from the quantum mechanical equations and the unprecedented access to high-performance computing clusters have empowered researchers to achieve ever more reliable insights into photovoltaic materials and devices. The adoption of machine learning approaches can accelerate the field further by facilitating the study of more realistic systems and the discovery of completely novel materials. By integrating quantum-mechanical observables into multiscale modelling of devices, computation will edge closer towards truly quantitative predictions of device performance.

**Acknowledgements**

A.M.G. was supported by an EPSRC Fellowship (no. EP/T033231/1). J.M.F. is supported by a Royal Society University Research Fellowship (no. URF-R1-191292). M.R.F acknowledges funding from EPSRC (no. EP/V010840/1, EP/X038777/1). We acknowledge the support of the ARCHER2 UK National Supercomputing Service (https://www.archer2.ac.uk) via our membership of the UK's HEC Materials Chemistry Consortium, funded by EPSRC (no. EP/L000202).

## 13. Other Important Factors – Sustainability and Solar Fuel Synthesis

### *13.1. Perspectives on the Sustainability of Photovoltaic Systems*

Jenny Baker,[1] Ruy S. Bonilla,[2] Pascal Kaienburg[3]

[1.] Faculty of Science and Engineering, Swansea University, UK

[2.] Department of Materials, University of Oxford, UK

[3.] Clarendon Laboratory, Department of Physics, University of Oxford, Oxford OX1 3PU, UK

In addition to the technology-specific challenges discussed in this roadmap, we must also consider photovoltaics as a system, and evaluate their environmental sustainability over their entire life cycle. In this section, we offer perspectives on this aspect.

Environmental impact assessments for photovoltaics can cover many impact factors and the effects will vary significantly depending on the location of manufacturing and the climate in which the photovoltaics are installed [264]. The climate change impact can be expressed in two ways. The first is $CO_{2eq}$ $kWp^{-1}$, representing the amount of $CO_{2eq}$ released during the production of PV panels which will produce 1 $kW_p$ of power under standard test conditions (STC). This allows PV panels to be compared and is independent of lifetime and installation conditions. The second is $CO_{2eq}$ $kWh^{-1}$, representing the amount of $CO_{2eq}$ released per kWh of energy generated and therefore considers lifetime and location allowing for comparison with other technologies such as wind power and fossil fuel electricity generation. Across nearly all impact factors, silicon-based photovoltaics are orders of magnitude lower impact than producing energy from fossil fuels, specifically the $CO_{2eq}$ emissions of c-Si photovoltaics is within the range of 14–73 g $CO_2$eq $kWh^{-1}$ [264]. As electricity grids in areas producing the panels reduce in carbon intensity then this will reduce even if there are no further technological improvements. The global electronics council has recently developed a standard to classify c-Si and CdTe modules including their frames where necessary, as low carbon (global warming potential (GWP) of 630 kg $CO_2$eq $kWp^{-1}$) and ultra-low carbon (GWP 400 kg $CO_2$eq $kWp^{-1}$) [265]. In a temperate climate, such as the UK with ~1000 kWh $m^{-2}$ $yr^{-1}$ insolation and assuming a 25 year lifetime, then this calculates as an impact of 24 and 16 g $CO_2$eq $kWh^{-1}$, respectively [266].

An accredited lifecycle assessment has been undertaken on organic photovoltaics (OPVs) commercialized by Heliatek, which concluded that with a 20 year module lifetime, the GWP impact will be between 3-15 g $CO_2$eq $kWh^{-1}$ [126], [267]. This includes emissions arising from end-of-life.

Lead-halide perovskite devices are still in the pre-commercialization stage and therefore any values will be estimates. Furthermore, there are several different architectures and material sets being used, and thus it is difficult to obtain precise values for the environmental impact. Leccisi and Fthenakis gave a summary of perovskite lab-scale devices, and these have widely differing impacts from ~100 g $CO_2$eq $kWh^{-1}$ to >10 000 g $CO_2$eq $kWh^{-1}$ [268]. This highlights the problems with assessing processes that are not at an industrial scale, as well as the significant impact that the predicted lifetime can have. Alberola-Borràs *et al.* assess a more industrially applicable manufacturing process for a screen-printed perovskite architecture with a carbon top contact rather than gold [269], [270]. Whilst the global warming impact of 149 g $CO_2$eq $kWh^{-1}$ calculated by Alberola-Borràs may appear high compared with c-Si, these values were obtained for an 11% efficiency module with a 10 000-hour lifetime [269]. If more realistic values for a commercialized device are used (*e.g.*, 20% efficiency and a modest 5 year lifetime), then this impact reduces to 18 g $CO_2$eq $kWh^{-1}$ [269].

Climate change is putting the water use during production and operation into the spotlight. Compared with fossil fuels, photovoltaics have a much lower water demand per kWh of electricity produced [271]. Water utilization is highest during the production of the PVs rather than during operation [271]. Operational water use is more geographically dependent, with water-scarce regions often experiencing higher soiling, and this has driven the development of self-cleaning systems [272].

Compared to fossil fuel electricity generation, solar PVs technologies have higher mineral resource scarcity potential (MRSP) [273]. Whilst it is not always included in environmental impact reviews of photovoltaics, MRSP is key to understanding the material constraints of different technologies and is typically measured in kg of $Cu_{eq}$ [274]. The UK roadmap to net zero 2050 [275] could lead to a doubling of MRSP per kWh electricity generated compared with the 2018 grid [276]. MRSP would be further exacerbated due to transport and heating becoming electrified under the net-zero 2050 scenario. However, the study by Zhao assumes that the production MRSP of all technologies remains constant [276]. In practice silicon usage for silicon PVs has been reduced significantly between 2008 and 2023 from around 16 g $Wp^{-1}$ to less than 2.5 g $Wp^{-1}$ due to increased manufacturing advances [18]. As discussed in section 2 for c-Si PVs, there is also a drive to reduce silver and indium use, and Müller *et al.* note that due to the fast development of these technologies, it is important to use the most recent data for both production methods and electricity grid impacts when undertaking a life cycle impact assessment [277]. Stamford and Azapagic compared MRSP across CIGS, CdTe and c-Si, with the latter two technologies having similar impacts ~90g $Cu_{eq}$ $W_p^{-1}$, whereas CIGS was 5 times higher with a MRSP 145 g $Cu_{eq}$ $W_p^{-1}$, due to a higher percentage of scarcer elements being used, such as indium and gallium [239].

When considering the balance of systems (BOS), the capacity factor (a measure of actual output divided by peak output, usually estimated over a year) is important, since it sets the utilization of the BOS. For this reason, it can be cost-effective in certain climates to undersize the inverter (reducing the cost of the BOS), this has the impact of reducing the peak output and increases the capacity factor of a system [278]. Planning to curtail some amount of solar production can be advantageous (environmentally and/or financially) rather than upgrading the grid or installing energy storage [278]. This is outside the scope of this roadmap, but modelling scenarios based on the capacity and demands of the grid local to any installation should be considered as part of any strategic plan and the properties of the PV technology (e.g. temperature dependence, wavelength, efficiency profile) is part of that model [279].

### *13.2. Bridging Photovoltaics with Solar Fuel Synthesis*

Virgil Andrei,[1,2] Ludmilla Steier[3]

[1.] Yusuf Hamied Department of Chemistry, University of Cambridge, Cambridge CB2 1EW, UK

[2.] Optoelectronics Group, Cavendish Laboratory, University of Cambridge, JJ Thomson Avenue, Cambridge CB3 0HE, UK

[3.] Inorganic Chemistry Laboratory, South Parks Road, Oxford OX1 3QR, UK

As described earlier, photovoltaic panels are mature technologies which find applications in solar and indoor light harvesting. However, the intermittency of sunlight limits energy production, placing strain on electricity grids. Accordingly, electricity only covers 20% of the global energy demand, as our current energy infrastructure is designed around high energy density fuels, which remain essential for industry, aviation, or heavy shipping. This extensive knowledge on light harvesting systems can be translated to the field of solar fuel synthesis, where sunlight is utilized to convert small molecules like water and carbon dioxide into sustainable fuels. In this case, light absorbers are interfaced with selective electrocatalysts, which can occur in a wide range of configurations, from wired PV-electrolysis systems to integrated photoelectrochemical (PEC) "artificial leaves" and photocatalyst powder suspensions [280].

The wide variety of light harvesting systems for solar fuel production results in specific challenges for each technology. For instance, PV-electrolysis is a mature technology for water splitting, with solar-to-hydrogen efficiencies reaching 12.3% by wiring a bifunctional NiFe layered double hydroxide catalyst with two perovskite solar cells [281], or 30% by wiring two Ir-Pt polymer electrolyte membrane (PEM) electrolyzers in series with an InGaP/GaAs/GaInNAsSb triple-junction PV cell [12]. However the additional cost of reactors, wiring, electronics, noble metal catalysts or corrosive electrolyte solutions has hindered their widespread implementation [282]. The advent of high-throughput $CO_2$ reduction to (multi)carbon products (*e.g.*, $C_1$ - $C_3$ hydrocarbons and alcohols) using copper-based catalysts within gas diffusion electrodes has resulted in current densities of 1 A $cm^{-2}$, which match the output of conventional PV modules. However, such catalysts tend to produce a broad mixture of products [283], which raise additional challenges in terms of product separation and purification. In addition, voltage fluctuations under intermittent daylight irradiation may lead to morphological changes and catalyst deactivation [284], [285].

In contrast to PV-electrolysis, PV-PEC and PEC systems possess a higher degree of integration between the light harvesting and catalysis components, which results in different operating conditions. Such tandem, or triple-junction devices can perform unassisted water splitting or $CO_2$ reduction at lower current densities of 0.1-10 mA $cm^{-2}$, whereas gas and liquid products are formed across the entire photoactive surface. This arrangement is beneficial for highly selective molecular or biological catalysts which require a benign pH, or may degrade at higher current densities. However, product collection over a wider area can be challenging in the absence of flow systems, as stagnant bubbles can induce optical losses or prevent diffusion processes near the catalytic surface, thereby contributing to mass transport limitations [280].

While traditional PEC systems involve a direct semiconductor-electrolyte interface, a significant increase in photocurrent and photovoltage can be attained by introducing photovoltaic concepts, including hole and electron transport layers for selective charge extraction, or by taking into account the band alignment of the individual components. These buried heterojunction systems, also known as buried PV systems, have shown great promise over the past few years, improving the activity of conventional PEC materials including silicon [16], $Cu_2O$ [286], or III-V semiconductors [287], while allowing new semiconductors like lead halide perovskites to operate in aqueous solutions [288].

Despite their promise, the performance of most PEC prototypes is only traced for several hours under laboratory conditions, whereas complex fabrication procedures may restrict photoactive areas below 1 $cm^2$. Accordingly, further progress is needed in terms of stability, scalability and reactor design, which can enable real-world benchmarking or detailed technoeconomic analyses. For instance, the scalability of PEC systems is often limited by resistive losses through the conductive FTO or ITO substrate, which can be addressed by introducing metal fingers, similar to those found in commercial PV cells (figure 21c) [289]. However, large area electrodes can also induce significant pH gradients across the solution, which contribute to additional ohmic losses [282]. Hence, an appropriate reactor design including ion exchange membranes or solution flow is just as essential for sustaining the small prototype performance on a larger scale.

In terms of stability, further progress is required on a material and device level. While robust oxide materials like $BiVO_4$, $WO_3$, or $Fe_2O_3$ are well established for PEC photoanodes, their wide bandgaps limit the total solar-to-hydrogen (STH) efficiency of PEC tandem devices to <5% [280]. In contrast, the interfaces of bare Si, III-V and perovskite semiconductors degrade within seconds when irradiated in solution, hence materials that balance a wider spectral absorption and intrinsic moisture stability still need to be developed. Alternatively, an increased stability under operation requires a rational design of buried-PV device structures, placing a strong focus on the hydrophobicity of functional layers and encapsulants [290].

Technologies such as light concentration can offset the higher costs associated with PV-electrolysis. The resulting heat can accelerate the reaction kinetics of electrocatalysis [291], whereas the temperature difference can be harvested by thermoelectric modules to provide an additional voltage [292], [293]. However, sunlight tracking capabilities are needed to sustain the performance throughout the day (figure 21a) [294], whereas diffuse light encountered on overcast days poses challenges, limiting the available irradiation to about 0.1 sun [295].

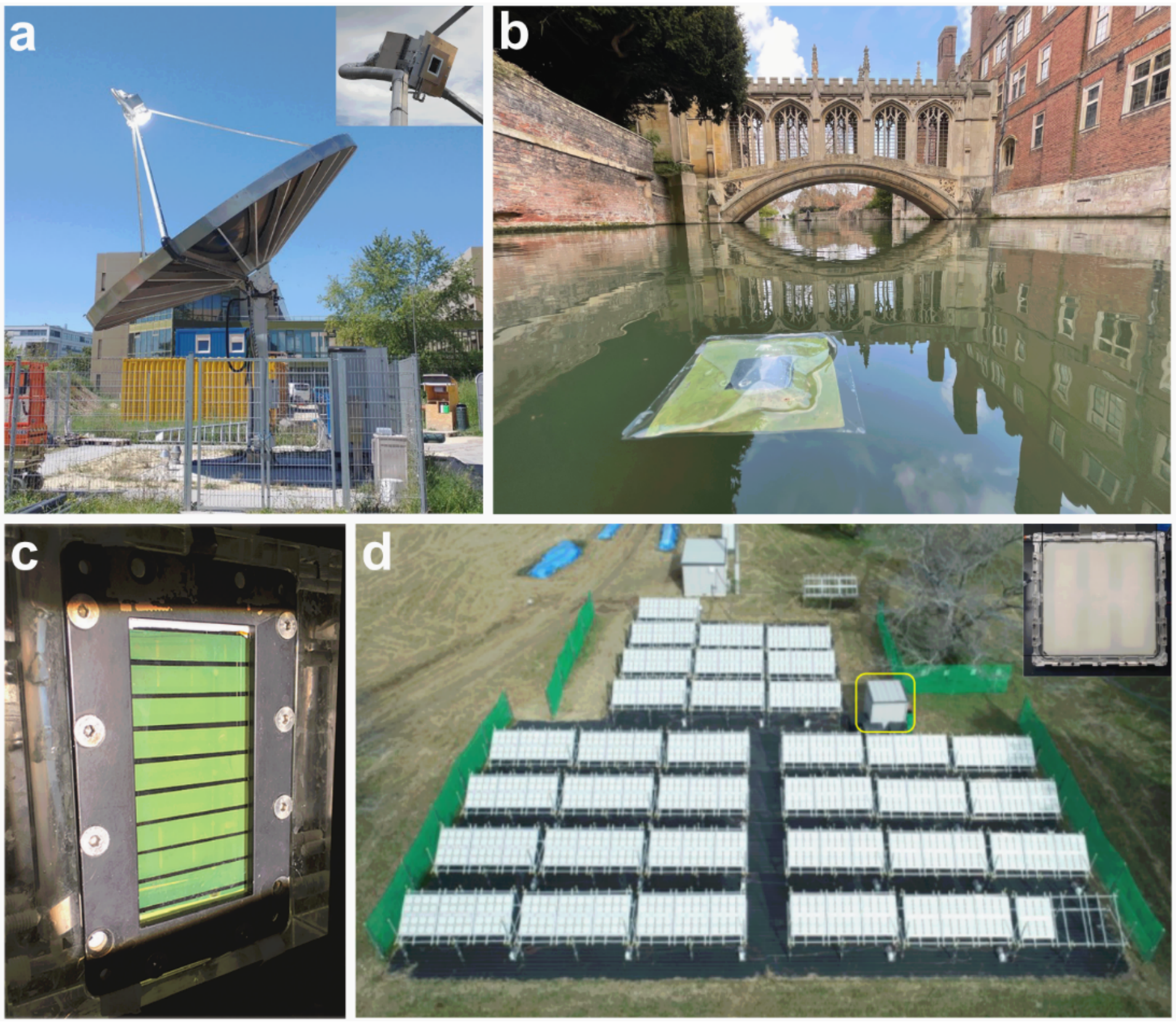

**Figure 21.** Examples of state-of-the-art systems for solar fuel production. (a) PV-electrolysis under 800× sunlight concentration using a 7 m-diameter dual-axis tracking solar parabolic dish. Light is focused on a triple-junction III–V PV module (see inset), which supplies power for two polymer electrolyte membrane electrolysers. Reproduced under the terms of a CC-BY license from Ref. [294]. Copyright 2023, The Authors. (b) Lightweight artificial leaf floating along River Cam in Cambridge, United Kingdom. The 100 $cm^2$ PEC device consists of a buried-PV perovskite photocathode for $H_2$ production and a $BiVO_4$ photoanode for $O_2$ evolution. Reproduced with permission from Ref. [11]. Copyright 2022, Springer Nature. (c) Tandem PV-PEC device assembled from a dual $BiVO_4$ photoanode with Ni metal fingers and two serially connected Si PV cells. Reproduced under the terms of a CC-BY-3.0 license from Ref. [289]. Copyright 2019, The Authors. (d) A 100 $m^2$ solar hydrogen farm contains 1600 photocatalytic sheets connected to a gas separating unit (highlighted in yellow). Each panel reactor hosts one 25×25 $cm^2$ $SrTiO_3$:Al PC sheet (see inset). Reproduced with permission from Ref. [296]. Copyright 2021, Springer Nature.

As outlined above, PEC systems still require further advances in terms of material development and overall system design. These challenges can be addressed using a combination of unconventional approaches and modern fabrication techniques, which make use of the particular benefits of integrated PEC panels. For instance, solar fuel panels may take advantage of the modularity of "artificial leaves". In contrast to PV modules, each standalone PEC unit operates independently from the others, hence if one degrades, the rest will continue to work. This design principle can also be applied on individual electrodes, as multiple-pixel photocathodes based on nanowire GaAs heterojunctions or BiOI solar cells can sustain the superior photovoltage and cumulative photocurrent of small-area photoactive units operating in parallel [10], [297].

In terms of PEC device stability, advances can be obtained by introducing hydrophobic layers into the device structure, in the form of oxide-based charge selective and passivating layers, or polymer encapsulants. For instance, a conductive graphite paste encapsulant increases the moisture stability of buried-junction lead halide perovskite and BiOI photocathodes from minutes to weeks, whereas $NiO_x$ and poly[bis(4-phenyl)(2,4,6-trimethylphenyl)amine] (PTAA) hole selective layers further improve performance through optimal band alignment [10], [11], [288].

The size and reliability of current prototypes are limited by manual deposition steps. Hence, device structures are needed that are compatible with modern, roll-to-roll fabrication techniques, similar to those encountered for thin-film PV cells. These can employ similar plastic and metal foils as conductive substrates, resulting in thin PEC devices with additional functionality. For instance, such lightweight perovskite-$BiVO_4$ "artificial leaves" were able to float along the River Cam, showcasing the potential applications of solar fuel farms (Fig. 12.2b). The high product rate per gram of these lightweight devices also stands out as an alternative metric to evaluate solar fuel production systems [11]. If PEC devices consisting of earth-abundant elements become as inexpensive, easy to manufacture and recycle as natural leaves, then lower STF efficiencies and stabilities become sufficient to obtain a return on investment.

Moving forward, automated setups are required to obtain the high-throughput, layer-by-layer deposition of the full PV device stack and corresponding buried-junction photoelectrodes, which will be essential for mass production and commercial benchmarking [298], [299]. This increased reproducibility of the deposition methods will also have a positive effect on the stability of devices, minimizing pinholes or defects which can promote internal interfacial degradation or provide moisture infiltration pathways. Ultimately, the product collection and separation of explosive $H_2$, $O_2$ mixtures may pose concerns, however recent studies have shown that circulating systems coupled to gas separating units can work effectively on a 100 $m^2$ scale (figure 21d) [296].

In conclusion, recent efforts showcase that PV concepts and materials can contribute significantly to the performance of solar fuels systems, from wired PV-electrolysers to integrated "artificial leaf" panels. This rapid progress demonstrates the potential of solar fuel technologies towards real-world applications, which is underpinned by fundamental and engineering studies on charge transfer, material degradation, device and reactor design, encapsulation methods, or scalability. This synergy in interests between photovoltaics and solar fuel communities can provide broad opportunities for interdisciplinary collaborations. Moving forward, solar fuel technologies can take advantage of the established light harvesting expertise within the PV community, while PV systems can benefit from the integrated designs of solar fuels systems, which aim to minimize costs and expand functionality. Hence, we welcome the PV community into exploring these questions together.

**Acknowledgements**

J.B. acknowledges support from EPSRC ECR Fellowship NoRESt (no. EP/S03711X/1), EPSRC Postdoctoral Fellowship (no. EP/V035770/1) and TREFCO (no. P/W019167/1). V.A. acknowledges funding from St John's College Cambridge (Title A Research Fellowship), the Winton Programme for the Physics of Sustainability, and the Royal Society of Chemistry (no. R23-5618510760). R.S.B was supported by the Royal Academy of Engineering under the Research Fellowship scheme (no. RF\201819\18\38). This work was supported by EPSRC (no. EP/V038605/1). P.K. thanks EPSRC for funding for a Postdoctoral Fellowship (no. EP/V035770/1) and Linacre College for the award of a Junior Research Fellowship.

## 14. Key Points from Roadmap

Over the past three years, photovoltaics have continued their rapid advance, both in terms of deployment and emerging technology development. Photovoltaics have proven to not just be a novel approach to producing electricity, but a central pillar in the world's transition to net-zero, and already provides the lowest-cost electricity in many parts of the world. This roadmap has discussed the steps that are needed to take photovoltaic technologies forward and grow in capacity. Some of the key takeaway messages are:

- *Photovoltaics research is far from being completed, and many exciting new opportunities are emerging, some of which will spur innovation in other technologies*. The new solar absorbers being developed are finding applications in solar fuels, memristors, radiation detectors, and light-emitting diodes, among many other applications. As such, photovoltaic technology development not only impacts on clean energy generation, but also contributes to green hydrogen production, low loss electronics, and many other sectors.

- *Increased investment in solar photovoltaics research and manufacturing is critical to accelerate the global installed capacity of PVs, and to overcome the barriers towards commercialization of emerging technologies that can deliver even lower LCOEs and new applications to market.* Specifically, investment is required in the following areas: 1) manufacturing facilities for established PV technologies, as well as facilities that can translate new technologies from the lab-scale to module-level, 2) research which investigates and develops ways to further improve the performance of established and emerging PVs, as well as 3) facilities, beyond manufacturing tools, required to support their development, namely advanced metrology instruments and exascale computing facilities. Although c-Si PVs are a mature technology and dominate the outdoor PV market, their deployment needs to increase by over an order of magnitude over the next two decades to meet net-zero targets [300]. It will be critical to not only expand the manufacturing of existing c-Si PV technologies, but to also maintain continuous improvements in module efficiency (0.2-0.5% absolute PCE per year) and stability (from 25-30 years currently to 40 years) in order to further continue to reduce the LCOE. Cell designs need to improve and reduce their dependence on scarce and expensive elements, such as Ag and In. Beyond c-Si, some thin film technologies have significant potential to achieve lower LCOEs. However, this will require investment to bring them to the multi-GW level of production capacity. With organic PVs and metal-halide perovskites, currently this technology is in the process of being transferred from lab to market. A variety of large-scale manufacturing methods being explored for both single-junction and tandem photovoltaics, and the first 100 $MW_p$ commercial production lines have come online recently [301], [302].

- *Terawatt-scale solar power is crucial to meeting carbon-neutrality targets across the world. For the UK, engaging with this challenge presents a strategic opportunity to take part in a growing multibillion pound supply chain.* The advancement and deployment of photovoltaic materials and technologies will contribute substantially to the generation of electricity in coming decades. For example, to achieve carbon neutrality, integrated assessment models predict that >7 $TW_p$ of PVs must be installed worldwide over the next 30 years [17], equivalent to a growth of over 8-fold. Such assessment models are based on projected changes and interactions between energy, economy, land, and climate. Although the potential of solar PVs has been underestimated, current technological progress and grid penetration show that it is possible for global solar capacity to grow to as much as 70 $TW_p$, equivalent to over 60 times more installed capacity than today [9], [57], [300]. In the UK, solar capacity is expected to grow by more than six-fold to >90 $GW_p$ [22], which is a conservative estimate given the sustained growth of PV installations [303]. The expertise that the UK holds in photovoltaic materials provides a competitive advantage to penetrate and expand a variety of segments of solar energy deployment, such as in perovskite-silicon tandem PV development and manufacturing conducting glass substrates. Linking the research and development of technologies to industrial and market needs can ensure that we exploit the local expertise and create high-tech businesses, create opportunities to generate a new manufacturing base and new IP, and increase exports. This is also supported by a global effort to foster effective and resilient decentralized supply chains, while simultaneously strengthening the energy security of individual nations.

- *Emerging photovoltaic technologies offer enormous potential to address some of the wider challenges facing society.* Inorganic and hybrid thin film PVs, as well as dye-sensitized PVs can enable the Internet of Things, infrastructure and agriculture to become more sustainable and resource-efficient through use of indoor and building-integrated PVs, as well as agrivoltaics. For applications in some of these sectors, c-Si is either not suitable or no longer the best option. For such applications, emerging inorganic, hybrid and organic materials have suitable properties, such as translucency, low dark currents, mechanical flexibility and tuneable absorption profiles. These applications unlock billions of square metres of surfaces for solar energy harvesting, potentially relaxing requirements on land use for photovoltaics. Furthermore, as missions to space are increasingly constrained by cost pressures, the lower cost per unit mass offered by thin film photovoltaics over traditional III-V semiconductors becomes appealing. Metal-halide perovskites particularly have shown promise because of their high efficiencies and radiation hardness.

- *The development and international adoption of standards for emerging PV technologies is critical to support their sustained development, faster commercialization and deployment.* The use of standards and ability to certify performance and stability is critical for market penetration. As emerging thin film technologies are commercialized, it is critical to understand their failure modes under real outdoor conditions to inform the design of effective accelerated degradation protocols, and deliver reliable and repeatable power rating methods within a short test duration that reflects real long-term field performance. As the indoor PVs sector rapidly increases in size and importance, it is crucial to develop and adopt standard test protocols (*e.g.*, light spectrum standardization). These will ensure comparability between groups and that increases in efficiency truly come about because of an improvement in technology rather than circumstantial factors, such as an improved match between the absorption profile of a material and the light source. This is also important for outdoor photovoltaic technologies in general, which are tested in the lab under AM1.5G illumination, but which can differ markedly from real-world light spectra, depending on where the devices are deployed and the time of day/season. Understanding how such devices behave under real-world conditions is therefore important in order to enable the development of approaches to rigorously test this in the laboratory.

- *More efforts and legislation are needed to close the lifecycle in photovoltaic products, and linking device end-of-life with the start of the lifecycle.* Currently, the cheapest option is to put used devices to landfill [27]. A sustainable technology, and especially given challenges with elements of limited supply (*e.g.*, Ag or Te), requires legislation for requiring photovoltaic modules to be recycled, (such as the WEEE regulation in the UK and Europe) [304] and the valuable elements extracted to be reused as precursors. Increasing material circularity can help reduce the environmental impacts associated with mining and manufacturing processes, as well as the cost. This will be crucial as the PV industry has, as of 2022, entered terawatt-scale global installation and future innovation must ensure material sustainability, especially in modules containing toxic Pb, Cd or Te. With emerging photovoltaic technologies, there is an opportunity to design for a circular economy from the outset following the lead of First Solar who can recover 90% of their modules for reuse.

- *More targeted collaboration and data sharing are needed, both between academia and industry, as well as between groups working on different technologies*. Current and future challenges in photovoltaic technology development are becoming more complex, and require a balance of skills in device engineering, materials development, simulations and machine learning in order to create optimal solutions in a timely manner. Across all technologies, it is important that researchers are addressing problems that industry faces in improving the performance, durability and value of photovoltaic materials. Funding mechanisms that encourage this will be important, as are mechanisms to encourage open sharing of data between groups and the resources to do so. This will accelerate the development of standards, and also the identification of solutions through the application of machine learning. International collaboration will be necessary for knowledge and skills transfer and ensuring relevant research in the latest technologies.
- *Investment in training and skills development is necessary to address the challenges in the field of PV technology development and accelerate deployment.* These skills span from new manufacturing and instrumentation development and deployment, understanding the process-property-structure-performance relationships of materials, through to computational abilities to predict the behavior of advanced materials and corresponding devices, or evaluate the performance of solar energy systems under real-life conditions. As such, these skills are highly transferrable and overlap with those required for developing other technologies, for example, in the field of semiconductors and power electronics.

- *Considering PV deployment at the systems level, rather than only at the individual modules level, is essential.* It is essential to consider the remainder of the ecosystem around photovoltaics, including the materials used for the inverter and encapsulant, as well as supply chains for the important materials required for photovoltaic devices, and the security of these supplies. The backlog for connecting new utility-scale PV plants to the electricity grid is considerable in many countries, and the cost of electricity for the consumer is much higher than the generation costs because of steadily increasing grid charges. Newly installed PV systems are often coupled with battery storage and in the future with electrolysis, mini-modules are integrated into IoT devices, and applications like agrivoltaics are inherently multipurpose. All of these challenges require an approach that does not stop at the PV module but accounts for the role of the module in a wider system. Such a systems approach might alter technical requirements and feed back into optimization targets from module to material design.
- *Policy and regulation for new applications.* Legislation often struggles to keep pace with new technologies and applications which can mean, for example solar PV is included as consumer electronics in the UK Waste Electronics regulations despite having a 25-30 year lifespan [304]. Multipurpose PV installations, such as building integrated/applied (including agrivoltaics), which emerging PV technologies often serve, require a different set of regulations, incentives, and re-financing schemes beyond feed-in compensation. While policy and regulation are crucial for the successful rollout of new PV technologies and applications and deserve great attention, they are beyond the scope of this technical roadmap.

Overall, research and innovation in photovoltaic materials, devices and systems are clearly active and rapidly growing. Fully realizing the future potential of photovoltaics will require increased engagement and partnership between policymakers, investors, industry and academic groups. Successfully bringing together the technological, economic, policy and finance components of this complex challenge will deliver many benefits in terms of economic prosperity, mitigating the effects of climate change, and improving energy security.

**Supplementary Materials for**

# Roadmap on Photovoltaic Absorber Materials for Sustainable Energy Conversion

*James C. Blakesley,*[1,†] *Ruy S. Bonilla,*[2,†] *Marina Freitag,*[3,†] *Alex M. Ganose,*[4, †] *Nicola Gasparini,*[4, †] *Pascal Kaienburg,*[5,†] *George Koutsourakis,*[1,†] *Jonathan D. Major,*[6, †] *Jenny Nelson,*[7,8 †] *Nakita K. Noel,*[5, †] *Bart Roose,*[9, †] *Jae Sung Yun,*[11, †] *Simon Aliwell,*[12] *Pietro P. Altermatt*[2,13] *Tayebeh Ameri,*[14,15] *Virgil Andrei,*[16,23] *Ardalan Armin,*[17] *Diego Bagnis,*[18] *Jenny Baker,*[19] *Hamish Beath,*[7,8] *Mathieu Bellanger,*[12] *Philippe Berrouard,*[20] *Jochen Blumberger,*[21] *Stuart A. Boden,*[22] *Hugo Bronstein,*[16,23] *Matthew J. Carnie,*[24,25] *Chris Case,*[26] *Fernando A. Castro,*[1] *Yi-Ming Chang,*[27] *Elmer Chao,*[28] *Tracey M. Clarke,*[29] *Graeme Cooke,*[30] *Pablo Docampo,*[30] *Ken Durose,*[6] *James R. Durrant,*[4,25] *Marina R. Filip,*[5] *Richard H. Friend,*[23] *Jarvist M. Frost,*[4] *Elizabeth A. Gibson,*[2] *Alexander J. Gillett,*[23] *Pooja Goddard,*[31] *Severin N. Habisreutinger,*[26] *Martin Heeney,*[4] *Arthur D. Hendsbee,*[20] *Louise C. Hirst,*[23,32] *M. Saiful Islam,*[2] *K. D. G. Imalka Jayawardena,*[11] *Michael B. Johnston,*[5] *Matthias Kauer,*[12] *Jeff Kettle,*[33] *Ji-Seon Kim,*[34] *Dan Lamb,*[35] *David Lidzey,*[36] *Jihoo Lim,*[11,37] *Roderick MacKenzie,*[38] *Nigel Mason,*[39] *Iain McCulloch,*[40] *Keith P. McKenna,*[41] *Sebastian B. Meier,*[42] *Paul Meredith,*[17] *Graham Morse,*[43] *John D. Murphy,*[44] *Chris Nicklin,*[45] *Paloma Ortega-Arriaga,*[7,8] *Thomas Osterberg,*[46] *Jay B. Patel,*[47] *Anthony Peaker,*[48] *Moritz Riede,*[5] *Martyn Rush,*[49] *James W. Ryan,*[50] *David O. Scanlon,*[29,51] *Peter J. Skabara,*[30] *Franky So,*[52,53] *Henry J. Snaith,*[5] *Ludmilla Steier,*[10] *Jarla Thiesbrummel,*[5] *Alessandro Troisi,*[54] *Craig Underwood,*[55] *Karsten Walzer,*[56] *Trystan Watson,*[25] *J. Michael Walls,*[57] *Aron Walsh,*[58] *Lucy D. Whalley,*[59] *Benedict Winchester,*[7,8] *Samuel D. Stranks*[9,*] *and Robert L. Z. Hoye*[10,58,*]

1. National Physical Laboratory, Hampton Rd, Teddington, UK

2. Department of Materials, University of Oxford, Oxford OX1 3PH, UK

3. School of Natural and Environmental Sciences, Bedson Building, Newcastle University, NE1 7RU, Newcastle upon Tyne, UK

[4.] Department of Chemistry, Imperial College London, 80 Wood Lane, London W12 0BZ, UK

[5.] Department of Physics, University of Oxford, Clarendon Laboratory, Parks Road, Oxford OX1 3PU, UK

[6.] Stephenson Institute for Renewable Energy, Department of Physics, University of Liverpool, Chadwick Building, Peach Street, Liverpool L69 7ZF, United Kingdom

[7.] Department of Physics and Centre for Processable Electronics, Imperial College London, London SW7 2AZ, UK

[8.] Grantham Institute – Climate Change and the Environment, Imperial College London, London SW7 2AZ, UK

[9.] Department of Chemical Engineering and Biotechnology, University of Cambridge, Phillipa Fawcett Road, Cambridge CB3 0AS, United Kingdom

[10.] Inorganic Chemistry Laboratory, Department of Chemistry, University of Oxford, South Parks Road, Oxford OX1 3QR, United Kingdom

[11.] School of Computer Science and Electronic Engineering, Advanced Technology Institute (ATI), University of Surrey, Guildford GU2 7XH, United Kingdom

[12.] Lightricity Ltd., Edmund Halley Road, Oxford OX4 4GB, United Kingdom

[13.] Trina Solar Co Ltd, Changzhou, China.

[14.] Institute for Materials and Processes, School of Engineering, University of Edinburgh, Sanderson Building, Robert Stevenson Road, Edinburgh EH9 3FB, UK

[15.] Chair for Composite Materials, Department of Materials Science, Faculty of Engineering, Christian-Albrechts-Universität zu Kiel, Kaiserstrasse 2, 24143 Kiel, Germany

[16.] Yusuf Hamied Department of Chemistry, University of Cambridge, Cambridge CB2 1EW, UK

[17] Centre for Integrative Semiconductor Materials (CISM), Department of Physics, Swansea University Bay Campus, Swansea SA1 8EN, UK

[18] Oninn – Centro de Inovação, Av. José Candido da Silveira 2000 - Horto Florestal, Belo Horizonte - MG, 31035-536, Brazil

[19] Faculty of Science and Engineering, Swansea University, UK

[20] Brilliant Matters, 237 Rue de Liverpool, Saint-Augustin-de-Desmaures, QC G3A 2C8, Canada

[21] Department of Physics and Astronomy and Thomas Young Centre, University College London, Gower Street, London WC1E 6BT, UK

[22] School of Electronics and Computer Science, University of Southampton, United Kingdom

[23] Cavendish Laboratory, University of Cambridge, JJ Thomson Ave, Cambridge CB3 0HE, UK

[24] Department of Materials Science and Engineering, Swansea University, Bay Campus, Fabian Way, Crymlyn Burrows, Swansea SA1 8EN, UK

[25] SPECIFIC, Swansea University, Bay Campus, Swansea SA1 8EN, UK

[26] Oxford Photovoltaics Ltd., Unit 7-8, Oxford Pioneer Park, Mead Road, Oxford OX5 1QU, United Kingdom

[27] Raynergy Tek Incorporation, 60, Park Ave. 2, Hsinchu Science Park, Hsinchu 30844, Taiwan

[28] Luminescence Technology Corp., Hsinchu 30076, Taiwan

[29] Department of Chemistry, University College London, Christopher Ingold Building, London WC1H 0AJ, UK

[30] School of Chemistry, University of Glasgow, Joseph Black Building, University Place, Glasgow G12 8QQ, United Kingdom

[31] School of Science, Loughborough University, LE11 3TU, UK

[32] Department of Materials Science and Metallurgy, University of Cambridge, 27 Charles Babbage Road, Cambridge CB3 0FS, UK

[33] James Watt School of Engineering, University of Glasgow, Glasgow G12 8QQ, UK

[35] Oxide and Chalcogenide Facility, Centre for Integrative Semiconductor Materials, Faculty of Science and Engineering, Swansea, University, SA1 8EN United Kingdom

[36] Department of Physics and Astronomy, The University of Sheffield, Hicks Building, Hounsfield Road, Sheffield S3 7RH, UK

[37] Australian Centre for Advanced Photovoltaics (ACAP), School of Photovoltaic and Renewable Energy Engineering, University of New South Wales, Sydney, NSW 2052, Australia

[38] Department of Engineering, Durham University, Lower Mount Joy, South Road, Durham DH1 3LE, UK

[39] Formerly BP Solar and PV Consulting, United Kingdom.

[40] Department of Chemistry, University of Oxford, Mansfield Road, Oxford OX1 3QR, UK

[41] School of Physics, Engineering and Technology, University of York, Heslington, York YO10 5DD, UK

[42] ASCA GmbH & Co. KG Steigweg 24, 97318 Kitzingen, Germany

[43] Merck Chemicals Ltd, Chilworth Technical Centre, University Parkway, Southampton SO16 7QD, UK

[44] School of Engineering, University of Warwick, Coventry CV4 7AL, United Kingdom

[45.] Diamond Light Source, Harwell Science and Innovation Campus, Didcot, Oxfordshire, OX11 0DE, UK

[46.] Epishine AB, 582 13 Linköping, Sweden

[47.] Department of Physics, King's College London, Strand, London WC2R 2LS, United Kingdom

[48.] School of Electrical Engineering and Electronics, University of Manchester, United Kingdom.

[49.] Polysolar Ltd, Aurora BAS, High Cross, Meadingley Road, Cambridge, CB3 0ET, UK

[50.] Department of Chemistry, Swansea University, Swansea SA2 8PP, UK

[51.] Thomas Young Centre, University College London, Gower St, London, UK

[52.] Materials Science and Engineering, North Carolina State University, Raleigh, NC, 27695 USA

[53.] Organic and Carbon Electronics Laboratories (ORaCEL), North Carolina State University, Raleigh, NC, 27695 USA

[54.] Department of Chemistry and Materials Innovation Factory, University of Liverpool, Liverpool L69 7ZD, UK

[55.] Surrey Space Centre, School of Computer Science and Electronic Engineering, University of Surrey, Guildford, Surrey, GU2 7XH, United Kingdom

[56.] Heliatek GmbH, Treidlerstrasse 3, 01139 Dresden, Germany

[57.] Centre for Renewable Energy Systems Technology (CREST), Loughborough University, Loughborough LE11 3TU, United Kingdom

[58.] Department of Materials, Imperial College London, Exhibition Road, London SW7 2AZ, United Kingdom

[59.] Department of Maths, Physics and Electrical Engineering, Northumbria University, Ellison Place, Newcastle Upon Tyne, NE1 8ST, UK

† These authors contributed equally to this roadmap article. Arranged alphabetically by surname. The remaining co-authors are also arranged alphabetically by surname

**Email:** sds65@cam.ac.uk (S.D.S), robert.hoye@chem.ox.ac.uk (R.L.Z.H.)

**Supplementary Table 1.** Champion reported power conversion efficiencies of lab- and commercial-scale indoor photovoltaics. WLED = white light emitting diode, FL = fluorescent light tube

| Year | Technology | Lab or commercial | PCE (%) | Illumination conditions | Ref. |
|---|---|---|---|---|---|
| 2012 | DSSC | Lab | 12.03 | 0.05 Sun, 5 mW $cm^{-2}$ | [1] |
| 2016 | DSSC | Lab | 19.5 | WLED, 350 lx | [2] |
| 2017 | DSSC | Lab | 28.9 | FL, 1000 lx | [3] |
| 2018 | DSSC | Lab | 32 | FL, 1000 lx | [4] |
| 2020 | DSSC | Lab | 34 | FL, 1000 lx | [5] |
| 2023 | DSSC | Lab | 38 | FL, 1000 lx | [6] |
| 2015 | OPV | Lab | 11.6 | WLED, 890 lx | [7] |
| 2016 | OPV | Lab | 21.04 | WLED, 20.5 mW $cm^{-2}$ | [8] |
| 2017 | OPV | Lab | 28.1 | FL, 1000 lx | [9] |
| 2020 | OPV | Lab | 31 | WLED (3000 K), 1650 lx | [10] |
| 2015 | LHP | Lab | 27.4 | FL, 1000 lx | [11] |
| 2018 | LHP | Lab | 35.2 | FL, 1000 lx | [12] |
| 2019 | LHP | Lab | 36.2 | FL, 1000 lx | [13] |
| 2020 | LHP | Lab | 37.2 | WLED (6500 K), 1000 lx | [14] |
| 2021 | LHP | Lab | 40.2 | FL, 1000 lx | [15] |
| 2023 | LHP | Lab | 41.2 | WLED (2956 K), 1062 lx | [16] |
| 2016 | III-V | Lab | 19.4 | WLED, 580 lx | [17] |
| 2011 | III-V | Commercial | 12 | FL, 0.001-0.003 Sun | [18] |
| 2017 | III-V | Commercial | 21 | FL, 1000 lx | [19] |
| 2015 | a-Si:H | Commercial (Sanyo) | 9.2 | FL, 200 lx | [20] |

N.B.: 1000 lux WLED or FL is typically 0.3-0.4 mW $cm^{-2}$ power density, but depends on the exact spectrum